\documentclass[reprint,superscriptaddress,amsmath,amssymb,aps,prl,floatfix]{revtex4-2}
\usepackage{graphicx}% Include figure files
\usepackage{dcolumn}% Align table columns on decimal point
\usepackage{bm}% bold math
\usepackage{lipsum}  % space filler
\usepackage{xcolor}
\usepackage{braket}% colored text
\usepackage{hyperref}% add hypertext capabilities
\usepackage{ulem} % for strikethrough text
\usepackage{booktabs}
\usepackage{mathtools}
\usepackage{amssymb}
\usepackage{datetime}
\newdate{date}{26}{08}{2024} % hardcode date for arxiv

\renewcommand{\thefigure}{\textbf{\arabic{figure}}}
\renewcommand{\figurename}{\textbf{Fig.}}

\makeatletter
\def\maketitle{
\@author@finish
\title@column\titleblock@produce
\suppressfloats[t]}
\makeatother

\AtBeginDocument{\let\oldcontentsline\contentsline}
\newcommand{\notoccontentsline}[4]{\oldcontentsline{}{}{}{}}
\newcommand{\droptocpage}{\addtocontents{toc}{\let\protect\contentsline\protect\notoccontentsline}}
\newcommand{\incltocpage}{\addtocontents{toc}{\let\protect\contentsline\protect\oldcontentsline}}

\begin{document}
\newcommand{\Rev }[1]{{\color{blue}{#1}\normalcolor}} % Revision
\newcommand{\diego }[1]{{\color{purple}{Diego: #1}\normalcolor}} % Revision
\newcommand{\Com}[1]{{\color{red}{#1}\normalcolor}} %Comment
\newcommand{\ketbra}[2]{\ket{#1}\hspace{-0.125cm}\bra{#2}}

\title{A dissipation-induced superradiant transition in a strontium cavity-QED system}

\author{Eric Yilun Song}
\affiliation{JILA, NIST, and Department of Physics, University of Colorado, Boulder, CO, USA}

\author{Diego Barberena}
\affiliation{JILA, NIST, and Department of Physics, University of Colorado, Boulder, CO, USA}
\affiliation{T.C.M. Group, Cavendish Laboratory, University of Cambridge, J.J. Thomson Avenue, Cambridge CB3 0HE, UK}
\affiliation{Center for Theory of Quantum Matter, University of Colorado, Boulder, CO, USA}

\author{Dylan J. Young}
\affiliation{JILA, NIST, and Department of Physics, University of Colorado, Boulder, CO, USA}

\author{Edwin Chaparro}
\affiliation{JILA, NIST, and Department of Physics, University of Colorado, Boulder, CO, USA}
\affiliation{Center for Theory of Quantum Matter, University of Colorado, Boulder, CO, USA}

\author{Anjun Chu}
\affiliation{JILA, NIST, and Department of Physics, University of Colorado, Boulder, CO, USA}
\affiliation{Center for Theory of Quantum Matter, University of Colorado, Boulder, CO, USA}

\author{\\Sanaa Agarwal}
\affiliation{JILA, NIST, and Department of Physics, University of Colorado, Boulder, CO, USA}
\affiliation{Center for Theory of Quantum Matter, University of Colorado, Boulder, CO, USA}

\author{Zhijing Niu}
\affiliation{JILA, NIST, and Department of Physics, University of Colorado, Boulder, CO, USA}

\author{Jeremy T. Young}
\affiliation{Institute of Physics, University of Amsterdam, 1098 XH Amsterdam, the Netherlands}
\affiliation{JILA, NIST, and Department of Physics, University of Colorado, Boulder, CO, USA}
\affiliation{Center for Theory of Quantum Matter, University of Colorado, Boulder, CO, USA}

\author{Ana Maria Rey}
\affiliation{JILA, NIST, and Department of Physics, University of Colorado, Boulder, CO, USA}
\affiliation{Center for Theory of Quantum Matter, University of Colorado, Boulder, CO, USA}

\author{James K. Thompson}
\affiliation{JILA, NIST, and Department of Physics, University of Colorado, Boulder, CO, USA}

\usdate
\date{\displaydate{date}}

\begin{abstract}
In cavity quantum electrodynamics (QED), emitters and a resonator are coupled together to enable precise studies of quantum light-matter interactions. Over the past few decades, this has led to a variety of quantum technologies such as more precise inertial sensors, clocks, memories, controllable qubits, and quantum simulators~\cite{Blais2021,Farokh2021,Schlawin2022,Bao2012,Greve2022, Kjaergaard2020, Young2024BCS}. Furthermore, the intrinsically dissipative nature of cavity-QED platforms makes them a natural testbed for exploring driven-dissipative phenomena in open quantum systems as well as equilibrium and non-equilibrium phase transitions in quantum optics~\cite{Baumann2010,Klinder2015,Kroeze2018,Zhiqiang2017,Marino2022}. One such model, the so-called cooperative resonance fluorescence (CRF) model, concerns the behavior of coherently driven emitters in the presence of collective dissipation (superradiance). Despite tremendous interest~\cite{Senitzky1972,Drummond1978,Narducci1978,Puri1979,Carmichael1980,Ferioli2023}, this model has yet to be realized in a clean experimental system.  Here we provide an observation of the continuous superradiant phase transition predicted in the CRF model using an ensemble of ultracold $^{88}$Sr atoms coupled to a driven high-finesse optical cavity on a long-lived optical transition. Below a critical drive, atoms quickly reach a steady state determined by the self-balancing of the drive and the collective dissipation. The steady state possesses a macroscopic dipole moment and corresponds to a superradiant phase. Above a critical drive strength, the atoms undergo persistent Rabi-like oscillations until other decoherence processes kick in. In fact, our platform also allows us to witness the change of this phase transition from second to first order induced by single-particle spontaneous emission, which pushes the system towards a different steady state. Our observations are a first step towards finer control of driven-dissipative systems, which have been predicted to generate quantum states that can be harnessed for quantum information processing and in particular quantum sensing~\cite{Pezze2018,Farokh2021,Lee2014,Young2024,Ye2024}.

\end{abstract}

\maketitle

\begin{figure*}[!t]
\centerline{\includegraphics[scale=1]{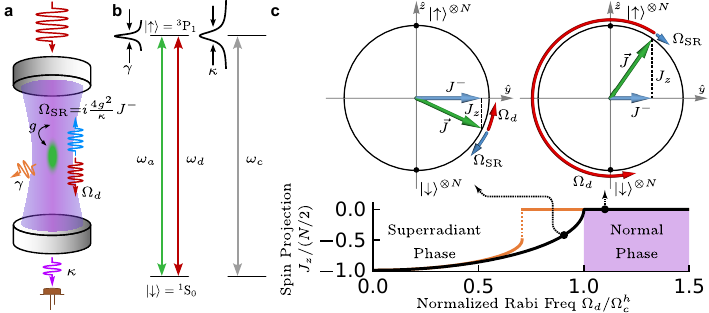}}
\caption{
    \textbf{Experimental schematics and the phase diagram of the CRF model.} 
        \textbf{a,} $^{88}\mathrm{Sr}$ atoms (green) are trapped in an optical lattice (omitted in the figure for simplicity) supported by a high-finesse optical cavity (grey mirrors) after laser cooling. The atoms couple to the cavity with a r.m.s.~single-photon Rabi frequency $2g$. The excited state has a spontaneous decay rate $\gamma$. The cavity has a FWHM linewidth of $\kappa$. We inject $689~$nm light into the cavity, which would establish a drive field (red) with Rabi frequency $\Omega_{d}$ inside the cavity. In response, the atoms emit a superradiant field (blue) at the same wavelength, characterized by Rabi frequency $\Omega_{\mathrm{SR}} = i\tfrac{4g^2}{\kappa}\langle {\hat{J}}^{-} \rangle$. The coherent addition of the two fields (purple) leaks through the cavity mirror at a rate $\kappa$ and is monitored on a detector (brown). 
        \textbf{b,} Energy level diagram of experimentally relevant frequency scales. The optical cavity resonance frequency $\omega_c$ and the atomic drive frequency $\omega_d$ are both resonant with the $^1\textrm{S}_0 -^3\!\text{P}_1 (m_J=0)$ transition. 
        \textbf{c,} Phase diagram of the CRF Model. The upper two panels show cuts of the Bloch sphere illustration in the y-z plane for the superradiant phase (left) and the normal phase (right), with the dashed lines indicating the representative points on the phase diagram on the lower row. The external drive $\Omega_d$ (red) acts like a torque that tries to rotate the collective Bloch vector $\Vec{J}$ (green) counter-clockwise. The superradiant field from the atoms $\Omega_{\mathrm{SR}}$ (blue) acts like a torque, with strength proportional to the collective atomic dipole moment $J^-$, that always tries to bring the Bloch vector back to the ground state (south pole). 
        In the superradiant phase ($\Omega_d < \Omega_c^h = N C\gamma/2 = 2 N g^2/\kappa$), the drive and the dissipation in the form of superradiance balance each other, and the atoms reach a steady state with a fixed spin-projection below the equator, $J_z<0$. Above the critical point ($\Omega_d>\Omega_c^h$), i.e., the normal phase, the superradiant field from the atoms is no longer strong enough to cancel the drive field, so the Bloch vector starts to Rabi flop, and the spin-projection $J_z$ time-averages to zero. The bottom row shows the steady state's normalized spin projection $J_z/(N/2)$ as a function of normalized Rabi frequency $\Omega_d/\Omega_c^h$. $J_z$ can be identified as an indicator of a second-order phase transition. Spontaneous emission modifies the steady state behaviour of the transition and changes it to a first order transition at a smaller critical drive (orange curve).}
        \label{fig1}
\end{figure*}

\droptocpage %hide sections from toc in SI
\section*{\label{sec:intro}Introduction}
Light-matter interactions lie at the heart of modern quantum technologies. The recent advances in quantum simulation~\cite{Georgescu2014,Altman2021,Daley2023}, computation~\cite{Bharti2022}, sensing~\cite{Ye2024} and communication~\cite{Couteau2023} rely on the precise control and engineering of single emitters by means of classical light. Another exciting yet challenging frontier is exploring the interaction between light and dense ensembles of emitters in which the dominant effect is collective emission or superradiance~\cite{Dicke1954,Gross1982}. Since the introduction of superradiance, there has been a continuous theoretical effort to understand this effect; superradiance has also been observed in a variety of experimental platforms such as cold atoms~\cite{Bohnet2012,Norcia2016_3}, matter waves~\cite{Inouye1999}, waveguides~\cite{Goban2015,Liedl2024}, circuit-QED~\cite{Fitzpatrick2017} and solid-state systems~\cite{Lei2023,Kersten2023}. Furthermore, it has been demonstrated that one can utilize superradiance for realizing improved optical atomic clocks~\cite{Norcia2018_2,Kristensen2023}.

Given this progress, a natural next step is to explore what happens when such a system is continuously driven. This scenario has been explored theoretically since the 1970s under the name of cooperative resonance fluorescence (CRF)~\cite{Senitzky1972,Drummond1978,Carmichael1980}. This iconic dissipative quantum optics model is known to feature a non-equilibrium second-order phase transition originating from the competition between the coherent drive and collective superradiant emission. This should be contrasted with open system implementations of the Dicke model~\cite{Hepp1973,Dimer2007,Baumann2010,Kroeze2018,Klinder2015,Zhiqiang2017,Ferri2021,Safavi-Naini2018}, in which dissipation modifies an already existing equilibrium phase transition, and with the lasing transition in a continuous superradiant laser~\cite{Meiser2009,Bohnet2012}, in which the drive is incoherent.

Over the years, CRF has gained considerable theoretical attention~\cite{Schneider2002,Morrison_2008,Hannukainen2018}, which has revealed intriguing connections to time crystals~\cite{Mattes2023,Iemini2018} and measurement-induced entanglement phase transitions~\cite{Pasarelli2024}. Despite this interest and the model's simplicity, experimental effort to demonstrate CRF has been scarce as it proved difficult to see the transition given the fully collective nature of the model and the fast timescales inherent to typical dipole-allowed optical transitions. A recent experiment in free space observed related physics~\cite{Ferioli2023}, though studies in the past year have found that near-field dipolar interactions in the system lead to a qualitative and quantitative departure from CRF and thus complicate the realization of the phase transition~\cite{Goncalves2024, Agarwal2024,Ruostekoski2024}.

Here we report an observation of the second-order superradiant transition in CRF by resonantly driving an ensemble of $^{88}$Sr atoms on a narrow linewidth transition inside of a resonant high-finesse cavity. We identify and distinguish the two predicted phases by observing both spin and light degrees of freedom as the drive strength is varied. Central to our observations are two key elements of the experiment. First, the use of a high-finesse optical cavity allows us to operate with a spatially dilute ensemble of atoms with negligible dipolar or contact interactions; the cavity also provides well-defined phase matching to a single optical mode, ensuring the atoms are always in a fully collective regime. Second, the interrogation of a narrow-linewidth atomic transition enables us to observe the collective physics of CRF on an experimentally accessible timescale. The clean separation of timescales between the collective decay and single-particle spontaneous decay also allows us to observe how spontaneous emission slowly steers the system towards a qualitatively different steady state that displays a discontinuous transition in both spin and light observables. This physics is related to prior studies in optical bistability~\cite{Bonifacio1978,Rosenberger1983,Rempe1993,Gripp1996,Rivero2023}, although past experimental works have focused mostly on the properties of the light, such as hysteretic transmission, and on timescales where spontaneous emission is important. Our work paves the way for experimental verification of new symmetries in open quantum systems~\cite{Buca_2012,Albert2014,Roberts2023,Young2024} and opens the door for exploring new ways of generating spin-squeezing~\cite{Young2024} and spectroscopy on ultranarrow clock transitions~\cite{Martin2011,Barberena2023}.

\section*{\label{sec:fig1}Experimental setup and theoretical model}
To explore the phase transition, we trap an ensemble of $N=10^3$ to $10^4$ $^{88}$Sr atoms to interact with a driven high-finesse optical cavity. The atoms are laser-cooled and trapped in the Lamb-Dicke regime by a one-dimensional optical lattice along the cavity axis at wavelength 813~nm. We work along the narrow-linewidth 689~nm transition and treat the ground state $\ket{\downarrow}=\ket{^1\textrm{S}_0,m_J=0}$ and the excited state $\ket{\uparrow}=\ket{^3\text{P}_1,m_J=0}$ of each atom as an effective spin-1/2 system with associated spin operators $\hat{s}_i^z=\frac{1}{2}\left(\ketbra{\uparrow}{\uparrow}_i-\ketbra{\downarrow}{\downarrow}_i\right)$ and $\hat{s}_i^-=\ketbra{\downarrow}{\uparrow}_i$, where $i$ labels the atoms. The excited state has a radiative linewidth $\gamma = 2\pi \times 7.5$~kHz, and the cavity has a FWHM linewidth of $\kappa=2\pi\times 153$ kHz. Atoms are inhomogenously coupled to the standing-wave cavity mode with a spatially averaged r.m.s.~single-photon Rabi frequency of $2 g =2\pi \times 15$ kHz. The r.m.s.~single-atom cooperativity is $C =4g^2/\kappa \gamma= 0.21(2)$, putting us in the collective strong coupling regime with $N C \gg 1$~\cite{Norcia2016}. In this experiment, we tune the cavity on resonance with the atomic transition and direct a laser beam, also resonant with the atoms, into the cavity to act as a drive (see Fig.~\ref{fig1}a).

Under these conditions, and assuming for simplicity that all the atoms couple to the cavity identically and with strength $g$, the system can be modelled by the following master equation (see SI): 
\begin{align}\begin{split}\label{eqn:MasterEquation}
    \hspace{-0.09cm}
    \frac{d\hat{\rho}}{dt}&=-\frac{i}{\hbar}\left[\hat{H},\hat{\rho}\right]+\kappa\,\mathcal{L}_{c}(\hat{\rho}), \\
    \hat{H}/\hbar&= g\left(\hat{a}\hat{J}^{+}+\hat{a}^\dagger\hat{J}^-\right)-\frac{i\kappa}{4 g}  \Omega_d\left(\hat{a}-\hat{a}^\dagger\right).
\end{split}\end{align}
Here, $\hat{\rho}$ is the density matrix of the full atom-cavity system, $\hat{J}^-=\sum_{i=1}^N\hat{s}^-_i$ and $\hat{J}^+=(\hat{J}^-)^\dagger$ are, respectively, collective spin lowering and raising operators that quantify the collective atomic coherence along the two-level system,  the operator $\hat{a}^\dagger$ ($\hat{a}$) creates (destroys) one photon in the cavity mode, and the Lindbladian superoperator $\mathcal{L}_{c}(\hat{\rho})=\hat{a}\hat{\rho}\hat{a}^\dagger-\{\hat{a}^\dagger\hat{a},\hat{\rho}\}/2$ describes photon emission through the cavity. The first term in the Hamiltonian $\hat{H}$ is the Tavis-Cummings atom-cavity interaction, and the second term represents the laser drive with a strength quantified by the Rabi frequency $\Omega_d =\alpha \times 2 g $, where $\alpha$ is the coherent state amplitude that would be established inside the cavity by the drive without atoms. For now, we have omitted single-particle spontaneous emission, which happens on a slower timescale set by $\gamma^{-1}$.

\begin{figure*}[!t]
\centerline{\includegraphics[scale=1]{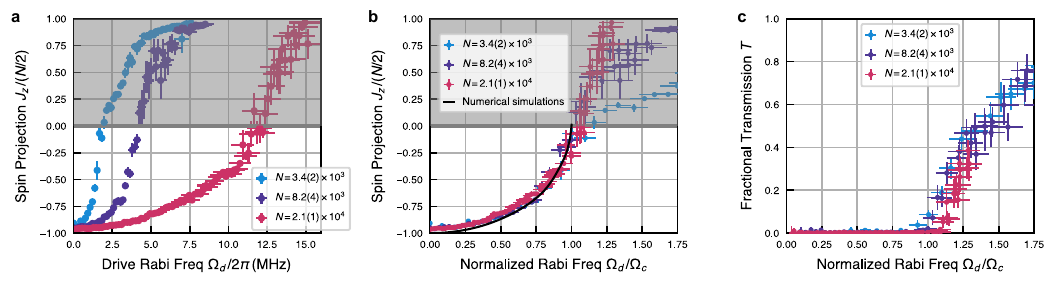}}
\caption{
    \textbf{Observing the second-order superradiant phase transition.}
        \textbf{a,b} Steady-state spin projection $J_z$ after a $9.3~\mu$s drive as a function of (\textbf{a}) Rabi frequency $\Omega_d$ and (\textbf{b}) normalized Rabi frequency $\Omega_d/\Omega_c$ with $\Omega_c = 0.35 NC\gamma$. Regions of gray hereafter indicate that our measurement of spin-projection $J_z$ does not accurately capture the real $J_z$ value of our system at the end of the drive duration (see main text). The three differently colored datasets are for three different total atom numbers $N$ with the error bars denoting the standard deviation. In \textbf{b}, below the critical drive, the three sets at different $N$ collapse on top of each other after normalizing to the collective decay rate $\Omega_c$, which at the same time align well with the theory prediction (solid black curve).
        %, \Rev{computed with $N=2\times 10^{4}$})   
        \textbf{c,} The transmission $T$ of the cavity with atoms inside normalized to that of a cavity without atoms as a function of normalized Rabi frequency $\Omega_d/\Omega_c$. The transmitted power is averaged over the last $5~\mu$s of the drive duration. The measured transmission is consistent with 0 below the critical Rabi drive $\Omega_c$, which indicates the complete cancellation of the external drive field by the superradiantly emitted field; the fact that transmission becomes finite above the critical drive $\Omega_c$ confirms that the superradiant field is no longer large enough to cancel the drive in the normal phase.
        Error bars in the plots, and hereafter, represent the standard error of the mean unless otherwise noted.
        }
        \label{fig2}
\end{figure*}

We refer to this model as cooperative resonance fluorescence (CRF) following the nomenclature starting in the 1970s~\cite{Walls1978,Carmichael1980}. Even though we are recasting the model in a broader setting, i.e.~including the intracavity field, Eq.~(\ref{eqn:MasterEquation}) gives rise to similar predictions for the steady state (see SI and \cite{Carmichael1980}). We note that this model has also been referred to both as the driven Dicke Model~\cite{Ferioli2023,Roberts2023,Goncalves2024} and as driven superradiance~\cite{Somech2023,Barberena2024}.

The properties of the steady state of the system, which satisfies $d\hat{\rho}_{ss}/dt =0$, can be understood by examining the mean-field equation of motion for the intracavity field $a = \langle \hat{a} \rangle$ and the atomic coherence $J^- = \langle \hat{J}^{-} \rangle$:
\begin{align}\begin{split}\label{eqn:cavityfieldEOM}
    \hspace{-0.09cm}
    \frac{d a}{dt} &= \frac{\kappa}{4 g}\Omega_d -i g  J^- - \frac{\kappa}{2} a,  \\
    \frac{d J^- }{dt} &= 2ig  a J_z ,\quad 
    \frac{d{J}_z}{dt}=-ig\left(aJ^+-a^*J^-\right)
\end{split}\end{align}
where $ J_z=\langle \hat{J}_z \rangle= \sum_i\langle \hat{s}^z_i\rangle$ represents the collective spin projection. In the equation for the field, the first term on the right hand side describes the external drive, the second term describes the superradiant field radiated into the cavity by the collective atomic dipole moment with strength proportional to $J^-$, and the last term accounts for the cavity field decay. The other two equations describe how the intracavity field rotates the collective Bloch vector. 

There is a unique mean-field stable steady state solution to Eq.~(\ref{eqn:cavityfieldEOM}) characterized by zero intracavity field $a=0$ and an atomic coherence given by $J^-=  -i  (\Omega_d/\Omega_c^h) (N/2)$, where we have defined the critical Rabi frequency $\Omega_c^h =N C\gamma/2=2 N  g^2/\kappa $ for homogeneous couplings. This solution does not exist when $\Omega_d>\Omega_c^h$ since the magnitude of $J^-$ must always be smaller than $N/2$. 

The nature of this steady state can be intuitively illustrated using Bloch spheres as shown in Fig.~\ref{fig1}c. In our experiment, the spins all start at the south pole ($\ket{\downarrow}^{\otimes N}$). The external field rotates the Bloch vector up, causing the transverse coherence $J^-$ to increase. This induces the atoms to emit a superradiant field into the cavity, characterized by an effective Rabi frequency $\Omega_{\mathrm{SR}} = i \tfrac{4g^2}{\kappa}J^-$, which destructively interferes with the drive. When $\Omega_d< \Omega_c^h$,  this field is strong enough to cancel the drive, resulting in a steady state with $a=0$ and spin projection $J_z<0$ that depends on the normalized Rabi frequency $\Omega_d/\Omega_c^h$. We call this the ``superradiant phase" because the atoms have a macroscopic dipole moment ($\lvert J^-\rvert>0$) even though the intracavity field is zero.

When $\Omega_d>\Omega_c^h$, the superradiant field emitted by the atoms is never large enough to cancel the drive field, leading to a non-zero total electric field inside the cavity. The Bloch vector will just keep rotating, and in the limit where $\Omega_d \gg \Omega_c^h$, the dynamics approach single-particle Rabi oscillation at a frequency $\Omega_d$. Hence, we call this the ``normal phase". Phase space diffusion arising from beyond mean-field dissipative effects leads to a steady state with $J_z = 0$~\cite{Carmichael1980} such that at $\Omega_d=\Omega_c^h$, the system exhibits a second-order phase transition, indicated by non-analytic behavior in $J_z$ as shown in Fig.~\ref{fig1}c. Since the steady state in this regime occurs beyond experimentally accessible timescales~\cite{Carmichael1980,Barberena2019}, instead we identify the normal phase with a measurement of a non-zero cavity field. 

\begin{figure*}[!t]
\label{Fig3}
\centerline{\includegraphics[scale=1]{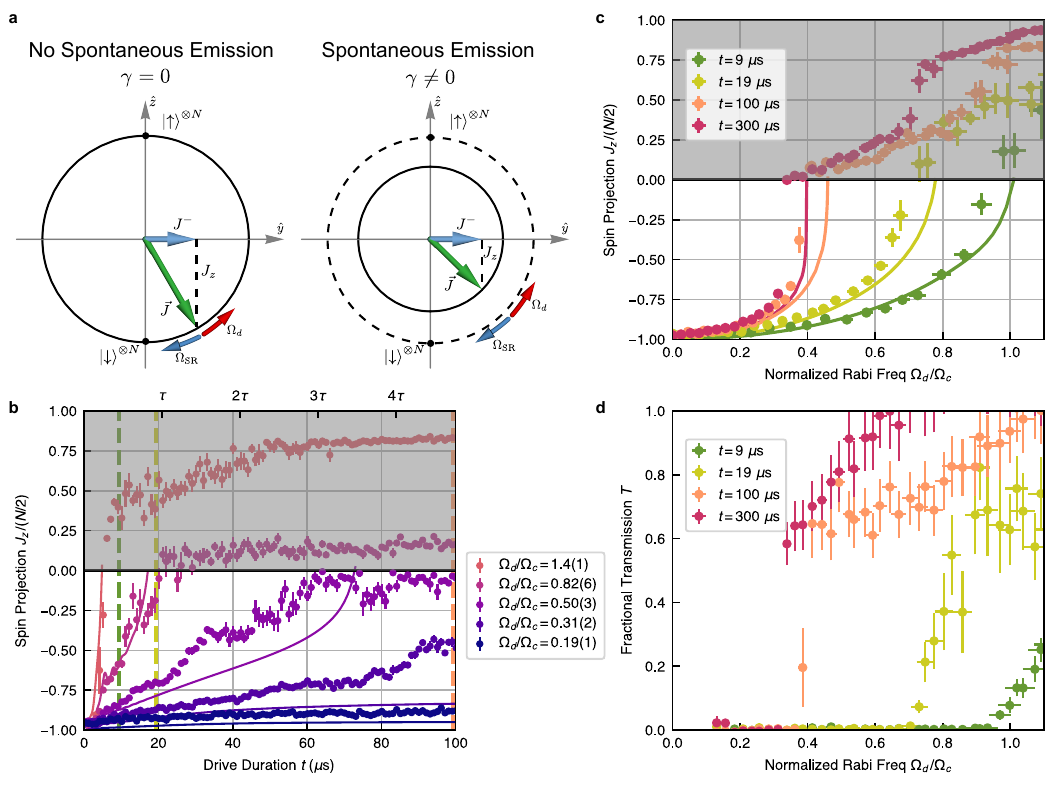}}
\caption{
    \textbf{Modification of critical behaviour by single-particle spontaneous emission.} 
        \textbf{a,} Bloch sphere illustration of the superradiant phase steady state without (left) and with (right) spontaneous emission. On the left, the steady state Bloch vector $\vec{J}$ stays in the collective manifold where the torque coming from the external drive $\Omega_d$ (red arrow) and superradiant field $\Omega_\textrm{SR}$ (blue arrow) can be balanced. On the right, the steady state Bloch vector $\vec{J}$ is no longer in the collective manifold. The Bloch vector is rotated up from the south pole to increase the coherence $J^-$ such that the superradiant field can still cancel the external drive. As a result, the steady state $J_z$ on the right is always higher than that on the left. 
        \textbf{b,} The spin projection $J_z$ as a function of drive duration $t$. Traces with different colors represent different drive strengths with fixed atom number. The solid lines represent the results of numerical simulation. The tick marks on the top horizontal axis indicate the drive duration as a function of multiples of $\tau=\gamma^{-1}=21~\mu$s, the 1/$e$ lifetime of the excited state. Traces with $\Omega_d/\Omega_c = 0.50, 0.82$ have $J_z$ slowly relax above 0 on a timescale set by the excited state lifetime, while the traces with $\Omega_d/\Omega_c=0.19,0.31$ have $J_z$ remain below 0 and stay in the superradiant phase. The coloured vertical dashed lines indicate the drive durations corresponding to the data with matching colours in \textbf{c, d}. 
        \textbf{c,} $J_z$ as a function of $\Omega_d/\Omega_c$ with fixed atom number. Different colors indicate different drive durations $t$ before the measurement of $J_z$. The critical point, where $J_z$ crosses 0, reduces to smaller drive values as drive duration increases, and the continuous second order phase transition ($t=9~\mu$s, green curve) turns into a first order transition ($t=300~\mu$s, red curve), where a clear jump is observed.
        \textbf{d,} The fractional transmission $T$ as a function of $\Omega_d/\Omega_c$. The drive strength at which transmission becomes non-zero gradually shifts to a lower value as the drive duration increases, which aligns with the critical point suggested by the $J_z$ measurement. The green curve ($t=9~\mu$s) displays a smooth transition while the red curve with a longer drive duration ($t=300~\mu$s) displays a discontinuity at the critical point.
    }
    \label{fig3}
\end{figure*}

\section*{Experimental results}
\subsection*{CRF superradiant phase transition}
We probe this transition by sending laser light resonant with the atomic dipoles through one of the cavity mirrors. To avoid transient effects, we linearly ramp the Rabi frequency of the drive from 0 to $\Omega_d$ over $5~\mu$s. We then hold the laser intensity fixed for $4.3~\mu$s to let the system reach its steady state. To measure the spin projection $J_z$ at the end of the hold period, we first shelve the atoms in $\ket{\uparrow}$ by flashing on a resonant 688 nm laser to optically pump them to metastable states while rapidly turning off the drive and then measure the number of atoms left in $\ket{\downarrow}$ (see Methods). We determine $J_z$ at the end of the hold period by comparing this result to the total number of atoms measured before the drive is turned on.

\begin{figure*}[!t]
\centerline{\includegraphics[scale=1]{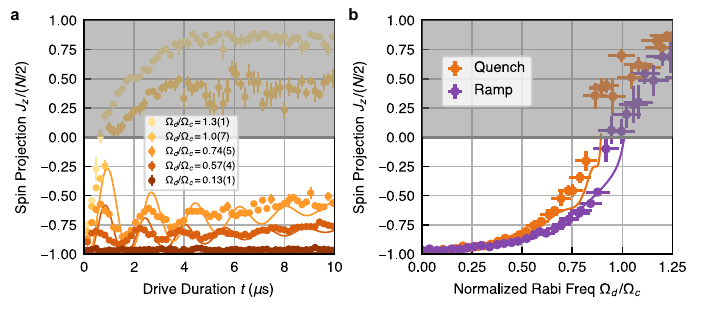}}
\caption{
    \textbf{Quench response of driven emitters in the resolved vacuum Rabi splitting regime.}
        \textbf{a,} The spin projection $J_z$ as a function of the total drive duration $t$ accompanied by theory predictions (solid curves). The data are taken by suddenly turning on the drive in $100~$ns and holding the drive constant for a variable amount of time, before extinguishing the drive and measuring the spin projection $J_z$. Traces with different colors represent different drive strengths with fixed atom number $N=8.0(8)\times 10^3$.
        %, $\Omega_d/\Omega_c$ \Com{, which is varied by varying the drive strength $\Omega_d$ while holding atom number to be constant around $N=8.0(8)\times10^3$}.
        The traces with $\Omega_d/\Omega_c=0.57, 0.74$ exhibit characteristic oscillations associated with the vacuum Rabi frequency $2g\sqrt{N}$. The trace with $\Omega_d/\Omega_c= 1.0$ should normally relax to $J_z=0$ as it is at the critical point of the CRF superradiant transition, but the quench causes $J_z$ quickly to shoot above $0$ in $1~\mu$s, after which it remains in the normal phase.
        \textbf{b,} The spin projection $J_z$ as a function of $\Omega_d/\Omega_c$ at the end of 9.3~$\mu$s drive for quenching (orange) and ramping on the drive (purple) with $N=8.3(3)\times 10^3$. The ramping procedure is the same as that for Fig.~\ref{fig2}. Both datasets agree well with numerical simulations (solid curves). However, for the same $\Omega_d$, the measured $J_z$ in quenched experiments is consistently above the $J_z$ obtained in ramped experiments. This suggests that with a quench the system can overshoot and fail to relax to the true steady state predicted by the mean-field theory (purple solid line) on experimental timescales. }
        % Under the same drive strength, the spin projection $J_z$ for the quenched experiments agrees with the full numerical simulation (orange solid line). However, they are consistently above or equal to $J_z$ obtained by ramping the drive,
    \label{fig4}
\end{figure*}

We show the (normalized) spin projection $J_z/(N/2)$ as a function of Rabi frequency $\Omega_d$ for three different atom numbers in Fig.~\ref{fig2}a. After normalizing the Rabi frequency by $\Omega_c = 0.70 \Omega_c^h = 0.35 NC\gamma$, we find that the curves in the region $\Omega_d/\Omega_c<1$ collapse on top of each other, certifying the collective nature of the steady state. The numerical factor 0.70 mainly accounts for inhomogenous atom-cavity coupling, which reduces the critical Rabi drive but does not change the second-order nature of the transition (see Methods, SI). The collapsed data is in good agreement with numerical simulations (solid black curve), consistent with the prediction of the superradiant phase. In the region where $\Omega_d>\Omega_c$, our shelving procedure does not properly capture the spin projection $J_z$ at the end of the hold period. This is because in this regime, the cavity is populated with a macroscopic field that rapidly transfers atoms from $\ket{\downarrow}$ to $\ket{\uparrow}$ where they are then subsequently shelved, leading to a mis-inferred value of $J_z>0$. To emphasize this, we shade the regions with $J_z>0$ in grey in Figures \ref{fig2} to \ref{fig5}.  Note, however, that for smaller atom number, we expect the intracavity field to be relatively small just above the transition, allowing the shelving method to provide a more accurate inference of $J_z$. Indeed, we observe in Fig.~\ref{fig2}b that for $N=3.4\times 10^3$, the measured value of $J_z$ exhibits an abrupt change of behavior at $\Omega_d = \Omega_c$.

To verify the presence of a phase transition at $\Omega_d = \Omega_c$, we also directly probe the light degree of freedom.  We infer the presence or absence of an intracavity field by measuring the transmitted power normalized to that of an empty cavity, averaged over the last $5~\mu$s of the drive duration, which we call the fractional transmission $T$.  This quantity establishes the extent to which the superradiant field is able to cancel the external drive. We measure $T$ while varying the drive strength $\Omega_d$ between different experimental shots and plot the results in Fig.~\ref{fig2}c. For $\Omega_d<\Omega_c$, we observe a value consistent with zero transmitted power, in accordance with the theoretical expectations. For $\Omega_d>\Omega_c$, the transmitted power rises very sharply, demonstrating that the superradiant field can no longer cancel the drive in the normal phase. Importantly, the value of the drive separating these two regions coincides with the point where the spin projection $J_z$ approaches 0 from below in our atomic observable. These two pieces of evidence together allow us to identify $\Omega_c$ as a critical point which separates the superradiant phase and the normal phase. 

% \clearpage
\subsection*{\label{sec:fig3} Melting into a first order transition}
By driving the system for longer times, we can explore the effect of a different type of dissipation: spontaneous emission, which is a non-collective effect described by another Lindblad term added to the master equation Eq.~(\ref{eqn:MasterEquation}),  $\gamma\mathcal{L}_{se}(\hat{\rho})=\gamma\sum_i (\hat{s}_i^-\hat{\rho}\hat{s}_i^+-\{\hat{s}_i^+\hat{s}_i^-,\hat{\rho}\}/2)$. Unlike superradiant decay, which preserves the total length of the Bloch vector $J\approx\sqrt{|J^-|^2+J_z^2}$, spontaneous emission shortens $J$. If the system starts in the steady state of the superradiant phase in the CRF model, the atomic coherence $J^-$ is shortened by spontaneous emission and in turn radiates a smaller field. This leads to a non-zero net field inside the cavity due to the imbalance of the drive and superradiant field. The Bloch vector then rotates upwards, which restores $J^-$ to its previous value but introduces an effective upwards force on $J_z$ that scales with the drive strength $\Omega_d$ (see SI). 
At the same time, the decay to the ground state induced by spontaneous emission leads to a competing process that pushes $J_z$ down towards $-N/2$. This competition leads to two different behaviours depending on the drive strength. For strong drive strengths, the upwards force causes $J_z$ to slowly relax towards 0 while holding $J^{-}$ constant. Once $J_z$ reaches 0, the atoms start to undergo Rabi-like oscillation, and the cavity gets quickly populated by photons, indicative of the normal phase despite having started in the CRF superradiant phase. However, when $\Omega_d$ is small enough, since the upwards force on $J_z$ scales with $\Omega_d$, $J_z$ will not relax all the way towards 0. Instead, the Bloch vector arrives at a new steady state where the upwards and downwards forces balance, and the superradiant and drive fields retain a near-perfect cancellation despite the reduction in spin length $J$ (see SI). This corresponds to the superradiant phase (see Fig.~\ref{fig3}a), which is analogous to the ``cooperative branch" in the bistability literature~\cite{Bonifacio1976,Bonifacio1978}. These two regimes are separated by a new critical point $\Omega_c'$, and the spin projection $J_z$ will now display a discontinuous jump at this point as opposed to the continuous transition in the CRF model~\cite{Carmichael1980,Leppenen2024} (also see Fig.~\ref{fig1}c).

We show traces of the spin projection $J_z$ as a function of the drive duration $t$ for different drive strengths in Fig.~\ref{fig3}b. At low drives ($\Omega_d/\Omega_c =0.19,0.31)$, $J_z$ reaches its fully collective superradiant steady state at short times, and at longer times the Bloch vector still remains below the south pole with a modified value. As the drive strength is increased ($\Omega_d/\Omega_c =0.50, 0.82$), the spin projection $J_z$ remains below the equator on a timescale shorter than the excited state lifetime $\tau = \gamma^{-1} = 21~\mu$s. For longer times, however, it slowly moves towards $0$, consistent with theoretical expectations. At even larger drives ($\Omega_d/\Omega_c=1.4$) the inversion quickly rises above 0, suggesting the systems are in the normal phase both at short and long times.

We characterize the gradual change in $J_z$ over time by providing snapshots of $J_z$ as a function of the Rabi frequency $\Omega_d$ for different drive durations in Fig.~\ref{fig3}c. At short times ($9~\mu$s), we observe a continuous superradiant phase transition, consistent with Fig.~\ref{fig2}b. At longer times ($t \gtrsim \tau$), spontaneous emission becomes important and noticeably shifts the transition point, characterized by the point where $J_z$ crosses 0, to a lower value of $\Omega_d$. By $t=100~\mu$s, the system has already equilibrated to a new steady state, and the spin projection clearly displays a discontinuous jump at $\Omega_d/\Omega_c=0.4$. These observations can be correlated with the results of fractional transmission, $T$, through the cavity, shown in Fig.~\ref{fig3}d as a function of $\Omega_d$ for different drive durations. Again, the transition points where the transmission rises above 0 coincide with the points at which $J_z$ reaches 0 in Fig.~\ref{fig3}c. 

\subsection*{Short-time dynamical response}
\begin{figure}[!t]
\centerline{\includegraphics[scale=1]{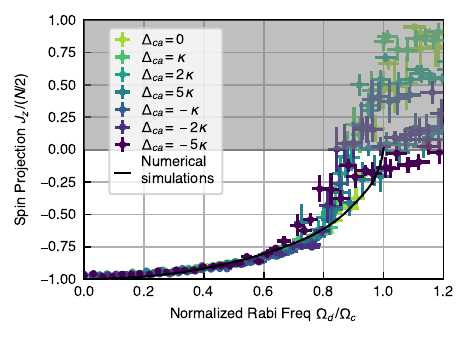}}
\caption{
    \textbf{Invariance of the critical point to detuning $\Delta_{ca}$ of the cavity from atomic resonance.} We show the measured spin projection $J_z$ as a function of $\Omega_d$, defined as the Rabi frequency that would be established in the bare cavity when $\Delta_{ca}=0$, normalized to the critical Rabi frequency $\Omega_c = 0.35 N C \gamma$. The different colours of data are taken at different cavity-atom detunings $\Delta_{ca}$ spanning many bare cavity linewidths (from $-5 \kappa$ to $+5 \kappa$) and with fixed atom number, while the drive is kept on resonance with the atoms ($\omega_d=\omega_a$). Despite such large changes, the datasets all overlap with each other below the transition, showing the predicted insensitivity to the cavity detuning. Full numerical simulations suggest the dependence on detuning on $\Delta_{ca}$ is small despite the presence of the inhomogeneous coupling (see SI), so for clarity here we only show simulations for $\Delta_{ca}=0$ (black curve).
    }
    \label{fig5}
\end{figure}

So far we have focused on the steady state and long-time behaviour of $J_z$ in our system, which follows the predictions of the original CRF model~\cite{Carmichael1980}. Nevertheless, our cavity system operates in a very different regime, characterized by a resolved vacuum Rabi splitting $2g\sqrt{N}>\kappa$ (``resolved VRS regime" hereafter). Under these conditions, excitations can be coherently exchanged between the atoms and the photon field, and the Bloch vector behaves like an underdamped oscillator~\cite{Norcia2016_2}. In contrast, the original CRF Lindbladian arises in the ``non-resolved VRS" regime ($2g\sqrt{N}\ll \kappa$) in which the cavity field adiabatically follows the atoms (\cite{Norcia2016_3} and see SI), corresponding to an overdamped oscillator. We explore this distinction by measuring the dynamical response of our system when we suddenly turn on the drive $\Omega_d$ rather than slowly ramping the amplitude of the drive as we did in Fig.~\ref{fig2}.

In this experiment, we quench the drive strength from $0$ to $\Omega_d$ in approximately 100~ns and then hold for a total drive duration $t$. After the quench, the atoms and the intracavity field will exhibit transient behaviour as they settle to the steady state. To illustrate this, we show time traces of the spin projection $J_z$ as a function of drive duration in Fig.~\ref{fig4}a. At low drive strengths, we observe oscillations in $J_z$ at a frequency consistent with $g\sqrt{N}$, with $N=8.0(8)\times 10^3$, characteristic of dynamics in the resolved VRS regime. The frequency of those oscillations aligns well with theoretical predictions, although the amplitude is observed to be smaller due to additional dephasing mechanisms in the experiment not accounted for in theory such as atomic motion~\cite{Young2024BCS,Muniz2020}. 

While the ramp employed in Fig.~\ref{fig2} is designed to be slow enough to prepare the system close to the superradiant steady state, if the system far away from this steady state is suddenly subject to a quenched drive, the atoms would not have enough time to generate a field to cancel the external drive and may start to oscillate. Hence, the resolved VRS regime features an additional region of bistability compared to the non-resolved VRS regime~\cite{Mattes2023} where the short-time dynamical behaviour of the system, either settling into the CRF steady state or oscillating persistently, is determined by the initial conditions. We can characterize this effect by comparing the $J_z$ values obtained after driving the system for $9.3~\mu$s following a quench or ramp of the drive. Fig.~\ref{fig4}b shows, consistently with theory predictions, that the post-quench $J_z$ is always above the post-ramp $J_z$, and the difference increases as we approach the critical point $\Omega_d/\Omega_c=1$. Moreover, for $\Omega_d/\Omega_c$ slightly smaller than 1, we observe that the system appears to enter the normal phase after a quench even though it remains in the superradiant phase after a ramp. This highlights how, in the resolved VRS regime, our ability to access the superradiant phase can be influenced by the dynamical response to the drive.

\subsection*{Invariance of the critical point under different cavity detunings}
In the non-resolved VRS limit, in which the cavity field can be adiabatically eliminated, detuning the cavity from resonance by $\Delta_{ca}=\omega_c-\omega_a$ gives rise to elastic spin exchange interactions $\chi \hat{J}^+\hat{J}^-$, with the ratio of elastic interactions to inelastic collective decay (with rate $\Gamma$) determined by $\chi/\Gamma = \Delta_{ca}/\kappa$~\cite{Norcia2018}.  Based off previous work in this limit~\cite{Barberena2019}, the elastic and inelastic interactions combine to keep the critical point separating the superradiant and normal phase independent of $\Delta_{ca}$ as long as the drive remains on resonance with the atomic transition (see SI). 

In fact, the invariance of the critical point to cavity detuning applies more generally, holding even in the resolved VRS limit studied in this work. At a high level, this can be understood by treating the cavity mode as a harmonic oscillator. Both the externally applied laser and the atomic dipole $J^-$ drive this oscillator with the same detuning $\Delta_{ca}$ such that the steady state field that builds up in response to these two drives is modified in a common-mode way and can therefore still cancel. Formally, the steady-state field established inside the bare cavity by the external drive is described by a modified Rabi frequency $\Omega_d(\Delta_{ca}) = \Omega_d / (1 + i\tfrac{\Delta_{ca}}{\kappa/2})$ where $\Omega_d$ is, as defined before, the Rabi frequency that would be established in the cavity when $\Delta_{ca}=0$.  Correspondingly, the field established by the radiating dipole without an external drive would have a Rabi frequency  $\Omega_{\mathrm{SR}}(\Delta_{ca}) = i \tfrac{4g^2 }{\kappa} J^-(\Delta_{ca})/ (1 + i\tfrac{\Delta_{ca}}{\kappa/2})$. Equating these two Rabi frequencies yields a steady-state solution $J^-(\Delta_{ca}) = -i (\Omega_d/\Omega_c^h) (N/2)$ which is independent of $\Delta_{ca}$, suggesting that the steady-state $J_z$ has no detuning dependence. Notably, the relative phase between the steady-state $J^-$ and the external applied drive $\Omega_d$ also does not change, even in the presence of a cavity detuning.

To validate these expectations, we ramp and hold the drive for $9.3~\mu$s and then measure the spin projection $J_z$ versus drive strength for different atom-cavity detunings $\Delta_{ca}$. As shown in Fig.~\ref{fig5}, we indeed see that, below the critical point, the different traces behave similarly, showing that the steady state $J_z$ does not depend strongly on the cavity detuning $\Delta_{ca}$. We highlight that when varying the cavity detuning over the range $\pm 5\kappa$,  the intracavity field established by the drive in the absence of atoms (or equivalently $\left| \Omega_d(\Delta_{ca}) \right|$) is reduced by a factor of 10 and therefore the fact that the steady state does not significantly change demonstrates its insensitivity to the cavity detuning over a wide parameter regime.

\section*{\label{sec:conclusion}Conclusion and Outlook}
In summary, we have observed the superradiant phase transition in the cooperative resonance fluorescence model predicted more than 40 years ago. We also witnessed how spontaneous emission melted the continuous transition into a first order transition. In the future, it will be interesting to explore the properties of the steady states in the presence of additional Hamiltonian interactions~\cite{Roberts2023} or at the level of quantum fluctuations since the spins get more squeezed as the drive strength approaches the critical point from below~\cite{Carmichael1980,Lee2014}. Alternatively, by introducing a third atomic level to our system, one can also potentially generate squeezed states between the ground state and the third state via the generation of a dissipative Berry phase~\cite{Young2024}. These schemes are all compatible with ultra-long-lived clock transitions in alkaline earth atoms and thus can be utilized to improve state-of-the-art atomic clocks~\cite{Pedrozo2020, Robinson2024}. Furthermore, carefully monitoring the light leaking out of the cavity, in combination with post-selection and feedback operations, opens up the possibility of implementing exotic out-of-equilibrium phenomena such as measurement-induced phase transitions~\cite{2020Ivanov,Pasarelli2024}.

\bibliographystyle{apsrev4-2} %
\bibliography{refs}{}

\begin{thebibliography}{80}%
\makeatletter
\providecommand \@ifxundefined [1]{%
 \@ifx{#1\undefined}
}%
\providecommand \@ifnum [1]{%
 \ifnum #1\expandafter \@firstoftwo
 \else \expandafter \@secondoftwo
 \fi
}%
\providecommand \@ifx [1]{%
 \ifx #1\expandafter \@firstoftwo
 \else \expandafter \@secondoftwo
 \fi
}%
\providecommand \natexlab [1]{#1}%
\providecommand \enquote  [1]{``#1''}%
\providecommand \bibnamefont  [1]{#1}%
\providecommand \bibfnamefont [1]{#1}%
\providecommand \citenamefont [1]{#1}%
\providecommand \href@noop [0]{\@secondoftwo}%
\providecommand \href [0]{\begingroup \@sanitize@url \@href}%
\providecommand \@href[1]{\@@startlink{#1}\@@href}%
\providecommand \@@href[1]{\endgroup#1\@@endlink}%
\providecommand \@sanitize@url [0]{\catcode `\\12\catcode `\$12\catcode `\&12\catcode `\#12\catcode `\^12\catcode `\_12\catcode `\%12\relax}%
\providecommand \@@startlink[1]{}%
\providecommand \@@endlink[0]{}%
\providecommand \url  [0]{\begingroup\@sanitize@url \@url }%
\providecommand \@url [1]{\endgroup\@href {#1}{\urlprefix }}%
\providecommand \urlprefix  [0]{URL }%
\providecommand \Eprint [0]{\href }%
\providecommand \doibase [0]{https://doi.org/}%
\providecommand \selectlanguage [0]{\@gobble}%
\providecommand \bibinfo  [0]{\@secondoftwo}%
\providecommand \bibfield  [0]{\@secondoftwo}%
\providecommand \translation [1]{[#1]}%
\providecommand \BibitemOpen [0]{}%
\providecommand \bibitemStop [0]{}%
\providecommand \bibitemNoStop [0]{.\EOS\space}%
\providecommand \EOS [0]{\spacefactor3000\relax}%
\providecommand \BibitemShut  [1]{\csname bibitem#1\endcsname}%
\let\auto@bib@innerbib\@empty
%</preamble>
\bibitem [{\citenamefont {Blais}\ \emph {et~al.}(2021)\citenamefont {Blais}, \citenamefont {Grimsmo}, \citenamefont {Girvin},\ and\ \citenamefont {Wallraff}}]{Blais2021}%
  \BibitemOpen
  \bibfield  {author} {\bibinfo {author} {\bibfnamefont {A.}~\bibnamefont {Blais}}, \bibinfo {author} {\bibfnamefont {A.~L.}\ \bibnamefont {Grimsmo}}, \bibinfo {author} {\bibfnamefont {S.~M.}\ \bibnamefont {Girvin}},\ and\ \bibinfo {author} {\bibfnamefont {A.}~\bibnamefont {Wallraff}},\ }\href {https://doi.org/10.1103/RevModPhys.93.025005} {\bibfield  {journal} {\bibinfo  {journal} {Rev. Mod. Phys.}\ }\textbf {\bibinfo {volume} {93}},\ \bibinfo {pages} {025005} (\bibinfo {year} {2021})}\BibitemShut {NoStop}%
\bibitem [{\citenamefont {Mivehvar}\ \emph {et~al.}(2021)\citenamefont {Mivehvar}, \citenamefont {Piazza}, \citenamefont {Donner},\ and\ \citenamefont {Ritsch}}]{Farokh2021}%
  \BibitemOpen
  \bibfield  {author} {\bibinfo {author} {\bibfnamefont {F.}~\bibnamefont {Mivehvar}}, \bibinfo {author} {\bibfnamefont {F.}~\bibnamefont {Piazza}}, \bibinfo {author} {\bibfnamefont {T.}~\bibnamefont {Donner}},\ and\ \bibinfo {author} {\bibfnamefont {H.}~\bibnamefont {Ritsch}},\ }\href {https://doi.org/10.1080/00018732.2021.1969727} {\bibfield  {journal} {\bibinfo  {journal} {Advances in Physics}\ }\textbf {\bibinfo {volume} {70}},\ \bibinfo {pages} {1} (\bibinfo {year} {2021})}\BibitemShut {NoStop}%
\bibitem [{\citenamefont {Schlawin}\ \emph {et~al.}(2022)\citenamefont {Schlawin}, \citenamefont {Kennes},\ and\ \citenamefont {Sentef}}]{Schlawin2022}%
  \BibitemOpen
  \bibfield  {author} {\bibinfo {author} {\bibfnamefont {F.}~\bibnamefont {Schlawin}}, \bibinfo {author} {\bibfnamefont {D.~M.}\ \bibnamefont {Kennes}},\ and\ \bibinfo {author} {\bibfnamefont {M.~A.}\ \bibnamefont {Sentef}},\ }\href {https://doi.org/10.1063/5.0083825} {\bibfield  {journal} {\bibinfo  {journal} {Applied Physics Reviews}\ }\textbf {\bibinfo {volume} {9}},\ \bibinfo {pages} {011312} (\bibinfo {year} {2022})}\BibitemShut {NoStop}%
\bibitem [{\citenamefont {Bao}\ \emph {et~al.}(2012)\citenamefont {Bao}, \citenamefont {Reingruber}, \citenamefont {Dietrich}, \citenamefont {Rui}, \citenamefont {D{\"u}ck}, \citenamefont {Strassel}, \citenamefont {Li}, \citenamefont {Liu}, \citenamefont {Zhao},\ and\ \citenamefont {Pan}}]{Bao2012}%
  \BibitemOpen
  \bibfield  {author} {\bibinfo {author} {\bibfnamefont {X.-H.}\ \bibnamefont {Bao}}, \bibinfo {author} {\bibfnamefont {A.}~\bibnamefont {Reingruber}}, \bibinfo {author} {\bibfnamefont {P.}~\bibnamefont {Dietrich}}, \bibinfo {author} {\bibfnamefont {J.}~\bibnamefont {Rui}}, \bibinfo {author} {\bibfnamefont {A.}~\bibnamefont {D{\"u}ck}}, \bibinfo {author} {\bibfnamefont {T.}~\bibnamefont {Strassel}}, \bibinfo {author} {\bibfnamefont {L.}~\bibnamefont {Li}}, \bibinfo {author} {\bibfnamefont {N.-L.}\ \bibnamefont {Liu}}, \bibinfo {author} {\bibfnamefont {B.}~\bibnamefont {Zhao}},\ and\ \bibinfo {author} {\bibfnamefont {J.-W.}\ \bibnamefont {Pan}},\ }\href {https://doi.org/10.1038/nphys2324} {\bibfield  {journal} {\bibinfo  {journal} {Nature Physics}\ }\textbf {\bibinfo {volume} {8}},\ \bibinfo {pages} {517} (\bibinfo {year} {2012})}\BibitemShut {NoStop}%
\bibitem [{\citenamefont {Greve}\ \emph {et~al.}(2022)\citenamefont {Greve}, \citenamefont {Luo}, \citenamefont {Wu},\ and\ \citenamefont {Thompson}}]{Greve2022}%
  \BibitemOpen
  \bibfield  {author} {\bibinfo {author} {\bibfnamefont {G.~P.}\ \bibnamefont {Greve}}, \bibinfo {author} {\bibfnamefont {C.}~\bibnamefont {Luo}}, \bibinfo {author} {\bibfnamefont {B.}~\bibnamefont {Wu}},\ and\ \bibinfo {author} {\bibfnamefont {J.~K.}\ \bibnamefont {Thompson}},\ }\href {https://doi.org/10.1038/s41586-022-05197-9} {\bibfield  {journal} {\bibinfo  {journal} {Nature}\ }\textbf {\bibinfo {volume} {610}},\ \bibinfo {pages} {472} (\bibinfo {year} {2022})}\BibitemShut {NoStop}%
\bibitem [{\citenamefont {Kjaergaard}\ \emph {et~al.}(2020)\citenamefont {Kjaergaard}, \citenamefont {Schwartz}, \citenamefont {Braumüller}, \citenamefont {Krantz}, \citenamefont {Wang}, \citenamefont {Gustavsson},\ and\ \citenamefont {Oliver}}]{Kjaergaard2020}%
  \BibitemOpen
  \bibfield  {author} {\bibinfo {author} {\bibfnamefont {M.}~\bibnamefont {Kjaergaard}}, \bibinfo {author} {\bibfnamefont {M.~E.}\ \bibnamefont {Schwartz}}, \bibinfo {author} {\bibfnamefont {J.}~\bibnamefont {Braumüller}}, \bibinfo {author} {\bibfnamefont {P.}~\bibnamefont {Krantz}}, \bibinfo {author} {\bibfnamefont {J.~I.-J.}\ \bibnamefont {Wang}}, \bibinfo {author} {\bibfnamefont {S.}~\bibnamefont {Gustavsson}},\ and\ \bibinfo {author} {\bibfnamefont {W.~D.}\ \bibnamefont {Oliver}},\ }\href {https://doi.org/https://doi.org/10.1146/annurev-conmatphys-031119-050605} {\bibfield  {journal} {\bibinfo  {journal} {Annual Review of Condensed Matter Physics}\ }\textbf {\bibinfo {volume} {11}},\ \bibinfo {pages} {369} (\bibinfo {year} {2020})}\BibitemShut {NoStop}%
\bibitem [{\citenamefont {Young}\ \emph {et~al.}(2024{\natexlab{a}})\citenamefont {Young}, \citenamefont {Chu}, \citenamefont {Song}, \citenamefont {Barberena}, \citenamefont {Wellnitz}, \citenamefont {Niu}, \citenamefont {Sch{\"a}fer}, \citenamefont {Lewis-Swan}, \citenamefont {Rey},\ and\ \citenamefont {Thompson}}]{Young2024BCS}%
  \BibitemOpen
  \bibfield  {author} {\bibinfo {author} {\bibfnamefont {D.~J.}\ \bibnamefont {Young}}, \bibinfo {author} {\bibfnamefont {A.}~\bibnamefont {Chu}}, \bibinfo {author} {\bibfnamefont {E.~Y.}\ \bibnamefont {Song}}, \bibinfo {author} {\bibfnamefont {D.}~\bibnamefont {Barberena}}, \bibinfo {author} {\bibfnamefont {D.}~\bibnamefont {Wellnitz}}, \bibinfo {author} {\bibfnamefont {Z.}~\bibnamefont {Niu}}, \bibinfo {author} {\bibfnamefont {V.~M.}\ \bibnamefont {Sch{\"a}fer}}, \bibinfo {author} {\bibfnamefont {R.~J.}\ \bibnamefont {Lewis-Swan}}, \bibinfo {author} {\bibfnamefont {A.~M.}\ \bibnamefont {Rey}},\ and\ \bibinfo {author} {\bibfnamefont {J.~K.}\ \bibnamefont {Thompson}},\ }\href {https://doi.org/10.1038/s41586-023-06911-x} {\bibfield  {journal} {\bibinfo  {journal} {Nature}\ }\textbf {\bibinfo {volume} {625}},\ \bibinfo {pages} {679} (\bibinfo {year} {2024}{\natexlab{a}})}\BibitemShut {NoStop}%
\bibitem [{\citenamefont {Baumann}\ \emph {et~al.}(2010)\citenamefont {Baumann}, \citenamefont {Guerlin}, \citenamefont {Brennecke},\ and\ \citenamefont {Esslinger}}]{Baumann2010}%
  \BibitemOpen
  \bibfield  {author} {\bibinfo {author} {\bibfnamefont {K.}~\bibnamefont {Baumann}}, \bibinfo {author} {\bibfnamefont {C.}~\bibnamefont {Guerlin}}, \bibinfo {author} {\bibfnamefont {F.}~\bibnamefont {Brennecke}},\ and\ \bibinfo {author} {\bibfnamefont {T.}~\bibnamefont {Esslinger}},\ }\href {https://doi.org/10.1038/nature09009} {\bibfield  {journal} {\bibinfo  {journal} {Nature}\ }\textbf {\bibinfo {volume} {464}},\ \bibinfo {pages} {1301} (\bibinfo {year} {2010})}\BibitemShut {NoStop}%
\bibitem [{\citenamefont {Klinder}\ \emph {et~al.}(2015)\citenamefont {Klinder}, \citenamefont {Keßler}, \citenamefont {Wolke}, \citenamefont {Mathey},\ and\ \citenamefont {Hemmerich}}]{Klinder2015}%
  \BibitemOpen
  \bibfield  {author} {\bibinfo {author} {\bibfnamefont {J.}~\bibnamefont {Klinder}}, \bibinfo {author} {\bibfnamefont {H.}~\bibnamefont {Keßler}}, \bibinfo {author} {\bibfnamefont {M.}~\bibnamefont {Wolke}}, \bibinfo {author} {\bibfnamefont {L.}~\bibnamefont {Mathey}},\ and\ \bibinfo {author} {\bibfnamefont {A.}~\bibnamefont {Hemmerich}},\ }\href {https://doi.org/10.1073/pnas.1417132112} {\bibfield  {journal} {\bibinfo  {journal} {Proceedings of the National Academy of Sciences}\ }\textbf {\bibinfo {volume} {112}},\ \bibinfo {pages} {3290} (\bibinfo {year} {2015})}\BibitemShut {NoStop}%
\bibitem [{\citenamefont {Kroeze}\ \emph {et~al.}(2018)\citenamefont {Kroeze}, \citenamefont {Guo}, \citenamefont {Vaidya}, \citenamefont {Keeling},\ and\ \citenamefont {Lev}}]{Kroeze2018}%
  \BibitemOpen
  \bibfield  {author} {\bibinfo {author} {\bibfnamefont {R.~M.}\ \bibnamefont {Kroeze}}, \bibinfo {author} {\bibfnamefont {Y.}~\bibnamefont {Guo}}, \bibinfo {author} {\bibfnamefont {V.~D.}\ \bibnamefont {Vaidya}}, \bibinfo {author} {\bibfnamefont {J.}~\bibnamefont {Keeling}},\ and\ \bibinfo {author} {\bibfnamefont {B.~L.}\ \bibnamefont {Lev}},\ }\href {https://doi.org/10.1103/PhysRevLett.121.163601} {\bibfield  {journal} {\bibinfo  {journal} {Phys. Rev. Lett.}\ }\textbf {\bibinfo {volume} {121}},\ \bibinfo {pages} {163601} (\bibinfo {year} {2018})}\BibitemShut {NoStop}%
\bibitem [{\citenamefont {Zhiqiang}\ \emph {et~al.}(2017)\citenamefont {Zhiqiang}, \citenamefont {Lee}, \citenamefont {Kumar}, \citenamefont {Arnold}, \citenamefont {Masson}, \citenamefont {Parkins},\ and\ \citenamefont {Barrett}}]{Zhiqiang2017}%
  \BibitemOpen
  \bibfield  {author} {\bibinfo {author} {\bibfnamefont {Z.}~\bibnamefont {Zhiqiang}}, \bibinfo {author} {\bibfnamefont {C.~H.}\ \bibnamefont {Lee}}, \bibinfo {author} {\bibfnamefont {R.}~\bibnamefont {Kumar}}, \bibinfo {author} {\bibfnamefont {K.~J.}\ \bibnamefont {Arnold}}, \bibinfo {author} {\bibfnamefont {S.~J.}\ \bibnamefont {Masson}}, \bibinfo {author} {\bibfnamefont {A.~S.}\ \bibnamefont {Parkins}},\ and\ \bibinfo {author} {\bibfnamefont {M.~D.}\ \bibnamefont {Barrett}},\ }\href {https://doi.org/10.1364/OPTICA.4.000424} {\bibfield  {journal} {\bibinfo  {journal} {Optica}\ }\textbf {\bibinfo {volume} {4}},\ \bibinfo {pages} {424} (\bibinfo {year} {2017})}\BibitemShut {NoStop}%
\bibitem [{\citenamefont {Marino}\ \emph {et~al.}(2022)\citenamefont {Marino}, \citenamefont {Eckstein}, \citenamefont {Foster},\ and\ \citenamefont {Rey}}]{Marino2022}%
  \BibitemOpen
  \bibfield  {author} {\bibinfo {author} {\bibfnamefont {J.}~\bibnamefont {Marino}}, \bibinfo {author} {\bibfnamefont {M.}~\bibnamefont {Eckstein}}, \bibinfo {author} {\bibfnamefont {M.~S.}\ \bibnamefont {Foster}},\ and\ \bibinfo {author} {\bibfnamefont {A.~M.}\ \bibnamefont {Rey}},\ }\href {https://doi.org/10.1088/1361-6633/ac906c} {\bibfield  {journal} {\bibinfo  {journal} {Reports on Progress in Physics}\ }\textbf {\bibinfo {volume} {85}},\ \bibinfo {pages} {116001} (\bibinfo {year} {2022})}\BibitemShut {NoStop}%
\bibitem [{\citenamefont {Senitzky}(1972)}]{Senitzky1972}%
  \BibitemOpen
  \bibfield  {author} {\bibinfo {author} {\bibfnamefont {I.~R.}\ \bibnamefont {Senitzky}},\ }\href {https://doi.org/10.1103/PhysRevA.6.1175} {\bibfield  {journal} {\bibinfo  {journal} {Phys. Rev. A}\ }\textbf {\bibinfo {volume} {6}},\ \bibinfo {pages} {1175} (\bibinfo {year} {1972})}\BibitemShut {NoStop}%
\bibitem [{\citenamefont {Drummond}\ and\ \citenamefont {Carmichael}(1978)}]{Drummond1978}%
  \BibitemOpen
  \bibfield  {author} {\bibinfo {author} {\bibfnamefont {P.}~\bibnamefont {Drummond}}\ and\ \bibinfo {author} {\bibfnamefont {H.}~\bibnamefont {Carmichael}},\ }\href {https://doi.org/https://doi.org/10.1016/0030-4018(78)90198-0} {\bibfield  {journal} {\bibinfo  {journal} {Optics Communications}\ }\textbf {\bibinfo {volume} {27}},\ \bibinfo {pages} {160} (\bibinfo {year} {1978})}\BibitemShut {NoStop}%
\bibitem [{\citenamefont {Narducci}\ \emph {et~al.}(1978)\citenamefont {Narducci}, \citenamefont {Feng}, \citenamefont {Gilmore},\ and\ \citenamefont {Agarwal}}]{Narducci1978}%
  \BibitemOpen
  \bibfield  {author} {\bibinfo {author} {\bibfnamefont {L.~M.}\ \bibnamefont {Narducci}}, \bibinfo {author} {\bibfnamefont {D.~H.}\ \bibnamefont {Feng}}, \bibinfo {author} {\bibfnamefont {R.}~\bibnamefont {Gilmore}},\ and\ \bibinfo {author} {\bibfnamefont {G.~S.}\ \bibnamefont {Agarwal}},\ }\href {https://doi.org/10.1103/PhysRevA.18.1571} {\bibfield  {journal} {\bibinfo  {journal} {Phys. Rev. A}\ }\textbf {\bibinfo {volume} {18}},\ \bibinfo {pages} {1571} (\bibinfo {year} {1978})}\BibitemShut {NoStop}%
\bibitem [{\citenamefont {Puri}\ and\ \citenamefont {Lawande}(1979)}]{Puri1979}%
  \BibitemOpen
  \bibfield  {author} {\bibinfo {author} {\bibfnamefont {R.}~\bibnamefont {Puri}}\ and\ \bibinfo {author} {\bibfnamefont {S.}~\bibnamefont {Lawande}},\ }\href {https://doi.org/https://doi.org/10.1016/0375-9601(79)90003-3} {\bibfield  {journal} {\bibinfo  {journal} {Physics Letters A}\ }\textbf {\bibinfo {volume} {72}},\ \bibinfo {pages} {200} (\bibinfo {year} {1979})}\BibitemShut {NoStop}%
\bibitem [{\citenamefont {Carmichael}(1980)}]{Carmichael1980}%
  \BibitemOpen
  \bibfield  {author} {\bibinfo {author} {\bibfnamefont {H.~J.}\ \bibnamefont {Carmichael}},\ }\href {https://doi.org/10.1088/0022-3700/13/18/009} {\bibfield  {journal} {\bibinfo  {journal} {Journal of Physics B: Atomic and Molecular Physics}\ }\textbf {\bibinfo {volume} {13}},\ \bibinfo {pages} {3551} (\bibinfo {year} {1980})}\BibitemShut {NoStop}%
\bibitem [{\citenamefont {Ferioli}\ \emph {et~al.}(2023)\citenamefont {Ferioli}, \citenamefont {Glicenstein}, \citenamefont {Ferrier-Barbut},\ and\ \citenamefont {Browaeys}}]{Ferioli2023}%
  \BibitemOpen
  \bibfield  {author} {\bibinfo {author} {\bibfnamefont {G.}~\bibnamefont {Ferioli}}, \bibinfo {author} {\bibfnamefont {A.}~\bibnamefont {Glicenstein}}, \bibinfo {author} {\bibfnamefont {I.}~\bibnamefont {Ferrier-Barbut}},\ and\ \bibinfo {author} {\bibfnamefont {A.}~\bibnamefont {Browaeys}},\ }\href {https://doi.org/10.1038/s41567-023-02064-w} {\bibfield  {journal} {\bibinfo  {journal} {Nature Physics}\ }\textbf {\bibinfo {volume} {19}},\ \bibinfo {pages} {1345} (\bibinfo {year} {2023})}\BibitemShut {NoStop}%
\bibitem [{\citenamefont {Pezz\`e}\ \emph {et~al.}(2018)\citenamefont {Pezz\`e}, \citenamefont {Smerzi}, \citenamefont {Oberthaler}, \citenamefont {Schmied},\ and\ \citenamefont {Treutlein}}]{Pezze2018}%
  \BibitemOpen
  \bibfield  {author} {\bibinfo {author} {\bibfnamefont {L.}~\bibnamefont {Pezz\`e}}, \bibinfo {author} {\bibfnamefont {A.}~\bibnamefont {Smerzi}}, \bibinfo {author} {\bibfnamefont {M.~K.}\ \bibnamefont {Oberthaler}}, \bibinfo {author} {\bibfnamefont {R.}~\bibnamefont {Schmied}},\ and\ \bibinfo {author} {\bibfnamefont {P.}~\bibnamefont {Treutlein}},\ }\href {https://doi.org/10.1103/RevModPhys.90.035005} {\bibfield  {journal} {\bibinfo  {journal} {Rev. Mod. Phys.}\ }\textbf {\bibinfo {volume} {90}},\ \bibinfo {pages} {035005} (\bibinfo {year} {2018})}\BibitemShut {NoStop}%
\bibitem [{\citenamefont {Lee}\ \emph {et~al.}(2014)\citenamefont {Lee}, \citenamefont {Chan},\ and\ \citenamefont {Yelin}}]{Lee2014}%
  \BibitemOpen
  \bibfield  {author} {\bibinfo {author} {\bibfnamefont {T.~E.}\ \bibnamefont {Lee}}, \bibinfo {author} {\bibfnamefont {C.-K.}\ \bibnamefont {Chan}},\ and\ \bibinfo {author} {\bibfnamefont {S.~F.}\ \bibnamefont {Yelin}},\ }\href {https://doi.org/10.1103/PhysRevA.90.052109} {\bibfield  {journal} {\bibinfo  {journal} {Phys. Rev. A}\ }\textbf {\bibinfo {volume} {90}},\ \bibinfo {pages} {052109} (\bibinfo {year} {2014})}\BibitemShut {NoStop}%
\bibitem [{\citenamefont {Young}\ \emph {et~al.}(2024{\natexlab{b}})\citenamefont {Young}, \citenamefont {Chaparro}, \citenamefont {Orioli}, \citenamefont {Thompson},\ and\ \citenamefont {Rey}}]{Young2024}%
  \BibitemOpen
  \bibfield  {author} {\bibinfo {author} {\bibfnamefont {J.~T.}\ \bibnamefont {Young}}, \bibinfo {author} {\bibfnamefont {E.}~\bibnamefont {Chaparro}}, \bibinfo {author} {\bibfnamefont {A.~P.}\ \bibnamefont {Orioli}}, \bibinfo {author} {\bibfnamefont {J.~K.}\ \bibnamefont {Thompson}},\ and\ \bibinfo {author} {\bibfnamefont {A.~M.}\ \bibnamefont {Rey}},\ }\href@noop {} {\bibinfo {title} {Engineering one axis twisting via a dissipative berry phase using strong symmetries}} (\bibinfo {year} {2024}{\natexlab{b}}),\ \Eprint {https://arxiv.org/abs/2401.06222} {arXiv:2401.06222 [quant-ph]} \BibitemShut {NoStop}%
\bibitem [{\citenamefont {Ye}\ and\ \citenamefont {Zoller}(2024)}]{Ye2024}%
  \BibitemOpen
  \bibfield  {author} {\bibinfo {author} {\bibfnamefont {J.}~\bibnamefont {Ye}}\ and\ \bibinfo {author} {\bibfnamefont {P.}~\bibnamefont {Zoller}},\ }\href {https://doi.org/10.1103/PhysRevLett.132.190001} {\bibfield  {journal} {\bibinfo  {journal} {Phys. Rev. Lett.}\ }\textbf {\bibinfo {volume} {132}},\ \bibinfo {pages} {190001} (\bibinfo {year} {2024})}\BibitemShut {NoStop}%
\bibitem [{\citenamefont {Georgescu}\ \emph {et~al.}(2014)\citenamefont {Georgescu}, \citenamefont {Ashhab},\ and\ \citenamefont {Nori}}]{Georgescu2014}%
  \BibitemOpen
  \bibfield  {author} {\bibinfo {author} {\bibfnamefont {I.~M.}\ \bibnamefont {Georgescu}}, \bibinfo {author} {\bibfnamefont {S.}~\bibnamefont {Ashhab}},\ and\ \bibinfo {author} {\bibfnamefont {F.}~\bibnamefont {Nori}},\ }\href {https://doi.org/10.1103/RevModPhys.86.153} {\bibfield  {journal} {\bibinfo  {journal} {Rev. Mod. Phys.}\ }\textbf {\bibinfo {volume} {86}},\ \bibinfo {pages} {153} (\bibinfo {year} {2014})}\BibitemShut {NoStop}%
\bibitem [{\citenamefont {Altman}\ \emph {et~al.}(2021)\citenamefont {Altman}, \citenamefont {Brown}, \citenamefont {Carleo}, \citenamefont {Carr}, \citenamefont {Demler}, \citenamefont {Chin}, \citenamefont {DeMarco}, \citenamefont {Economou}, \citenamefont {Eriksson}, \citenamefont {Fu}, \citenamefont {Greiner}, \citenamefont {Hazzard}, \citenamefont {Hulet}, \citenamefont {Koll\'ar}, \citenamefont {Lev}, \citenamefont {Lukin}, \citenamefont {Ma}, \citenamefont {Mi}, \citenamefont {Misra}, \citenamefont {Monroe}, \citenamefont {Murch}, \citenamefont {Nazario}, \citenamefont {Ni}, \citenamefont {Potter}, \citenamefont {Roushan}, \citenamefont {Saffman}, \citenamefont {Schleier-Smith}, \citenamefont {Siddiqi}, \citenamefont {Simmonds}, \citenamefont {Singh}, \citenamefont {Spielman}, \citenamefont {Temme}, \citenamefont {Weiss}, \citenamefont {Vu\ifmmode \check{c}\else \v{c}\fi{}kovi\ifmmode~\acute{c}\else \'{c}\fi{}}, \citenamefont {Vuleti\ifmmode~\acute{c}\else \'{c}\fi{}}, \citenamefont {Ye},\ and\
  \citenamefont {Zwierlein}}]{Altman2021}%
  \BibitemOpen
  \bibfield  {author} {\bibinfo {author} {\bibfnamefont {E.}~\bibnamefont {Altman}}, \bibinfo {author} {\bibfnamefont {K.~R.}\ \bibnamefont {Brown}}, \bibinfo {author} {\bibfnamefont {G.}~\bibnamefont {Carleo}}, \bibinfo {author} {\bibfnamefont {L.~D.}\ \bibnamefont {Carr}}, \bibinfo {author} {\bibfnamefont {E.}~\bibnamefont {Demler}}, \bibinfo {author} {\bibfnamefont {C.}~\bibnamefont {Chin}}, \bibinfo {author} {\bibfnamefont {B.}~\bibnamefont {DeMarco}}, \bibinfo {author} {\bibfnamefont {S.~E.}\ \bibnamefont {Economou}}, \bibinfo {author} {\bibfnamefont {M.~A.}\ \bibnamefont {Eriksson}}, \bibinfo {author} {\bibfnamefont {K.-M.~C.}\ \bibnamefont {Fu}}, \bibinfo {author} {\bibfnamefont {M.}~\bibnamefont {Greiner}}, \bibinfo {author} {\bibfnamefont {K.~R.}\ \bibnamefont {Hazzard}}, \bibinfo {author} {\bibfnamefont {R.~G.}\ \bibnamefont {Hulet}}, \bibinfo {author} {\bibfnamefont {A.~J.}\ \bibnamefont {Koll\'ar}}, \bibinfo {author} {\bibfnamefont {B.~L.}\ \bibnamefont {Lev}}, \bibinfo {author} {\bibfnamefont
  {M.~D.}\ \bibnamefont {Lukin}}, \bibinfo {author} {\bibfnamefont {R.}~\bibnamefont {Ma}}, \bibinfo {author} {\bibfnamefont {X.}~\bibnamefont {Mi}}, \bibinfo {author} {\bibfnamefont {S.}~\bibnamefont {Misra}}, \bibinfo {author} {\bibfnamefont {C.}~\bibnamefont {Monroe}}, \bibinfo {author} {\bibfnamefont {K.}~\bibnamefont {Murch}}, \bibinfo {author} {\bibfnamefont {Z.}~\bibnamefont {Nazario}}, \bibinfo {author} {\bibfnamefont {K.-K.}\ \bibnamefont {Ni}}, \bibinfo {author} {\bibfnamefont {A.~C.}\ \bibnamefont {Potter}}, \bibinfo {author} {\bibfnamefont {P.}~\bibnamefont {Roushan}}, \bibinfo {author} {\bibfnamefont {M.}~\bibnamefont {Saffman}}, \bibinfo {author} {\bibfnamefont {M.}~\bibnamefont {Schleier-Smith}}, \bibinfo {author} {\bibfnamefont {I.}~\bibnamefont {Siddiqi}}, \bibinfo {author} {\bibfnamefont {R.}~\bibnamefont {Simmonds}}, \bibinfo {author} {\bibfnamefont {M.}~\bibnamefont {Singh}}, \bibinfo {author} {\bibfnamefont {I.}~\bibnamefont {Spielman}}, \bibinfo {author} {\bibfnamefont {K.}~\bibnamefont
  {Temme}}, \bibinfo {author} {\bibfnamefont {D.~S.}\ \bibnamefont {Weiss}}, \bibinfo {author} {\bibfnamefont {J.}~\bibnamefont {Vu\ifmmode \check{c}\else \v{c}\fi{}kovi\ifmmode~\acute{c}\else \'{c}\fi{}}}, \bibinfo {author} {\bibfnamefont {V.}~\bibnamefont {Vuleti\ifmmode~\acute{c}\else \'{c}\fi{}}}, \bibinfo {author} {\bibfnamefont {J.}~\bibnamefont {Ye}},\ and\ \bibinfo {author} {\bibfnamefont {M.}~\bibnamefont {Zwierlein}},\ }\href {https://doi.org/10.1103/PRXQuantum.2.017003} {\bibfield  {journal} {\bibinfo  {journal} {PRX Quantum}\ }\textbf {\bibinfo {volume} {2}},\ \bibinfo {pages} {017003} (\bibinfo {year} {2021})}\BibitemShut {NoStop}%
\bibitem [{\citenamefont {Daley}(2023)}]{Daley2023}%
  \BibitemOpen
  \bibfield  {author} {\bibinfo {author} {\bibfnamefont {A.~J.}\ \bibnamefont {Daley}},\ }\href {https://doi.org/10.1038/s42254-023-00666-0} {\bibfield  {journal} {\bibinfo  {journal} {Nature Reviews Physics}\ }\textbf {\bibinfo {volume} {5}},\ \bibinfo {pages} {702} (\bibinfo {year} {2023})}\BibitemShut {NoStop}%
\bibitem [{\citenamefont {Bharti}\ \emph {et~al.}(2022)\citenamefont {Bharti}, \citenamefont {Cervera-Lierta}, \citenamefont {Kyaw}, \citenamefont {Haug}, \citenamefont {Alperin-Lea}, \citenamefont {Anand}, \citenamefont {Degroote}, \citenamefont {Heimonen}, \citenamefont {Kottmann}, \citenamefont {Menke}, \citenamefont {Mok}, \citenamefont {Sim}, \citenamefont {Kwek},\ and\ \citenamefont {Aspuru-Guzik}}]{Bharti2022}%
  \BibitemOpen
  \bibfield  {author} {\bibinfo {author} {\bibfnamefont {K.}~\bibnamefont {Bharti}}, \bibinfo {author} {\bibfnamefont {A.}~\bibnamefont {Cervera-Lierta}}, \bibinfo {author} {\bibfnamefont {T.~H.}\ \bibnamefont {Kyaw}}, \bibinfo {author} {\bibfnamefont {T.}~\bibnamefont {Haug}}, \bibinfo {author} {\bibfnamefont {S.}~\bibnamefont {Alperin-Lea}}, \bibinfo {author} {\bibfnamefont {A.}~\bibnamefont {Anand}}, \bibinfo {author} {\bibfnamefont {M.}~\bibnamefont {Degroote}}, \bibinfo {author} {\bibfnamefont {H.}~\bibnamefont {Heimonen}}, \bibinfo {author} {\bibfnamefont {J.~S.}\ \bibnamefont {Kottmann}}, \bibinfo {author} {\bibfnamefont {T.}~\bibnamefont {Menke}}, \bibinfo {author} {\bibfnamefont {W.-K.}\ \bibnamefont {Mok}}, \bibinfo {author} {\bibfnamefont {S.}~\bibnamefont {Sim}}, \bibinfo {author} {\bibfnamefont {L.-C.}\ \bibnamefont {Kwek}},\ and\ \bibinfo {author} {\bibfnamefont {A.}~\bibnamefont {Aspuru-Guzik}},\ }\href {https://doi.org/10.1103/RevModPhys.94.015004} {\bibfield  {journal} {\bibinfo  {journal}
  {Rev. Mod. Phys.}\ }\textbf {\bibinfo {volume} {94}},\ \bibinfo {pages} {015004} (\bibinfo {year} {2022})}\BibitemShut {NoStop}%
\bibitem [{\citenamefont {Couteau}\ \emph {et~al.}(2023)\citenamefont {Couteau}, \citenamefont {Barz}, \citenamefont {Durt}, \citenamefont {Gerrits}, \citenamefont {Huwer}, \citenamefont {Prevedel}, \citenamefont {Rarity}, \citenamefont {Shields},\ and\ \citenamefont {Weihs}}]{Couteau2023}%
  \BibitemOpen
  \bibfield  {author} {\bibinfo {author} {\bibfnamefont {C.}~\bibnamefont {Couteau}}, \bibinfo {author} {\bibfnamefont {S.}~\bibnamefont {Barz}}, \bibinfo {author} {\bibfnamefont {T.}~\bibnamefont {Durt}}, \bibinfo {author} {\bibfnamefont {T.}~\bibnamefont {Gerrits}}, \bibinfo {author} {\bibfnamefont {J.}~\bibnamefont {Huwer}}, \bibinfo {author} {\bibfnamefont {R.}~\bibnamefont {Prevedel}}, \bibinfo {author} {\bibfnamefont {J.}~\bibnamefont {Rarity}}, \bibinfo {author} {\bibfnamefont {A.}~\bibnamefont {Shields}},\ and\ \bibinfo {author} {\bibfnamefont {G.}~\bibnamefont {Weihs}},\ }\href {https://doi.org/10.1038/s42254-023-00583-2} {\bibfield  {journal} {\bibinfo  {journal} {Nature Reviews Physics}\ }\textbf {\bibinfo {volume} {5}},\ \bibinfo {pages} {326} (\bibinfo {year} {2023})}\BibitemShut {NoStop}%
\bibitem [{\citenamefont {Dicke}(1954)}]{Dicke1954}%
  \BibitemOpen
  \bibfield  {author} {\bibinfo {author} {\bibfnamefont {R.~H.}\ \bibnamefont {Dicke}},\ }\href {https://doi.org/10.1103/PhysRev.93.99} {\bibfield  {journal} {\bibinfo  {journal} {Phys. Rev.}\ }\textbf {\bibinfo {volume} {93}},\ \bibinfo {pages} {99} (\bibinfo {year} {1954})}\BibitemShut {NoStop}%
\bibitem [{\citenamefont {Gross}\ and\ \citenamefont {Haroche}(1982)}]{Gross1982}%
  \BibitemOpen
  \bibfield  {author} {\bibinfo {author} {\bibfnamefont {M.}~\bibnamefont {Gross}}\ and\ \bibinfo {author} {\bibfnamefont {S.}~\bibnamefont {Haroche}},\ }\href {https://doi.org/https://doi.org/10.1016/0370-1573(82)90102-8} {\bibfield  {journal} {\bibinfo  {journal} {Physics Reports}\ }\textbf {\bibinfo {volume} {93}},\ \bibinfo {pages} {301} (\bibinfo {year} {1982})}\BibitemShut {NoStop}%
\bibitem [{\citenamefont {Bohnet}\ \emph {et~al.}(2012)\citenamefont {Bohnet}, \citenamefont {Chen}, \citenamefont {Weiner}, \citenamefont {Meiser}, \citenamefont {Holland},\ and\ \citenamefont {Thompson}}]{Bohnet2012}%
  \BibitemOpen
  \bibfield  {author} {\bibinfo {author} {\bibfnamefont {J.~G.}\ \bibnamefont {Bohnet}}, \bibinfo {author} {\bibfnamefont {Z.}~\bibnamefont {Chen}}, \bibinfo {author} {\bibfnamefont {J.~M.}\ \bibnamefont {Weiner}}, \bibinfo {author} {\bibfnamefont {D.}~\bibnamefont {Meiser}}, \bibinfo {author} {\bibfnamefont {M.~J.}\ \bibnamefont {Holland}},\ and\ \bibinfo {author} {\bibfnamefont {J.~K.}\ \bibnamefont {Thompson}},\ }\href {https://doi.org/10.1038/nature10920} {\bibfield  {journal} {\bibinfo  {journal} {Nature}\ }\textbf {\bibinfo {volume} {484}},\ \bibinfo {pages} {78} (\bibinfo {year} {2012})}\BibitemShut {NoStop}%
\bibitem [{\citenamefont {Norcia}\ \emph {et~al.}(2016)\citenamefont {Norcia}, \citenamefont {Winchester}, \citenamefont {Cline},\ and\ \citenamefont {Thompson}}]{Norcia2016_3}%
  \BibitemOpen
  \bibfield  {author} {\bibinfo {author} {\bibfnamefont {M.~A.}\ \bibnamefont {Norcia}}, \bibinfo {author} {\bibfnamefont {M.~N.}\ \bibnamefont {Winchester}}, \bibinfo {author} {\bibfnamefont {J.~R.~K.}\ \bibnamefont {Cline}},\ and\ \bibinfo {author} {\bibfnamefont {J.~K.}\ \bibnamefont {Thompson}},\ }\href {https://doi.org/10.1126/sciadv.1601231} {\bibfield  {journal} {\bibinfo  {journal} {Science Advances}\ }\textbf {\bibinfo {volume} {2}},\ \bibinfo {pages} {e1601231} (\bibinfo {year} {2016})}\BibitemShut {NoStop}%
\bibitem [{\citenamefont {Inouye}\ \emph {et~al.}(1999)\citenamefont {Inouye}, \citenamefont {Chikkatur}, \citenamefont {Stamper-Kurn}, \citenamefont {Stenger}, \citenamefont {Pritchard},\ and\ \citenamefont {Ketterle}}]{Inouye1999}%
  \BibitemOpen
  \bibfield  {author} {\bibinfo {author} {\bibfnamefont {S.}~\bibnamefont {Inouye}}, \bibinfo {author} {\bibfnamefont {A.~P.}\ \bibnamefont {Chikkatur}}, \bibinfo {author} {\bibfnamefont {D.~M.}\ \bibnamefont {Stamper-Kurn}}, \bibinfo {author} {\bibfnamefont {J.}~\bibnamefont {Stenger}}, \bibinfo {author} {\bibfnamefont {D.~E.}\ \bibnamefont {Pritchard}},\ and\ \bibinfo {author} {\bibfnamefont {W.}~\bibnamefont {Ketterle}},\ }\href {https://doi.org/10.1126/science.285.5427.571} {\bibfield  {journal} {\bibinfo  {journal} {Science}\ }\textbf {\bibinfo {volume} {285}},\ \bibinfo {pages} {571} (\bibinfo {year} {1999})}\BibitemShut {NoStop}%
\bibitem [{\citenamefont {Goban}\ \emph {et~al.}(2015)\citenamefont {Goban}, \citenamefont {Hung}, \citenamefont {Hood}, \citenamefont {Yu}, \citenamefont {Muniz}, \citenamefont {Painter},\ and\ \citenamefont {Kimble}}]{Goban2015}%
  \BibitemOpen
  \bibfield  {author} {\bibinfo {author} {\bibfnamefont {A.}~\bibnamefont {Goban}}, \bibinfo {author} {\bibfnamefont {C.-L.}\ \bibnamefont {Hung}}, \bibinfo {author} {\bibfnamefont {J.~D.}\ \bibnamefont {Hood}}, \bibinfo {author} {\bibfnamefont {S.-P.}\ \bibnamefont {Yu}}, \bibinfo {author} {\bibfnamefont {J.~A.}\ \bibnamefont {Muniz}}, \bibinfo {author} {\bibfnamefont {O.}~\bibnamefont {Painter}},\ and\ \bibinfo {author} {\bibfnamefont {H.~J.}\ \bibnamefont {Kimble}},\ }\href {https://doi.org/10.1103/PhysRevLett.115.063601} {\bibfield  {journal} {\bibinfo  {journal} {Phys. Rev. Lett.}\ }\textbf {\bibinfo {volume} {115}},\ \bibinfo {pages} {063601} (\bibinfo {year} {2015})}\BibitemShut {NoStop}%
\bibitem [{\citenamefont {Liedl}\ \emph {et~al.}(2024)\citenamefont {Liedl}, \citenamefont {Tebbenjohanns}, \citenamefont {Bach}, \citenamefont {Pucher}, \citenamefont {Rauschenbeutel},\ and\ \citenamefont {Schneeweiss}}]{Liedl2024}%
  \BibitemOpen
  \bibfield  {author} {\bibinfo {author} {\bibfnamefont {C.}~\bibnamefont {Liedl}}, \bibinfo {author} {\bibfnamefont {F.}~\bibnamefont {Tebbenjohanns}}, \bibinfo {author} {\bibfnamefont {C.}~\bibnamefont {Bach}}, \bibinfo {author} {\bibfnamefont {S.}~\bibnamefont {Pucher}}, \bibinfo {author} {\bibfnamefont {A.}~\bibnamefont {Rauschenbeutel}},\ and\ \bibinfo {author} {\bibfnamefont {P.}~\bibnamefont {Schneeweiss}},\ }\href {https://doi.org/10.1103/PhysRevX.14.011020} {\bibfield  {journal} {\bibinfo  {journal} {Phys. Rev. X}\ }\textbf {\bibinfo {volume} {14}},\ \bibinfo {pages} {011020} (\bibinfo {year} {2024})}\BibitemShut {NoStop}%
\bibitem [{\citenamefont {Fitzpatrick}\ \emph {et~al.}(2017)\citenamefont {Fitzpatrick}, \citenamefont {Sundaresan}, \citenamefont {Li}, \citenamefont {Koch},\ and\ \citenamefont {Houck}}]{Fitzpatrick2017}%
  \BibitemOpen
  \bibfield  {author} {\bibinfo {author} {\bibfnamefont {M.}~\bibnamefont {Fitzpatrick}}, \bibinfo {author} {\bibfnamefont {N.~M.}\ \bibnamefont {Sundaresan}}, \bibinfo {author} {\bibfnamefont {A.~C.~Y.}\ \bibnamefont {Li}}, \bibinfo {author} {\bibfnamefont {J.}~\bibnamefont {Koch}},\ and\ \bibinfo {author} {\bibfnamefont {A.~A.}\ \bibnamefont {Houck}},\ }\href {https://doi.org/10.1103/PhysRevX.7.011016} {\bibfield  {journal} {\bibinfo  {journal} {Phys. Rev. X}\ }\textbf {\bibinfo {volume} {7}},\ \bibinfo {pages} {011016} (\bibinfo {year} {2017})}\BibitemShut {NoStop}%
\bibitem [{\citenamefont {Lei}\ \emph {et~al.}(2023)\citenamefont {Lei}, \citenamefont {Fukumori}, \citenamefont {Rochman}, \citenamefont {Zhu}, \citenamefont {Endres}, \citenamefont {Choi},\ and\ \citenamefont {Faraon}}]{Lei2023}%
  \BibitemOpen
  \bibfield  {author} {\bibinfo {author} {\bibfnamefont {M.}~\bibnamefont {Lei}}, \bibinfo {author} {\bibfnamefont {R.}~\bibnamefont {Fukumori}}, \bibinfo {author} {\bibfnamefont {J.}~\bibnamefont {Rochman}}, \bibinfo {author} {\bibfnamefont {B.}~\bibnamefont {Zhu}}, \bibinfo {author} {\bibfnamefont {M.}~\bibnamefont {Endres}}, \bibinfo {author} {\bibfnamefont {J.}~\bibnamefont {Choi}},\ and\ \bibinfo {author} {\bibfnamefont {A.}~\bibnamefont {Faraon}},\ }\href {https://doi.org/10.1038/s41586-023-05884-1} {\bibfield  {journal} {\bibinfo  {journal} {Nature}\ }\textbf {\bibinfo {volume} {617}},\ \bibinfo {pages} {271} (\bibinfo {year} {2023})}\BibitemShut {NoStop}%
\bibitem [{\citenamefont {Kersten}\ \emph {et~al.}(2023)\citenamefont {Kersten}, \citenamefont {de~Zordo}, \citenamefont {Diekmann}, \citenamefont {Reiter}, \citenamefont {Zens}, \citenamefont {Kanagin}, \citenamefont {Rotter}, \citenamefont {Schmiedmayer},\ and\ \citenamefont {Angerer}}]{Kersten2023}%
  \BibitemOpen
  \bibfield  {author} {\bibinfo {author} {\bibfnamefont {W.}~\bibnamefont {Kersten}}, \bibinfo {author} {\bibfnamefont {N.}~\bibnamefont {de~Zordo}}, \bibinfo {author} {\bibfnamefont {O.}~\bibnamefont {Diekmann}}, \bibinfo {author} {\bibfnamefont {T.}~\bibnamefont {Reiter}}, \bibinfo {author} {\bibfnamefont {M.}~\bibnamefont {Zens}}, \bibinfo {author} {\bibfnamefont {A.~N.}\ \bibnamefont {Kanagin}}, \bibinfo {author} {\bibfnamefont {S.}~\bibnamefont {Rotter}}, \bibinfo {author} {\bibfnamefont {J.}~\bibnamefont {Schmiedmayer}},\ and\ \bibinfo {author} {\bibfnamefont {A.}~\bibnamefont {Angerer}},\ }\href {https://doi.org/10.1103/PhysRevLett.131.043601} {\bibfield  {journal} {\bibinfo  {journal} {Phys. Rev. Lett.}\ }\textbf {\bibinfo {volume} {131}},\ \bibinfo {pages} {043601} (\bibinfo {year} {2023})}\BibitemShut {NoStop}%
\bibitem [{\citenamefont {Norcia}\ \emph {et~al.}(2018{\natexlab{a}})\citenamefont {Norcia}, \citenamefont {Cline}, \citenamefont {Muniz}, \citenamefont {Robinson}, \citenamefont {Hutson}, \citenamefont {Goban}, \citenamefont {Marti}, \citenamefont {Ye},\ and\ \citenamefont {Thompson}}]{Norcia2018_2}%
  \BibitemOpen
  \bibfield  {author} {\bibinfo {author} {\bibfnamefont {M.~A.}\ \bibnamefont {Norcia}}, \bibinfo {author} {\bibfnamefont {J.~R.~K.}\ \bibnamefont {Cline}}, \bibinfo {author} {\bibfnamefont {J.~A.}\ \bibnamefont {Muniz}}, \bibinfo {author} {\bibfnamefont {J.~M.}\ \bibnamefont {Robinson}}, \bibinfo {author} {\bibfnamefont {R.~B.}\ \bibnamefont {Hutson}}, \bibinfo {author} {\bibfnamefont {A.}~\bibnamefont {Goban}}, \bibinfo {author} {\bibfnamefont {G.~E.}\ \bibnamefont {Marti}}, \bibinfo {author} {\bibfnamefont {J.}~\bibnamefont {Ye}},\ and\ \bibinfo {author} {\bibfnamefont {J.~K.}\ \bibnamefont {Thompson}},\ }\href {https://doi.org/10.1103/PhysRevX.8.021036} {\bibfield  {journal} {\bibinfo  {journal} {Phys. Rev. X}\ }\textbf {\bibinfo {volume} {8}},\ \bibinfo {pages} {021036} (\bibinfo {year} {2018}{\natexlab{a}})}\BibitemShut {NoStop}%
\bibitem [{\citenamefont {Kristensen}\ \emph {et~al.}(2023)\citenamefont {Kristensen}, \citenamefont {Bohr}, \citenamefont {Robinson-Tait}, \citenamefont {Zelevinsky}, \citenamefont {Thomsen},\ and\ \citenamefont {M\"uller}}]{Kristensen2023}%
  \BibitemOpen
  \bibfield  {author} {\bibinfo {author} {\bibfnamefont {S.~L.}\ \bibnamefont {Kristensen}}, \bibinfo {author} {\bibfnamefont {E.}~\bibnamefont {Bohr}}, \bibinfo {author} {\bibfnamefont {J.}~\bibnamefont {Robinson-Tait}}, \bibinfo {author} {\bibfnamefont {T.}~\bibnamefont {Zelevinsky}}, \bibinfo {author} {\bibfnamefont {J.~W.}\ \bibnamefont {Thomsen}},\ and\ \bibinfo {author} {\bibfnamefont {J.~H.}\ \bibnamefont {M\"uller}},\ }\href {https://doi.org/10.1103/PhysRevLett.130.223402} {\bibfield  {journal} {\bibinfo  {journal} {Phys. Rev. Lett.}\ }\textbf {\bibinfo {volume} {130}},\ \bibinfo {pages} {223402} (\bibinfo {year} {2023})}\BibitemShut {NoStop}%
\bibitem [{\citenamefont {Hepp}\ and\ \citenamefont {Lieb}(1973)}]{Hepp1973}%
  \BibitemOpen
  \bibfield  {author} {\bibinfo {author} {\bibfnamefont {K.}~\bibnamefont {Hepp}}\ and\ \bibinfo {author} {\bibfnamefont {E.~H.}\ \bibnamefont {Lieb}},\ }\href {https://doi.org/10.1103/PhysRevA.8.2517} {\bibfield  {journal} {\bibinfo  {journal} {Phys. Rev. A}\ }\textbf {\bibinfo {volume} {8}},\ \bibinfo {pages} {2517} (\bibinfo {year} {1973})}\BibitemShut {NoStop}%
\bibitem [{\citenamefont {Dimer}\ \emph {et~al.}(2007)\citenamefont {Dimer}, \citenamefont {Estienne}, \citenamefont {Parkins},\ and\ \citenamefont {Carmichael}}]{Dimer2007}%
  \BibitemOpen
  \bibfield  {author} {\bibinfo {author} {\bibfnamefont {F.}~\bibnamefont {Dimer}}, \bibinfo {author} {\bibfnamefont {B.}~\bibnamefont {Estienne}}, \bibinfo {author} {\bibfnamefont {A.~S.}\ \bibnamefont {Parkins}},\ and\ \bibinfo {author} {\bibfnamefont {H.~J.}\ \bibnamefont {Carmichael}},\ }\href {https://doi.org/10.1103/PhysRevA.75.013804} {\bibfield  {journal} {\bibinfo  {journal} {Phys. Rev. A}\ }\textbf {\bibinfo {volume} {75}},\ \bibinfo {pages} {013804} (\bibinfo {year} {2007})}\BibitemShut {NoStop}%
\bibitem [{\citenamefont {Ferri}\ \emph {et~al.}(2021)\citenamefont {Ferri}, \citenamefont {Rosa-Medina}, \citenamefont {Finger}, \citenamefont {Dogra}, \citenamefont {Soriente}, \citenamefont {Zilberberg}, \citenamefont {Donner},\ and\ \citenamefont {Esslinger}}]{Ferri2021}%
  \BibitemOpen
  \bibfield  {author} {\bibinfo {author} {\bibfnamefont {F.}~\bibnamefont {Ferri}}, \bibinfo {author} {\bibfnamefont {R.}~\bibnamefont {Rosa-Medina}}, \bibinfo {author} {\bibfnamefont {F.}~\bibnamefont {Finger}}, \bibinfo {author} {\bibfnamefont {N.}~\bibnamefont {Dogra}}, \bibinfo {author} {\bibfnamefont {M.}~\bibnamefont {Soriente}}, \bibinfo {author} {\bibfnamefont {O.}~\bibnamefont {Zilberberg}}, \bibinfo {author} {\bibfnamefont {T.}~\bibnamefont {Donner}},\ and\ \bibinfo {author} {\bibfnamefont {T.}~\bibnamefont {Esslinger}},\ }\href {https://doi.org/10.1103/PhysRevX.11.041046} {\bibfield  {journal} {\bibinfo  {journal} {Phys. Rev. X}\ }\textbf {\bibinfo {volume} {11}},\ \bibinfo {pages} {041046} (\bibinfo {year} {2021})}\BibitemShut {NoStop}%
\bibitem [{\citenamefont {Safavi-Naini}\ \emph {et~al.}(2018)\citenamefont {Safavi-Naini}, \citenamefont {Lewis-Swan}, \citenamefont {Bohnet}, \citenamefont {G\"arttner}, \citenamefont {Gilmore}, \citenamefont {Jordan}, \citenamefont {Cohn}, \citenamefont {Freericks}, \citenamefont {Rey},\ and\ \citenamefont {Bollinger}}]{Safavi-Naini2018}%
  \BibitemOpen
  \bibfield  {author} {\bibinfo {author} {\bibfnamefont {A.}~\bibnamefont {Safavi-Naini}}, \bibinfo {author} {\bibfnamefont {R.~J.}\ \bibnamefont {Lewis-Swan}}, \bibinfo {author} {\bibfnamefont {J.~G.}\ \bibnamefont {Bohnet}}, \bibinfo {author} {\bibfnamefont {M.}~\bibnamefont {G\"arttner}}, \bibinfo {author} {\bibfnamefont {K.~A.}\ \bibnamefont {Gilmore}}, \bibinfo {author} {\bibfnamefont {J.~E.}\ \bibnamefont {Jordan}}, \bibinfo {author} {\bibfnamefont {J.}~\bibnamefont {Cohn}}, \bibinfo {author} {\bibfnamefont {J.~K.}\ \bibnamefont {Freericks}}, \bibinfo {author} {\bibfnamefont {A.~M.}\ \bibnamefont {Rey}},\ and\ \bibinfo {author} {\bibfnamefont {J.~J.}\ \bibnamefont {Bollinger}},\ }\href {https://doi.org/10.1103/PhysRevLett.121.040503} {\bibfield  {journal} {\bibinfo  {journal} {Phys. Rev. Lett.}\ }\textbf {\bibinfo {volume} {121}},\ \bibinfo {pages} {040503} (\bibinfo {year} {2018})}\BibitemShut {NoStop}%
\bibitem [{\citenamefont {Meiser}\ \emph {et~al.}(2009)\citenamefont {Meiser}, \citenamefont {Ye}, \citenamefont {Carlson},\ and\ \citenamefont {Holland}}]{Meiser2009}%
  \BibitemOpen
  \bibfield  {author} {\bibinfo {author} {\bibfnamefont {D.}~\bibnamefont {Meiser}}, \bibinfo {author} {\bibfnamefont {J.}~\bibnamefont {Ye}}, \bibinfo {author} {\bibfnamefont {D.~R.}\ \bibnamefont {Carlson}},\ and\ \bibinfo {author} {\bibfnamefont {M.~J.}\ \bibnamefont {Holland}},\ }\href {https://doi.org/10.1103/PhysRevLett.102.163601} {\bibfield  {journal} {\bibinfo  {journal} {Phys. Rev. Lett.}\ }\textbf {\bibinfo {volume} {102}},\ \bibinfo {pages} {163601} (\bibinfo {year} {2009})}\BibitemShut {NoStop}%
\bibitem [{\citenamefont {Schneider}\ and\ \citenamefont {Milburn}(2002)}]{Schneider2002}%
  \BibitemOpen
  \bibfield  {author} {\bibinfo {author} {\bibfnamefont {S.}~\bibnamefont {Schneider}}\ and\ \bibinfo {author} {\bibfnamefont {G.~J.}\ \bibnamefont {Milburn}},\ }\href {https://doi.org/10.1103/PhysRevA.65.042107} {\bibfield  {journal} {\bibinfo  {journal} {Phys. Rev. A}\ }\textbf {\bibinfo {volume} {65}},\ \bibinfo {pages} {042107} (\bibinfo {year} {2002})}\BibitemShut {NoStop}%
\bibitem [{\citenamefont {Morrison}\ and\ \citenamefont {Parkins}(2008)}]{Morrison_2008}%
  \BibitemOpen
  \bibfield  {author} {\bibinfo {author} {\bibfnamefont {S.}~\bibnamefont {Morrison}}\ and\ \bibinfo {author} {\bibfnamefont {A.~S.}\ \bibnamefont {Parkins}},\ }\href {https://doi.org/10.1088/0953-4075/41/19/195502} {\bibfield  {journal} {\bibinfo  {journal} {Journal of Physics B: Atomic, Molecular and Optical Physics}\ }\textbf {\bibinfo {volume} {41}},\ \bibinfo {pages} {195502} (\bibinfo {year} {2008})}\BibitemShut {NoStop}%
\bibitem [{\citenamefont {Hannukainen}\ and\ \citenamefont {Larson}(2018)}]{Hannukainen2018}%
  \BibitemOpen
  \bibfield  {author} {\bibinfo {author} {\bibfnamefont {J.}~\bibnamefont {Hannukainen}}\ and\ \bibinfo {author} {\bibfnamefont {J.}~\bibnamefont {Larson}},\ }\href {https://doi.org/10.1103/PhysRevA.98.042113} {\bibfield  {journal} {\bibinfo  {journal} {Phys. Rev. A}\ }\textbf {\bibinfo {volume} {98}},\ \bibinfo {pages} {042113} (\bibinfo {year} {2018})}\BibitemShut {NoStop}%
\bibitem [{\citenamefont {Mattes}\ \emph {et~al.}(2023)\citenamefont {Mattes}, \citenamefont {Lesanovsky},\ and\ \citenamefont {Carollo}}]{Mattes2023}%
  \BibitemOpen
  \bibfield  {author} {\bibinfo {author} {\bibfnamefont {R.}~\bibnamefont {Mattes}}, \bibinfo {author} {\bibfnamefont {I.}~\bibnamefont {Lesanovsky}},\ and\ \bibinfo {author} {\bibfnamefont {F.}~\bibnamefont {Carollo}},\ }\href {https://doi.org/10.1103/PhysRevA.108.062216} {\bibfield  {journal} {\bibinfo  {journal} {Phys. Rev. A}\ }\textbf {\bibinfo {volume} {108}},\ \bibinfo {pages} {062216} (\bibinfo {year} {2023})}\BibitemShut {NoStop}%
\bibitem [{\citenamefont {Iemini}\ \emph {et~al.}(2018)\citenamefont {Iemini}, \citenamefont {Russomanno}, \citenamefont {Keeling}, \citenamefont {Schir\`o}, \citenamefont {Dalmonte},\ and\ \citenamefont {Fazio}}]{Iemini2018}%
  \BibitemOpen
  \bibfield  {author} {\bibinfo {author} {\bibfnamefont {F.}~\bibnamefont {Iemini}}, \bibinfo {author} {\bibfnamefont {A.}~\bibnamefont {Russomanno}}, \bibinfo {author} {\bibfnamefont {J.}~\bibnamefont {Keeling}}, \bibinfo {author} {\bibfnamefont {M.}~\bibnamefont {Schir\`o}}, \bibinfo {author} {\bibfnamefont {M.}~\bibnamefont {Dalmonte}},\ and\ \bibinfo {author} {\bibfnamefont {R.}~\bibnamefont {Fazio}},\ }\href {https://doi.org/10.1103/PhysRevLett.121.035301} {\bibfield  {journal} {\bibinfo  {journal} {Phys. Rev. Lett.}\ }\textbf {\bibinfo {volume} {121}},\ \bibinfo {pages} {035301} (\bibinfo {year} {2018})}\BibitemShut {NoStop}%
\bibitem [{\citenamefont {Passarelli}\ \emph {et~al.}(2024)\citenamefont {Passarelli}, \citenamefont {Turkeshi}, \citenamefont {Russomanno}, \citenamefont {Lucignano}, \citenamefont {Schir\`o},\ and\ \citenamefont {Fazio}}]{Pasarelli2024}%
  \BibitemOpen
  \bibfield  {author} {\bibinfo {author} {\bibfnamefont {G.}~\bibnamefont {Passarelli}}, \bibinfo {author} {\bibfnamefont {X.}~\bibnamefont {Turkeshi}}, \bibinfo {author} {\bibfnamefont {A.}~\bibnamefont {Russomanno}}, \bibinfo {author} {\bibfnamefont {P.}~\bibnamefont {Lucignano}}, \bibinfo {author} {\bibfnamefont {M.}~\bibnamefont {Schir\`o}},\ and\ \bibinfo {author} {\bibfnamefont {R.}~\bibnamefont {Fazio}},\ }\href {https://doi.org/10.1103/PhysRevLett.132.163401} {\bibfield  {journal} {\bibinfo  {journal} {Phys. Rev. Lett.}\ }\textbf {\bibinfo {volume} {132}},\ \bibinfo {pages} {163401} (\bibinfo {year} {2024})}\BibitemShut {NoStop}%
\bibitem [{\citenamefont {Goncalves}\ \emph {et~al.}(2024)\citenamefont {Goncalves}, \citenamefont {Bombieri}, \citenamefont {Ferioli}, \citenamefont {Pancaldi}, \citenamefont {Ferrier-Barbut}, \citenamefont {Browaeys}, \citenamefont {Shahmoon},\ and\ \citenamefont {Chang}}]{Goncalves2024}%
  \BibitemOpen
  \bibfield  {author} {\bibinfo {author} {\bibfnamefont {D.}~\bibnamefont {Goncalves}}, \bibinfo {author} {\bibfnamefont {L.}~\bibnamefont {Bombieri}}, \bibinfo {author} {\bibfnamefont {G.}~\bibnamefont {Ferioli}}, \bibinfo {author} {\bibfnamefont {S.}~\bibnamefont {Pancaldi}}, \bibinfo {author} {\bibfnamefont {I.}~\bibnamefont {Ferrier-Barbut}}, \bibinfo {author} {\bibfnamefont {A.}~\bibnamefont {Browaeys}}, \bibinfo {author} {\bibfnamefont {E.}~\bibnamefont {Shahmoon}},\ and\ \bibinfo {author} {\bibfnamefont {D.~E.}\ \bibnamefont {Chang}},\ }\href@noop {} {\bibinfo {title} {Driven-dissipative phase separation in free-space atomic ensembles}} (\bibinfo {year} {2024}),\ \Eprint {https://arxiv.org/abs/2403.15237} {arXiv:2403.15237 [quant-ph]} \BibitemShut {NoStop}%
\bibitem [{\citenamefont {Agarwal}\ \emph {et~al.}(2024)\citenamefont {Agarwal}, \citenamefont {Chaparro}, \citenamefont {Barberena}, \citenamefont {Orioli}, \citenamefont {Ferioli}, \citenamefont {Pancaldi}, \citenamefont {Ferrier-Barbut}, \citenamefont {Browaeys},\ and\ \citenamefont {Rey}}]{Agarwal2024}%
  \BibitemOpen
  \bibfield  {author} {\bibinfo {author} {\bibfnamefont {S.}~\bibnamefont {Agarwal}}, \bibinfo {author} {\bibfnamefont {E.}~\bibnamefont {Chaparro}}, \bibinfo {author} {\bibfnamefont {D.}~\bibnamefont {Barberena}}, \bibinfo {author} {\bibfnamefont {A.~P.}\ \bibnamefont {Orioli}}, \bibinfo {author} {\bibfnamefont {G.}~\bibnamefont {Ferioli}}, \bibinfo {author} {\bibfnamefont {S.}~\bibnamefont {Pancaldi}}, \bibinfo {author} {\bibfnamefont {I.}~\bibnamefont {Ferrier-Barbut}}, \bibinfo {author} {\bibfnamefont {A.}~\bibnamefont {Browaeys}},\ and\ \bibinfo {author} {\bibfnamefont {A.~M.}\ \bibnamefont {Rey}},\ }\href@noop {} {\bibinfo {title} {Directional superradiance in a driven ultracold atomic gas in free-space}} (\bibinfo {year} {2024}),\ \Eprint {https://arxiv.org/abs/2403.15556} {arXiv:2403.15556 [cond-mat.quant-gas]} \BibitemShut {NoStop}%
\bibitem [{\citenamefont {Ruostekoski}(2024)}]{Ruostekoski2024}%
  \BibitemOpen
  \bibfield  {author} {\bibinfo {author} {\bibfnamefont {J.}~\bibnamefont {Ruostekoski}},\ }\href@noop {} {\bibinfo {title} {Superradiant phase transition in a large interacting driven atomic ensemble in free space}} (\bibinfo {year} {2024}),\ \Eprint {https://arxiv.org/abs/2404.12939} {arXiv:2404.12939 [quant-ph]} \BibitemShut {NoStop}%
\bibitem [{\citenamefont {Bonifacio}\ and\ \citenamefont {Lugiato}(1978)}]{Bonifacio1978}%
  \BibitemOpen
  \bibfield  {author} {\bibinfo {author} {\bibfnamefont {R.}~\bibnamefont {Bonifacio}}\ and\ \bibinfo {author} {\bibfnamefont {L.~A.}\ \bibnamefont {Lugiato}},\ }\href {https://doi.org/10.1103/PhysRevA.18.1129} {\bibfield  {journal} {\bibinfo  {journal} {Phys. Rev. A}\ }\textbf {\bibinfo {volume} {18}},\ \bibinfo {pages} {1129} (\bibinfo {year} {1978})}\BibitemShut {NoStop}%
\bibitem [{\citenamefont {Rosenberger}\ \emph {et~al.}(1983)\citenamefont {Rosenberger}, \citenamefont {Orozco},\ and\ \citenamefont {Kimble}}]{Rosenberger1983}%
  \BibitemOpen
  \bibfield  {author} {\bibinfo {author} {\bibfnamefont {A.~T.}\ \bibnamefont {Rosenberger}}, \bibinfo {author} {\bibfnamefont {L.~A.}\ \bibnamefont {Orozco}},\ and\ \bibinfo {author} {\bibfnamefont {H.~J.}\ \bibnamefont {Kimble}},\ }\href {https://doi.org/10.1103/PhysRevA.28.2569} {\bibfield  {journal} {\bibinfo  {journal} {Phys. Rev. A}\ }\textbf {\bibinfo {volume} {28}},\ \bibinfo {pages} {2569} (\bibinfo {year} {1983})}\BibitemShut {NoStop}%
\bibitem [{\citenamefont {Rempe}\ \emph {et~al.}(1991)\citenamefont {Rempe}, \citenamefont {Thompson}, \citenamefont {Brecha}, \citenamefont {Lee},\ and\ \citenamefont {Kimble}}]{Rempe1993}%
  \BibitemOpen
  \bibfield  {author} {\bibinfo {author} {\bibfnamefont {G.}~\bibnamefont {Rempe}}, \bibinfo {author} {\bibfnamefont {R.~J.}\ \bibnamefont {Thompson}}, \bibinfo {author} {\bibfnamefont {R.~J.}\ \bibnamefont {Brecha}}, \bibinfo {author} {\bibfnamefont {W.~D.}\ \bibnamefont {Lee}},\ and\ \bibinfo {author} {\bibfnamefont {H.~J.}\ \bibnamefont {Kimble}},\ }\href {https://doi.org/10.1103/PhysRevLett.67.1727} {\bibfield  {journal} {\bibinfo  {journal} {Phys. Rev. Lett.}\ }\textbf {\bibinfo {volume} {67}},\ \bibinfo {pages} {1727} (\bibinfo {year} {1991})}\BibitemShut {NoStop}%
\bibitem [{\citenamefont {Gripp}\ \emph {et~al.}(1996)\citenamefont {Gripp}, \citenamefont {Mielke}, \citenamefont {Orozco},\ and\ \citenamefont {Carmichael}}]{Gripp1996}%
  \BibitemOpen
  \bibfield  {author} {\bibinfo {author} {\bibfnamefont {J.}~\bibnamefont {Gripp}}, \bibinfo {author} {\bibfnamefont {S.~L.}\ \bibnamefont {Mielke}}, \bibinfo {author} {\bibfnamefont {L.~A.}\ \bibnamefont {Orozco}},\ and\ \bibinfo {author} {\bibfnamefont {H.~J.}\ \bibnamefont {Carmichael}},\ }\href {https://doi.org/10.1103/PhysRevA.54.R3746} {\bibfield  {journal} {\bibinfo  {journal} {Phys. Rev. A}\ }\textbf {\bibinfo {volume} {54}},\ \bibinfo {pages} {R3746} (\bibinfo {year} {1996})}\BibitemShut {NoStop}%
\bibitem [{\citenamefont {Rivero}\ \emph {et~al.}(2023)\citenamefont {Rivero}, \citenamefont {Jr}, \citenamefont {de~França}, \citenamefont {Teixeira}, \citenamefont {Slama},\ and\ \citenamefont {Courteille}}]{Rivero2023}%
  \BibitemOpen
  \bibfield  {author} {\bibinfo {author} {\bibfnamefont {D.}~\bibnamefont {Rivero}}, \bibinfo {author} {\bibfnamefont {C.~A.~P.}\ \bibnamefont {Jr}}, \bibinfo {author} {\bibfnamefont {G.~H.}\ \bibnamefont {de~França}}, \bibinfo {author} {\bibfnamefont {R.~C.}\ \bibnamefont {Teixeira}}, \bibinfo {author} {\bibfnamefont {S.}~\bibnamefont {Slama}},\ and\ \bibinfo {author} {\bibfnamefont {P.~W.}\ \bibnamefont {Courteille}},\ }\href {https://doi.org/10.1088/1367-2630/acf954} {\bibfield  {journal} {\bibinfo  {journal} {New Journal of Physics}\ }\textbf {\bibinfo {volume} {25}},\ \bibinfo {pages} {093053} (\bibinfo {year} {2023})}\BibitemShut {NoStop}%
\bibitem [{\citenamefont {Buča}\ and\ \citenamefont {Prosen}(2012)}]{Buca_2012}%
  \BibitemOpen
  \bibfield  {author} {\bibinfo {author} {\bibfnamefont {B.}~\bibnamefont {Buča}}\ and\ \bibinfo {author} {\bibfnamefont {T.}~\bibnamefont {Prosen}},\ }\href {https://doi.org/10.1088/1367-2630/14/7/073007} {\bibfield  {journal} {\bibinfo  {journal} {New Journal of Physics}\ }\textbf {\bibinfo {volume} {14}},\ \bibinfo {pages} {073007} (\bibinfo {year} {2012})}\BibitemShut {NoStop}%
\bibitem [{\citenamefont {Albert}\ and\ \citenamefont {Jiang}(2014)}]{Albert2014}%
  \BibitemOpen
  \bibfield  {author} {\bibinfo {author} {\bibfnamefont {V.~V.}\ \bibnamefont {Albert}}\ and\ \bibinfo {author} {\bibfnamefont {L.}~\bibnamefont {Jiang}},\ }\href {https://doi.org/10.1103/PhysRevA.89.022118} {\bibfield  {journal} {\bibinfo  {journal} {Phys. Rev. A}\ }\textbf {\bibinfo {volume} {89}},\ \bibinfo {pages} {022118} (\bibinfo {year} {2014})}\BibitemShut {NoStop}%
\bibitem [{\citenamefont {Roberts}\ and\ \citenamefont {Clerk}(2023)}]{Roberts2023}%
  \BibitemOpen
  \bibfield  {author} {\bibinfo {author} {\bibfnamefont {D.}~\bibnamefont {Roberts}}\ and\ \bibinfo {author} {\bibfnamefont {A.~A.}\ \bibnamefont {Clerk}},\ }\href {https://doi.org/10.1103/PhysRevLett.131.190403} {\bibfield  {journal} {\bibinfo  {journal} {Phys. Rev. Lett.}\ }\textbf {\bibinfo {volume} {131}},\ \bibinfo {pages} {190403} (\bibinfo {year} {2023})}\BibitemShut {NoStop}%
\bibitem [{\citenamefont {Martin}\ \emph {et~al.}(2011)\citenamefont {Martin}, \citenamefont {Meiser}, \citenamefont {Thomsen}, \citenamefont {Ye},\ and\ \citenamefont {Holland}}]{Martin2011}%
  \BibitemOpen
  \bibfield  {author} {\bibinfo {author} {\bibfnamefont {M.~J.}\ \bibnamefont {Martin}}, \bibinfo {author} {\bibfnamefont {D.}~\bibnamefont {Meiser}}, \bibinfo {author} {\bibfnamefont {J.~W.}\ \bibnamefont {Thomsen}}, \bibinfo {author} {\bibfnamefont {J.}~\bibnamefont {Ye}},\ and\ \bibinfo {author} {\bibfnamefont {M.~J.}\ \bibnamefont {Holland}},\ }\href {https://doi.org/10.1103/PhysRevA.84.063813} {\bibfield  {journal} {\bibinfo  {journal} {Phys. Rev. A}\ }\textbf {\bibinfo {volume} {84}},\ \bibinfo {pages} {063813} (\bibinfo {year} {2011})}\BibitemShut {NoStop}%
\bibitem [{\citenamefont {Barberena}\ \emph {et~al.}(2023)\citenamefont {Barberena}, \citenamefont {Lewis-Swan}, \citenamefont {Rey},\ and\ \citenamefont {Thompson}}]{Barberena2023}%
  \BibitemOpen
  \bibfield  {author} {\bibinfo {author} {\bibfnamefont {D.}~\bibnamefont {Barberena}}, \bibinfo {author} {\bibfnamefont {R.~J.}\ \bibnamefont {Lewis-Swan}}, \bibinfo {author} {\bibfnamefont {A.~M.}\ \bibnamefont {Rey}},\ and\ \bibinfo {author} {\bibfnamefont {J.~K.}\ \bibnamefont {Thompson}},\ }\href {https://doi.org/10.5802/crphys.146} {\bibfield  {journal} {\bibinfo  {journal} {Comptes Rendus. Physique}\ }\textbf {\bibinfo {volume} {24}},\ \bibinfo {pages} {55} (\bibinfo {year} {2023})}\BibitemShut {NoStop}%
\bibitem [{\citenamefont {Norcia}\ and\ \citenamefont {Thompson}(2016{\natexlab{a}})}]{Norcia2016}%
  \BibitemOpen
  \bibfield  {author} {\bibinfo {author} {\bibfnamefont {M.~A.}\ \bibnamefont {Norcia}}\ and\ \bibinfo {author} {\bibfnamefont {J.~K.}\ \bibnamefont {Thompson}},\ }\href {https://doi.org/10.1103/PhysRevA.93.023804} {\bibfield  {journal} {\bibinfo  {journal} {Phys. Rev. A}\ }\textbf {\bibinfo {volume} {93}},\ \bibinfo {pages} {023804} (\bibinfo {year} {2016}{\natexlab{a}})}\BibitemShut {NoStop}%
\bibitem [{\citenamefont {Walls}\ \emph {et~al.}(1978)\citenamefont {Walls}, \citenamefont {Drummond}, \citenamefont {Hassan},\ and\ \citenamefont {Carmichael}}]{Walls1978}%
  \BibitemOpen
  \bibfield  {author} {\bibinfo {author} {\bibfnamefont {D.~F.}\ \bibnamefont {Walls}}, \bibinfo {author} {\bibfnamefont {P.~D.}\ \bibnamefont {Drummond}}, \bibinfo {author} {\bibfnamefont {S.~S.}\ \bibnamefont {Hassan}},\ and\ \bibinfo {author} {\bibfnamefont {H.~J.}\ \bibnamefont {Carmichael}},\ }\href {https://doi.org/10.1143/PTPS.64.307} {\bibfield  {journal} {\bibinfo  {journal} {Progress of Theoretical Physics Supplement}\ }\textbf {\bibinfo {volume} {64}},\ \bibinfo {pages} {307} (\bibinfo {year} {1978})}\BibitemShut {NoStop}%
\bibitem [{\citenamefont {Somech}\ \emph {et~al.}(2023)\citenamefont {Somech}, \citenamefont {Shimshi},\ and\ \citenamefont {Shahmoon}}]{Somech2023}%
  \BibitemOpen
  \bibfield  {author} {\bibinfo {author} {\bibfnamefont {O.}~\bibnamefont {Somech}}, \bibinfo {author} {\bibfnamefont {Y.}~\bibnamefont {Shimshi}},\ and\ \bibinfo {author} {\bibfnamefont {E.}~\bibnamefont {Shahmoon}},\ }\href {https://doi.org/10.1103/PhysRevA.108.023725} {\bibfield  {journal} {\bibinfo  {journal} {Phys. Rev. A}\ }\textbf {\bibinfo {volume} {108}},\ \bibinfo {pages} {023725} (\bibinfo {year} {2023})}\BibitemShut {NoStop}%
\bibitem [{\citenamefont {Barberena}\ and\ \citenamefont {Rey}(2024)}]{Barberena2024}%
  \BibitemOpen
  \bibfield  {author} {\bibinfo {author} {\bibfnamefont {D.}~\bibnamefont {Barberena}}\ and\ \bibinfo {author} {\bibfnamefont {A.~M.}\ \bibnamefont {Rey}},\ }\href {https://doi.org/10.1103/PhysRevA.109.013709} {\bibfield  {journal} {\bibinfo  {journal} {Phys. Rev. A}\ }\textbf {\bibinfo {volume} {109}},\ \bibinfo {pages} {013709} (\bibinfo {year} {2024})}\BibitemShut {NoStop}%
\bibitem [{\citenamefont {Barberena}\ \emph {et~al.}(2019)\citenamefont {Barberena}, \citenamefont {Lewis-Swan}, \citenamefont {Thompson},\ and\ \citenamefont {Rey}}]{Barberena2019}%
  \BibitemOpen
  \bibfield  {author} {\bibinfo {author} {\bibfnamefont {D.}~\bibnamefont {Barberena}}, \bibinfo {author} {\bibfnamefont {R.~J.}\ \bibnamefont {Lewis-Swan}}, \bibinfo {author} {\bibfnamefont {J.~K.}\ \bibnamefont {Thompson}},\ and\ \bibinfo {author} {\bibfnamefont {A.~M.}\ \bibnamefont {Rey}},\ }\href {https://doi.org/10.1103/PhysRevA.99.053411} {\bibfield  {journal} {\bibinfo  {journal} {Phys. Rev. A}\ }\textbf {\bibinfo {volume} {99}},\ \bibinfo {pages} {053411} (\bibinfo {year} {2019})}\BibitemShut {NoStop}%
\bibitem [{\citenamefont {Bonifacio}\ and\ \citenamefont {Lugiato}(1976)}]{Bonifacio1976}%
  \BibitemOpen
  \bibfield  {author} {\bibinfo {author} {\bibfnamefont {R.}~\bibnamefont {Bonifacio}}\ and\ \bibinfo {author} {\bibfnamefont {L.}~\bibnamefont {Lugiato}},\ }\href {https://doi.org/https://doi.org/10.1016/0030-4018(76)90335-7} {\bibfield  {journal} {\bibinfo  {journal} {Optics Communications}\ }\textbf {\bibinfo {volume} {19}},\ \bibinfo {pages} {172} (\bibinfo {year} {1976})}\BibitemShut {NoStop}%
\bibitem [{\citenamefont {Leppenen}\ and\ \citenamefont {Shahmoon}(2024)}]{Leppenen2024}%
  \BibitemOpen
  \bibfield  {author} {\bibinfo {author} {\bibfnamefont {N.}~\bibnamefont {Leppenen}}\ and\ \bibinfo {author} {\bibfnamefont {E.}~\bibnamefont {Shahmoon}},\ }\href@noop {} {\bibinfo {title} {Quantum bistability at the interplay between collective and individual decay}} (\bibinfo {year} {2024}),\ \Eprint {https://arxiv.org/abs/2404.02134} {arXiv:2404.02134 [quant-ph]} \BibitemShut {NoStop}%
\bibitem [{\citenamefont {Norcia}\ and\ \citenamefont {Thompson}(2016{\natexlab{b}})}]{Norcia2016_2}%
  \BibitemOpen
  \bibfield  {author} {\bibinfo {author} {\bibfnamefont {M.~A.}\ \bibnamefont {Norcia}}\ and\ \bibinfo {author} {\bibfnamefont {J.~K.}\ \bibnamefont {Thompson}},\ }\href {https://doi.org/10.1103/PhysRevX.6.011025} {\bibfield  {journal} {\bibinfo  {journal} {Phys. Rev. X}\ }\textbf {\bibinfo {volume} {6}},\ \bibinfo {pages} {011025} (\bibinfo {year} {2016}{\natexlab{b}})}\BibitemShut {NoStop}%
\bibitem [{\citenamefont {Muniz}\ \emph {et~al.}(2020)\citenamefont {Muniz}, \citenamefont {Barberena}, \citenamefont {Lewis-Swan}, \citenamefont {Young}, \citenamefont {Cline}, \citenamefont {Rey},\ and\ \citenamefont {Thompson}}]{Muniz2020}%
  \BibitemOpen
  \bibfield  {author} {\bibinfo {author} {\bibfnamefont {J.~A.}\ \bibnamefont {Muniz}}, \bibinfo {author} {\bibfnamefont {D.}~\bibnamefont {Barberena}}, \bibinfo {author} {\bibfnamefont {R.~J.}\ \bibnamefont {Lewis-Swan}}, \bibinfo {author} {\bibfnamefont {D.~J.}\ \bibnamefont {Young}}, \bibinfo {author} {\bibfnamefont {J.~R.~K.}\ \bibnamefont {Cline}}, \bibinfo {author} {\bibfnamefont {A.~M.}\ \bibnamefont {Rey}},\ and\ \bibinfo {author} {\bibfnamefont {J.~K.}\ \bibnamefont {Thompson}},\ }\href {https://doi.org/10.1038/s41586-020-2224-x} {\bibfield  {journal} {\bibinfo  {journal} {Nature}\ }\textbf {\bibinfo {volume} {580}},\ \bibinfo {pages} {602} (\bibinfo {year} {2020})}\BibitemShut {NoStop}%
\bibitem [{\citenamefont {Norcia}\ \emph {et~al.}(2018{\natexlab{b}})\citenamefont {Norcia}, \citenamefont {Lewis-Swan}, \citenamefont {Cline}, \citenamefont {Zhu}, \citenamefont {Rey},\ and\ \citenamefont {Thompson}}]{Norcia2018}%
  \BibitemOpen
  \bibfield  {author} {\bibinfo {author} {\bibfnamefont {M.~A.}\ \bibnamefont {Norcia}}, \bibinfo {author} {\bibfnamefont {R.~J.}\ \bibnamefont {Lewis-Swan}}, \bibinfo {author} {\bibfnamefont {J.~R.~K.}\ \bibnamefont {Cline}}, \bibinfo {author} {\bibfnamefont {B.}~\bibnamefont {Zhu}}, \bibinfo {author} {\bibfnamefont {A.~M.}\ \bibnamefont {Rey}},\ and\ \bibinfo {author} {\bibfnamefont {J.~K.}\ \bibnamefont {Thompson}},\ }\href {https://doi.org/10.1126/science.aar3102} {\bibfield  {journal} {\bibinfo  {journal} {Science}\ }\textbf {\bibinfo {volume} {361}},\ \bibinfo {pages} {259} (\bibinfo {year} {2018}{\natexlab{b}})}\BibitemShut {NoStop}%
\bibitem [{\citenamefont {Pedrozo-Pe{\~{n}}afiel}\ \emph {et~al.}(2020)\citenamefont {Pedrozo-Pe{\~{n}}afiel}, \citenamefont {Colombo}, \citenamefont {Shu}, \citenamefont {Adiyatullin}, \citenamefont {Li}, \citenamefont {Mendez}, \citenamefont {Braverman}, \citenamefont {Kawasaki}, \citenamefont {Akamatsu}, \citenamefont {Xiao},\ and\ \citenamefont {Vuleti{\'{c}}}}]{Pedrozo2020}%
  \BibitemOpen
  \bibfield  {author} {\bibinfo {author} {\bibfnamefont {E.}~\bibnamefont {Pedrozo-Pe{\~{n}}afiel}}, \bibinfo {author} {\bibfnamefont {S.}~\bibnamefont {Colombo}}, \bibinfo {author} {\bibfnamefont {C.}~\bibnamefont {Shu}}, \bibinfo {author} {\bibfnamefont {A.~F.}\ \bibnamefont {Adiyatullin}}, \bibinfo {author} {\bibfnamefont {Z.}~\bibnamefont {Li}}, \bibinfo {author} {\bibfnamefont {E.}~\bibnamefont {Mendez}}, \bibinfo {author} {\bibfnamefont {B.}~\bibnamefont {Braverman}}, \bibinfo {author} {\bibfnamefont {A.}~\bibnamefont {Kawasaki}}, \bibinfo {author} {\bibfnamefont {D.}~\bibnamefont {Akamatsu}}, \bibinfo {author} {\bibfnamefont {Y.}~\bibnamefont {Xiao}},\ and\ \bibinfo {author} {\bibfnamefont {V.}~\bibnamefont {Vuleti{\'{c}}}},\ }\href {https://doi.org/10.1038/s41586-020-3006-1} {\bibfield  {journal} {\bibinfo  {journal} {Nature}\ }\textbf {\bibinfo {volume} {588}},\ \bibinfo {pages} {414} (\bibinfo {year} {2020})}\BibitemShut {NoStop}%
\bibitem [{\citenamefont {Robinson}\ \emph {et~al.}(2024)\citenamefont {Robinson}, \citenamefont {Miklos}, \citenamefont {Tso}, \citenamefont {Kennedy}, \citenamefont {Bothwell}, \citenamefont {Kedar}, \citenamefont {Thompson},\ and\ \citenamefont {Ye}}]{Robinson2024}%
  \BibitemOpen
  \bibfield  {author} {\bibinfo {author} {\bibfnamefont {J.~M.}\ \bibnamefont {Robinson}}, \bibinfo {author} {\bibfnamefont {M.}~\bibnamefont {Miklos}}, \bibinfo {author} {\bibfnamefont {Y.~M.}\ \bibnamefont {Tso}}, \bibinfo {author} {\bibfnamefont {C.~J.}\ \bibnamefont {Kennedy}}, \bibinfo {author} {\bibfnamefont {T.}~\bibnamefont {Bothwell}}, \bibinfo {author} {\bibfnamefont {D.}~\bibnamefont {Kedar}}, \bibinfo {author} {\bibfnamefont {J.~K.}\ \bibnamefont {Thompson}},\ and\ \bibinfo {author} {\bibfnamefont {J.}~\bibnamefont {Ye}},\ }\href {https://doi.org/10.1038/s41567-023-02310-1} {\bibfield  {journal} {\bibinfo  {journal} {Nature Physics}\ }\textbf {\bibinfo {volume} {20}},\ \bibinfo {pages} {208} (\bibinfo {year} {2024})}\BibitemShut {NoStop}%
\bibitem [{\citenamefont {Ivanov}\ \emph {et~al.}(2020)\citenamefont {Ivanov}, \citenamefont {Ivanova}, \citenamefont {Caballero-Benitez},\ and\ \citenamefont {Mekhov}}]{2020Ivanov}%
  \BibitemOpen
  \bibfield  {author} {\bibinfo {author} {\bibfnamefont {D.~A.}\ \bibnamefont {Ivanov}}, \bibinfo {author} {\bibfnamefont {T.~Y.}\ \bibnamefont {Ivanova}}, \bibinfo {author} {\bibfnamefont {S.~F.}\ \bibnamefont {Caballero-Benitez}},\ and\ \bibinfo {author} {\bibfnamefont {I.~B.}\ \bibnamefont {Mekhov}},\ }\href {https://doi.org/10.1103/PhysRevLett.124.010603} {\bibfield  {journal} {\bibinfo  {journal} {Phys. Rev. Lett.}\ }\textbf {\bibinfo {volume} {124}},\ \bibinfo {pages} {010603} (\bibinfo {year} {2020})}\BibitemShut {NoStop}%
\bibitem [{\citenamefont {Ido}\ and\ \citenamefont {Katori}(2003)}]{Ido2003}%
  \BibitemOpen
  \bibfield  {author} {\bibinfo {author} {\bibfnamefont {T.}~\bibnamefont {Ido}}\ and\ \bibinfo {author} {\bibfnamefont {H.}~\bibnamefont {Katori}},\ }\href {https://doi.org/10.1103/PhysRevLett.91.053001} {\bibfield  {journal} {\bibinfo  {journal} {Phys. Rev. Lett.}\ }\textbf {\bibinfo {volume} {91}},\ \bibinfo {pages} {053001} (\bibinfo {year} {2003})}\BibitemShut {NoStop}%
\bibitem [{\citenamefont {Thompson}\ \emph {et~al.}(1992)\citenamefont {Thompson}, \citenamefont {Rempe},\ and\ \citenamefont {Kimble}}]{Thompson1992}%
  \BibitemOpen
  \bibfield  {author} {\bibinfo {author} {\bibfnamefont {R.~J.}\ \bibnamefont {Thompson}}, \bibinfo {author} {\bibfnamefont {G.}~\bibnamefont {Rempe}},\ and\ \bibinfo {author} {\bibfnamefont {H.~J.}\ \bibnamefont {Kimble}},\ }\href {https://doi.org/10.1103/PhysRevLett.68.1132} {\bibfield  {journal} {\bibinfo  {journal} {Phys. Rev. Lett.}\ }\textbf {\bibinfo {volume} {68}},\ \bibinfo {pages} {1132} (\bibinfo {year} {1992})}\BibitemShut {NoStop}%
\bibitem [{\citenamefont {Bonifacio}\ \emph {et~al.}(1971)\citenamefont {Bonifacio}, \citenamefont {Schwendimann},\ and\ \citenamefont {Haake}}]{Bonifacio1971}%
  \BibitemOpen
  \bibfield  {author} {\bibinfo {author} {\bibfnamefont {R.}~\bibnamefont {Bonifacio}}, \bibinfo {author} {\bibfnamefont {P.}~\bibnamefont {Schwendimann}},\ and\ \bibinfo {author} {\bibfnamefont {F.}~\bibnamefont {Haake}},\ }\href {https://doi.org/10.1103/PhysRevA.4.302} {\bibfield  {journal} {\bibinfo  {journal} {Phys. Rev. A}\ }\textbf {\bibinfo {volume} {4}},\ \bibinfo {pages} {302} (\bibinfo {year} {1971})}\BibitemShut {NoStop}%
\bibitem [{\citenamefont {Holstein}\ and\ \citenamefont {Primakoff}(1940)}]{Holstein1940}%
  \BibitemOpen
  \bibfield  {author} {\bibinfo {author} {\bibfnamefont {T.}~\bibnamefont {Holstein}}\ and\ \bibinfo {author} {\bibfnamefont {H.}~\bibnamefont {Primakoff}},\ }\href {https://doi.org/10.1103/PhysRev.58.1098} {\bibfield  {journal} {\bibinfo  {journal} {Phys. Rev.}\ }\textbf {\bibinfo {volume} {58}},\ \bibinfo {pages} {1098} (\bibinfo {year} {1940})}\BibitemShut {NoStop}%
\end{thebibliography}%

\clearpage

% Reset figure naming/numbering for Methods
\setcounter{figure}{0}
\renewcommand{\figurename}{\textbf{Extended Data Fig.}}

\section*{Methods}
\subsection*{Experimental procedure}
\begin{figure}[t!]
\centerline{\includegraphics[scale=1]{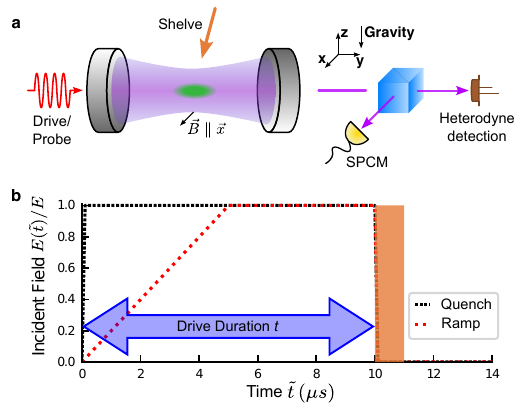}}
\caption{
    \textbf{Detailed experimental setup.} 
        \textbf{a,} $^{88}\mathrm{Sr}$ atoms (green) are trapped inside a high finesse optical cavity. During the experiment, a horizontally polarized resonant 689~nm is sent towards the cavity. The intracavity field (purple) is monitored by directing 50\% of the cavity transmission to a balanced heterodyne detection (brown). At the end of the drive duration, we shine a $\hat{y}$-polarized 688~nm beam from above (orange), resonant with the $^{3}P_{1}\!-\,\!\!^{3}S_1$ transition to optically pump the atoms in the excited state $|^3\mathrm{P}_1, m_{J}=0\rangle$ to the metastable states $^3\mathrm{P}_0$ and $^3\mathrm{P}_2$, a procedure which we call ``shelving" in the main text. We send weak $\hat{x}$-polarized 689~nm probe light before and after the drive and detect 50\% of the cavity transmission on a single-photon counting module (SPCM) to perform a vacuum-Rabi splitting (VRS) measurement to count atom numbers in the ground state $|^1\mathrm{S}_0\rangle$. 
        \textbf{b,} An example of the temporal profile of the incident field for a $10~\mu$s drive duration when ramping (red) and quenching (black) the drive. The y-axis is normalized to the maximum incident field $E$ applied within a single shot. In both cases, the drive duration $t$ is defined as the period from when the drive is initially turned on to when it is suddenly turned off. The orange shaded area indicates when the shelving of the excited state is performed.
        }
    \label{ExtendedFig1}
\end{figure}

\noindent Each shot of the experiment starts by loading $N=10^3$ to $10^4$ $^{88}\mathrm{Sr}$ atoms into a 1D 813~nm optical lattice inside a high finesse optical cavity. The thermal cloud has a temperature of about $15~\mu$K. After loading into the lattice as shown in Extended Data Fig.~\ref{ExtendedFig1}a, we apply a magnetic field of $
\vec{B}= 1.5\,$G$\,\hat{x}$. %, nominally 0 in the other two directions.
The lattice is linear-polarized in the $xz$-plane and its polarization is adjusted to 52~degrees with respect to $\vec{B}$ to reduce the differential AC stark shift on the $|^1\mathrm{S}_0 \rangle - |^3\mathrm{P}_1, m_{J}=0\rangle$ transition (the $\pi$ transition) \cite{Ido2003,Muniz2020}. We observe a residual (lattice-induced) maximum differential AC shift of $125(25)~$kHz between the ground $\ket{\downarrow}$ and the excited state $\ket{\uparrow}$. 

We then measure the initial total atom number via the magnitude of the vacuum Rabi splitting (VRS). This is accomplished by ramping the frequency of a $\vec{x}$-polarized probe over 5~MHz in $10$~ms and recording the transmitted power on a single photon counting module (SPCM) \cite{Thompson1992, Norcia2016} (see Extended Data Fig.~\ref{ExtendedFig1}).
%measurements along the $\pi$ transition for $10~$ms to calibrate atom number by probing the cavity with horizontally polarized light and recording the transmission via a single photon counting module (SPCM) \cite{Thompson1992, Norcia2016} (see Fig~\ref{ExtendedFig1}). 

We begin the actual experiment by coherently driving the cavity with $\vec{x}$-polarized light resonant with the $\pi$ transition. The drive incident on the cavity has a temporal profile as shown in Extended Data Fig.~\ref{ExtendedFig1}b. For the experiment where we ramp the drive, we linearly ramp the incident field $E(\tilde{t})$ over time $\tilde{t}$ to the desired maximum incident field $E$, such that this maximum field would establish a Rabi frequency of $\Omega_d$ inside a bare cavity in a steady state, and then hold the field strength fixed. We refer to the total length of the drive and hold operation as the drive duration $t$.

To track the intracavity field, we monitor the cavity transmission using heterodyne detection. For each drive strength, we separately measure the transmitted power with no atoms in the cavity to obtain the fractional (power) transmission $T$. 

To measure the spin projection $J_z$ at the end of the drive duration, we both turn off the incident drive and rapidly freeze the dynamics by flashing on a $688~$nm beam resonant with the excited state $^{3}P_{1}\!-\,\!\!^{3}S_1$ transition. This optically pumps or shelves the excited state atoms into the metastable states $^3\mathrm{P}_0$ and $^3\mathrm{P}_2$, where they do not interact with the cavity (see next section for details).  We can then count the number of atoms left in the ground state via another measurement of the vacuum Rabi splitting as described previously.  Combined with knowledge of the total atom number, we can then infer $J_z$ at the end of the drive duration. 

%in the subsequent VRS measurement  (see the next section for details) as soon as we start to snap off the incident drive. 
For the data in Fig.~\ref{fig5} where the cavity-atom detuning satisfies $\Delta_{ca} \neq 0$, we account for the dispersive shift when extracting atom numbers from the normal mode splitting acquired before and after the actual experiment. For each data point of the spin projection $J_z$ and fractional transmission $T$ presented in the main text, we average over at least 8 shots of the experiment. 

\subsection*{Shelving protocol}
\begin{figure}[hbt!]
\centerline{\includegraphics[scale=1]{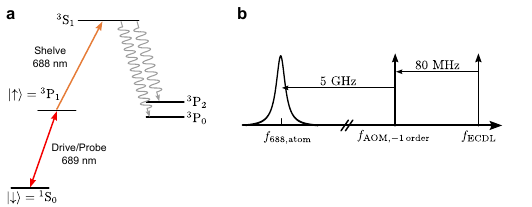}}
\caption{
    \textbf{Energy level diagram associated with the 688~nm shelving beam.} 
        \textbf{a,} Strontium level diagram. 
        \textbf{b,} 688 laser frequency shifting diagram. 50~ns before the 689~nm drive is turned off, we modulate a 688~nm AOM to turn on a -$1^{\mathrm{st}}$ order sideband. Simultaneously with turning off the 689~nm drive, we turn on the 5~GHz sideband using a fiber EOM to generate a 688~nm tone resonant with the $^{3}P_{1}-\,\!\!^{3}S_1$ transition.
        }
    \label{ExtendedFig2}
\end{figure}

\noindent We shelve the atoms that are in $|^3\mathrm{P}_1, m_{J}=0\rangle$ at the end of the drive duration by shining 688~nm laser resonant with the $^{3}P_{1}\!-\,\!\!^{3}S_1$ transition that radiatively decays into the metastable state at rate $\gamma_\mathrm{shelve}/2\pi=8.1~$MHz. To ensure a high on/off extinction ratio of the 688~nm beam, as well as a rapid switching time, we send the 688~nm beam first through an acoustic-optic modulator (AOM) and then a fiber electro-optic modulator (EOM) to control the amplitude and frequency of the 688~nm laser. The frequency shifting diagram of the 688~nm beam is shown in Extended Data Fig.~\ref{ExtendedFig2}\textbf{b}. We turn on the 688~nm AOM 50~ns before we turn off the 689 resonant drive. The 688~nm AOM has a rise time of 40~ns. As the 689~nm drive is turned off via switching off the RF signal to the AOM, we send a 5~GHz tone to the fiber EOM, placing a sideband of 688~nm light on resonance with $|^3\mathrm{S}_1\rangle$ and obtaining a shelving beam with a Rabi frequency of 25~MHz. The whole shelving procedure lasts for 1~$\mu$s.

\subsection*{Model and simulations}
\noindent We model the experimental system using a modified version of the master equation given in Eq.~(\ref{eqn:MasterEquation}):
\begin{align}
\begin{split}\label{eqn:Methods:FullMasterEquation}
    \hspace{-0.09cm}
    \frac{d\hat{\rho}}{dt}&=-\frac{i}{\hbar}\left[\hat{H}_{\text{tot}},\hat{\rho}\right]+\kappa\,\mathcal{L}_c(\hat{\rho})+\gamma\,\mathcal{L}_{se}(\hat{\rho}),
\end{split}
\end{align}
where the Hamiltonian can be expressed as the sum of three pieces  $\hat{H}_{\text{tot}}=\hat{H}_{a}+\hat{H}_c+\hat{H}_{\text{int}}$, defined by
\begin{align}\begin{split}
    \hat{H}_a&=\sum_{k=1}^N \hbar (\omega_a +\delta_k)\hat{s}_z^k\\
    \hat{H}_c&=\hbar\omega_c\hat{a}^\dagger\hat{a}-i\hbar\sqrt{\kappa}\alpha_{\text{in}}(\hat{a}e^{i\omega_d t}-\hat{a}^\dagger e^{-i\omega_d t})\\
    \hat{H}_{\text{int}}&=\sum_{k=1}^N \hbar g_k(\hat{a}\hat{s}_k^++\hat{a}^\dagger\hat{s}_k^-),
\end{split}\end{align}
which account for the atoms, the driven cavity, and the atom-light interactions, respectively. Here $\omega_a$ is the transition frequency between the two atomic states, $\omega_c$ is the cavity resonance frequency, and $\omega_d$ is the frequency of the driving laser, which carries $\alpha_{\text{in}}^2$ photons per second and we have chosen $\alpha_{\text{in}}$ to be real. The spatial dependence of the cavity electric field gives rise to inhomogeneity in the couplings, given by $g_k=g_0\cos(k\delta \phi)$ where $\delta \phi= \pi \lambda_l/\lambda_c$, and $2g_0$ is the single photon Rabi frequency at an antinode of the cavity mode. This inhomogeneity is caused by the incommensurability of the lattice ($\lambda_l=813$~nm) and cavity mode ($\lambda_c=689$~nm) wavelengths. In reality, each coupling $g_k$ is associated with a lattice site instead of an atom, but the large number of lattice sites ($\approx 10^3$) makes this distinction unnecessary. For definiteness, we will express all the coupling in terms of the r.m.s.~coupling $g\equiv g_{\text{rms}}=g_0/\sqrt{2}$. This is the convention followed in the main text.
We also included inhomogeneous broadening of the atomic transition, accounted for by the $\delta_k$ terms, which is caused by small AC Stark
shifts due to the trapping lattice.

Dissipation is modelled by the Lindblad superoperators
\begin{align}
    \begin{split}
        \kappa\,\mathcal{L}_c(\hat{\rho})&=\frac{\kappa}{2}\left(2\hat{a}\hat{\rho}\hat{a}^\dagger-\hat{a}^\dagger\hat{a}\hat{\rho}+\hat{\rho}\hat{a}^\dagger\hat{a}\right)\\
        \gamma\,\mathcal{L}_{se}(\hat{\rho})&=\frac{\gamma}{2}\sum_{k=1}^N\left(2\hat{s}^-_k\hat{\rho}\hat{s}^+_k-\hat{s}^+_k\hat{s}^-_k\hat{\rho}+\hat{\rho}\hat{s}^+_k\hat{s}^-_k\right),
    \end{split}
\end{align}
which account for photon leakage through the cavity mirrors with rate $\kappa=2\pi\times 153~\text{kHz}$ and spontaneous emission of photons into free space with rate $\gamma=2\pi\times 7.5~\text{kHz}$.

In the rotating frame of the drive, assuming that $\omega_d=\omega_a$ (atom-drive resonance) and that there is no broadening ($\delta_k=0$), the Hamiltonian is given by
\begin{equation}
    \hat{H}'_{\text{tot}}/\hbar=\Delta_{ca} \hat{a}^\dagger\hat{a}+\sum_{k=1}^N g_k(\hat{a}\hat{s}_k^++\hat{a}^\dagger\hat{s}_k^-)-i\sqrt{\kappa}{\alpha}_{\text{in}}(\hat{a}-\hat{a}^\dagger),
\end{equation}
where $\Delta_{ca}=\omega_c-\omega_a$ is the atom-cavity detuning. We recover Eq.~(\ref{eqn:MasterEquation}) by setting $\Delta_{ca}=0$, omitting $\gamma$ and replacing $g_k\to g$. In the absence of atoms, these equations lead to a steady-state cavity field $\braket{\hat{a}}=2\alpha_{\text{in}}/\sqrt{\kappa}$, which would establish an intracavity Rabi frequency (for the r.m.s.~coupler) equal to $\Omega_d=4g\alpha_{\text{in}}/\sqrt{\kappa}$. In the main text, we express the strength of the laser drive in terms of $\Omega_d$ instead of $\alpha_{\text{in}}$.

Numerical simulations are done by solving Eq.~(\ref{eqn:Methods:FullMasterEquation}) in the mean field approximation. More concretely, we calculate the exact equations of motion for $\braket{\hat{s}_z^k}$, $\braket{\hat{s}_k^-}$ and $\braket{\hat{a}}$, and factorize the operator product $\braket{\hat{a}\hat{s}_z^k}\to \braket{\hat{a}}\braket{\hat{s}_z^k}$. To reduce the computational complexity of the simulation, we group the $N\approx 10^4$ atoms into $N_{\text{eff}}$ groups of $N/N_{\text{eff}}$ particles with the $k'$ ensemble of $N_{\text{eff}}$ atoms having a cavity coupling $g_{k'}=g_0\cos[2\pi k'/N_{\text{eff}}]$, $k'=1,... N_{\text{eff}}$. By varying $N_{\text{eff}}$, we find empirically that for $N_{\text{eff}}=30$ the simulations in the superradiant phase have converged, and there is no further need to increase $N_{\text{eff}}$.

The presence of inhomogeneous coupling means that the result of the measurement of atom numbers and hence $J_z$ is a weighted average of the individual atomic inversions
\begin{equation}
    {J}_z/N\to \frac{1}{\sum_k g_k^2}\sum_k g_k^2 \braket{\hat{s}_z^k},
\end{equation}
instead of the total inversion. This weighted inversion is also $-1$ when all the atoms are in the ground state (note that $\braket{\hat{s}_z^k}=-1/2$ when atom $k$ is in the ground state).

As shown in the Supplementary Information, the modifications introduced by inhomogeneity in the couplings ($g_k$) and by the broadening of the atomic transition ($\delta_k$, with r.m.s.~variation of the same size as $\gamma$ in our system) do not change the qualitative nature of the second order superradiant transition nor of the first order transition at longer times, but they do introduce a renormalization of the critical drives, as accounted for in the main text. 

% \begin{acknowledgments}
\section*{Acknowledgments}
\noindent We acknowledge stimulating discussions with Alexander Baumgärtner, Murray J. Holland, Sofus L. Kristensen, Klaus Mølmer, and Cameron Wagner. This material is based upon work supported by the U.S. Department of Energy, Office of Science, National Quantum Information Science Research Centers, Quantum Systems Accelerator. We acknowledge additional funding support from the VBFF, the  National Science Foundation under Grant Numbers PFC PHY-2317149  (Physics Frontier Center) and OMA-2016244 (QLCI Q-SEnSE) and  NIST. J.T.Y.~was supported by the NWO Talent Programme (project number VI.Veni.222.312), which is (partly) financed by the Dutch Research Council (NWO). D.B. was supported by the Simons collaboration on Ultra-Quantum Matter (UQM)  which is funded by
grants from the Simons Foundation (Grant No. 651440), and acknowledges the hospitality of the KITP while parts of this work were completed.

\section*{Data availability}
\noindent The data that support the plots within this paper and other findings of this study are available from the corresponding author upon reasonable request.

% SI hereafter
\clearpage
% \newpage

\renewcommand{\theequation}{S\arabic{equation}}
\setcounter{equation}{0}
\renewcommand{\thesection}{S\arabic{section}}
\setcounter{section}{0}
\renewcommand{\thetable}{S\arabic{table}}
\setcounter{table}{0}
\renewcommand{\figurename}{\textbf{Fig.}}
\renewcommand{\thefigure}{\textbf{S\arabic{figure}}}
\setcounter{figure}{0}
\renewcommand{\figurename}{\textbf{Fig.}}
\renewcommand{\tocname}{Contents}
\renewcommand{\appendixname}{Supplement}
% \makeatletter

% reset page count
\setcounter{page}{1}

\clearpage

\title{Supplementary information: A dissipation-induced superradiant transition in a strontium cavity-QED system}

\date{\displaydate{date}}

\begin{abstract}
In this Supplementary Information we (1) describe the relation between our experiment and the original models of cooperative resonance fluorescence, (2) write down the theoretical model describing our experiment including all relevant technical imperfections and (3) provide theoretical analysis, at various degrees of theoretical complexity, of the steady state/dynamical behaviour of the system in the regimes explored in our experiment.
\end{abstract}
\maketitle

\onecolumngrid

\tableofcontents
\setcounter{secnumdepth}{2}
\incltocpage

\clearpage
\section{Relation to cooperative resonance fluorescence}
In this section, we discuss the relation between our experimental system and the original models of cooperative resonance fluorescence~\cite{Carmichael1980}. Our experimental system is described by the following master equation \begin{equation}\label{eqn:SuppExperimentMasterEq}
    \dot{\hat{\rho}}=-i\left[g\left(\hat{a}^\dagger\hat{J}^-+\hat{a}\hat{J}^+\right)-\frac{i\kappa\Omega_d}{4g}(\hat{a}-\hat{a}^\dagger),\hat{\rho}\right]+\kappa\left(\hat{a}\hat{\rho}\hat{a}^\dagger-\frac{\{\hat{a}^\dagger\hat{a},\hat{\rho}\}}{2}\right),
\end{equation}
where $\hat{a}$ ($\hat{a}^\dagger$) are annihilation (creation) operators and $\hat{J}^{\pm,z}$ are collective spin $N/2$ operators. Eq.~(\ref{eqn:SuppExperimentMasterEq}) has three ingredients: the Tavis-Cummings interaction ($\propto 2g$, the single-photon Rabi frequency), the laser pump through the cavity mirror ($\propto\Omega_d$) and loss of photons through the cavity mirror ($\propto\kappa$, the cavity power linewidth). Cooperative resonance fluorescence is described by a different master equation
\begin{equation}\label{eqn:SuppCRFMasterEq}
    \dot{\hat{\rho}}=-\frac{i\Omega_d}{2}\left[\hat{J}^++\hat{J}^-,\hat{\rho}\right]+\Gamma\left(\hat{J}^-\hat{\rho}\hat{J}^+-\frac{\{\hat{J}^+\hat{J}^-,\hat{\rho}\}}{2}\right),
\end{equation}
which involves only the atomic degrees of freedom and describes the competition between a Rabi drive ($\propto \Omega_d$) and superradiant decay ($\propto \Gamma$).

Eq.~(\ref{eqn:SuppExperimentMasterEq}) reduces to Eq.~(\ref{eqn:SuppCRFMasterEq}) (with $\Gamma=4g^2/\kappa$) when $2g\sqrt{N}\ll \kappa$, but it turns out that for low drives (superradiant phase) and $N$ large, the atomic steady states of both equations are identical and independent of the relation between $\kappa$ and $g\sqrt{N}$. For larger drives (normal phase) both Eq.~(\ref{eqn:SuppExperimentMasterEq}) and Eq.~(\ref{eqn:SuppCRFMasterEq}) predict persistent oscillations but the details do depend on the ratio between $g\sqrt{N}$ and $\kappa$. In the following subsections we include
\begin{enumerate}
    \item A schematic derivation of Eq.~(\ref{eqn:SuppCRFMasterEq}) from Eq.~(\ref{eqn:SuppExperimentMasterEq}) when $g\sqrt{N}\ll\kappa$ (this is not the regime we operate in, but we include this for completeness)
    \item The analytical solution of the steady state Eq.~(\ref{eqn:SuppExperimentMasterEq}) in the superradiant phase ($\Omega_d<2g^2N/\kappa)$ for $N$ large
    \item A discussion of the normal [Rabi flopping, ($\Omega_d>2g^2N/\kappa)$] phase in both models
\end{enumerate}

\subsection{Adiabatic elimination of the cavity mode}
We include this derivation for completeness, although it is valid only when $g\sqrt{N}\ll\kappa$, which is not the regime we operate in [in our experiment $g\sqrt{N}\sim (7-10)\kappa$]. As discussed in the main text, in the absence of the atoms the laser pump would establish an intracavity field of size $\Omega_d/2g$. The connection to cooperative resonance fluorescence can be made manifest by defining the operator $\hat{b}$
\begin{equation}
    \hat{b}=\hat{a}-\frac{\Omega_d}{2g},
\end{equation}
which measures the field with respect to $\Omega_d/2g$. The master equation becomes
\begin{equation}\label{eqn:SuppExperimentMasterEq2}
    \dot{\hat{\rho}}=-i\left[g\left(\hat{b}^\dagger\hat{J}^-+\hat{b}\hat{J}^+\right)+\frac{\Omega_d}{2}(\hat{J}^++\hat{J}^-),\hat{\rho}\right]+\kappa\left(\hat{b}\hat{\rho}\hat{b}^\dagger-\frac{\{\hat{b}^\dagger\hat{b},\hat{\rho}\}}{2}\right).
\end{equation}
In these variables, the Rabi drive acting on the atoms is now explicit. If $\kappa$ is large compared with the typical timescales of the atomic dynamics, the cavity field $\hat{b}$ will follow adiabatically the atoms. This can be seen from the equation of motion for $\braket{\hat{b}}$:
\begin{equation}
    \frac{d\braket{\hat{b}}}{dt}=-\frac{\kappa}{2}\braket{\hat{b}}-ig\braket{\hat J^-}.
\end{equation}
For $\kappa$ large we can neglect the time derivative and are led to the schematic equivalence $\hat{b}\to -2ig\hat{J}^-/\kappa$. Performing this replacement in Eq.~(\ref{eqn:SuppExperimentMasterEq2}) leads to (see Ref.~\cite{Bonifacio1971} for a rigorous derivation)
\begin{equation}\label{eqn:SuppCRFMasterEq.2}
    \dot{\hat{\rho}}=-i\left[\frac{\Omega_d}{2}(\hat{J}^++\hat{J}^-),\hat{\rho}\right]+\frac{4g^2}{\kappa}\left(\hat{J}^-\hat{\rho}\hat{J}^+-\frac{\{\hat{J}^+\hat{J}^-,\hat{\rho}\}}{2}\right),
\end{equation}
which is Eq.~(\ref{eqn:SuppCRFMasterEq}). The evolution described by Eq.~(\ref{eqn:SuppCRFMasterEq.2}) occurs on timescales $\Omega_d,\,4g^2N/\kappa$, so self-consistency requires both $\Omega_d\ll\kappa$ and $g\sqrt{N}\ll\kappa$.

% Finally, notice that Eq.~(\ref{eqn:SuppExperimentMasterEq2}) would be obtained directly in our experimental system if we were to drive the atoms directly by sending a laser through the side of the cavity instead of through the cavity mirrors. This has implications for the properties of the steady state, as we demonstrate in the next subsection.
\subsection{Steady state in superradiant phase}\label{subsec:SuppSuperradiant}
In this subsection, we show that the steady states of Eq.~(\ref{eqn:SuppExperimentMasterEq}) and Eq.~(\ref{eqn:SuppCRFMasterEq}) are identical below the transition point when $N$ is large independent of the relative sizes of $g\sqrt{N}$ and $\kappa$. When $N$ is large, mean field theory may provide an accurate description of the steady state, though this assumption needs to be checked by computing fluctuations about the mean field solutions and verifying that they are small. We do both in this subsection. In the case of Eq.~(\ref{eqn:SuppExperimentMasterEq}), the equations of motion are
\begin{align}
    \begin{split}
        \frac{d\braket{\hat{a}}}{dt}&=-\frac{\kappa}{2}\braket{\hat{a}}-ig\braket{\hat{J}^-}+\frac{\kappa\Omega_d}{4g}\\
        \frac{d\braket{\hat{J}^-}}{dt}&=2ig\braket{\hat{a}\hat{J}_z}\\
        \frac{d\braket{\hat{J}_z}}{dt}&=-ig\left(\braket{\hat{a}\hat{J}^+}-\braket{\hat{a}^\dagger\hat{J}^-}\right)
    \end{split}
\end{align} 
and the mean field steady state solutions are obtained by factorizing the expectation values (e.g. $\braket{\hat{a}\hat{J}_z}\to\braket{\hat{a}}\braket{\hat{J}_z}$) and setting the time derivatives to 0. The superradiant phase is given by $\braket{\hat{a}}=\braket{\hat{J}_x}=0$ as well as
\begin{align}\begin{split}\label{eqn:SuppMeanFieldSpin}
% \braket{\hat{J}_x}&=0\\[5pt]
    \braket{\hat{J}_y}&=\frac{\Omega_d}{(4g^2N/\kappa)}\equiv\frac{N}{2}\sin\theta\\
    \braket{\hat{J}_z}&=-\frac{N}{2}\cos\theta,
\end{split}\end{align}
where the angle $\theta$ is measured from the $-z$ axis towards the $+y$ axis. The mean field solution describes a spin of length $N/2$ pointing at an angle $\theta$ in the $zy$ plane. Since spin length is conserved, the relevant fluctuations occur in directions transverse to the mean field spin and behave like boson quadratures for $N$ large~\cite{Holstein1940}. We can thus write 
\begin{align}\begin{split}
    \hat{J}_x&=\sqrt{\frac{N}{2}}\hat{x}+O(N^{-1/2})\\
    \hat{J}_y\cos\theta+\hat{J}_z\sin\theta&=-\sqrt{\frac{N}{2}}\hat{p}+O(N^{-1/2})\\
    \hat{J}_y\sin\theta-\hat{J}_z\cos\theta&=\frac{N}{2}+O(N^{0}),\\
\end{split}\end{align}
where $[\hat{x},\hat{p}]=i$ are a pair of auxiliary Holstein-Primakoff bosons. The first two equations describe the transverse fluctuations and the third describes the variable along the Bloch vector direction, whose fluctuations we neglect because they are smaller due to spin length conservation. Plugging this decomposition into Eq.~(\ref{eqn:SuppExperimentMasterEq}) and keeping only the leading order non-vanishing terms (in a $1/N$ expansion) yields
\begin{equation}\label{eqn:SuppLargeNMasterEq}
    \frac{d\hat{\rho}}{dt}=-\frac{ig\sqrt{N}}{\sqrt{2}}\big[\hat{a}^\dagger(\hat{x}+i\hat{p}\cos\theta)+\hat{a}(\hat{x}-i\hat{p}\cos\theta),\hat{\rho}\Big]+\kappa\left(\hat{a}\hat{\rho}\hat{a}^\dagger-\frac{\{\hat{a}^\dagger\hat{a},\hat{\rho}\}}{2}\right)+O(gN^{0}\hat{p}^3\hat{a}).
\end{equation}
At this level of approximation, it can be checked that the steady state $\hat{\rho}_{ss}=\ketbra{ss}{ss}$ is given by
\begin{equation}
    \hat{a}\ket{ss}=\left(\hat{x}+i\hat{p}\cos\theta\right)\ket{ss}=0.
\end{equation}
The cavity field is truly in the vacuum state while spin fluctuations along the $\hat{S}_x$ direction are squeezed
\begin{equation}
    \braket{\hat{J}_x^2}_{ss}\approx\frac{N}{2}\braket{\hat{x}^2}_{ss}=\frac{N}{4}\cos\theta.
\end{equation}
These results are valid away from the transition point, when fluctuations in $\hat{p}$ are small [see neglected terms in Eq.~(\ref{eqn:SuppLargeNMasterEq})]. This derivation is valid for $N\to\infty$ at fixed $g\sqrt{N}/\kappa$, but it didn't rely on any assumption about the actual value of $g\sqrt{N}/\kappa$, so the results should also be valid for Eq.~(\ref{eqn:SuppCRFMasterEq}). In fact, the mean field spin direction Eq.~(\ref{eqn:SuppMeanFieldSpin}) and the spin fluctuations are also identical to the results for the master equation of cooperative resonance fluorescence Eq.~(\ref{eqn:SuppCRFMasterEq}), the latter of which has been studied extensively in the literature~\cite{Carmichael1980,Hannukainen2018,Barberena2024}.

We finalize this section with a comment pertaining to experimental implementations. The experiment we report in the main text is described by Eq.~(\ref{eqn:SuppExperimentMasterEq}). Alternatively, if the atoms were driven directly through the side of the cavity (instead of through the cavity mirrors) the experiment would be directly described by Eq.~(\ref{eqn:SuppExperimentMasterEq2}). The atomic response would be identical but the cavity state would differ because now the superradiant phase ($\Omega_d<2g^2 N/\kappa$) would be populated by a large number of photons ($\braket{\hat{b}}=\Omega_d/2g$). Thus, the presence or absence of a macroscopically large cavity field is context dependent.
\subsection{Normal phase}
The steady state solution to CRF Eq.~(\ref{eqn:SuppCRFMasterEq}) can be written down analytically~\cite{Carmichael1980}
\begin{equation}\label{eqn:SuppSolution}
    \hat{\rho}=\left(\frac{1}{\hat{J}^-+\frac{i\Omega_d}{4g^2/\kappa}}\right)\times \left(\frac{1}{\hat{J}^+-\frac{i\Omega_d}{4g^2/\kappa}}\right).
\end{equation}
This formula is valid in both the superradiant and the normal phase, but in the normal phase ($\Omega>2g^2N/\kappa$) this expression for the steady state can be re-interpreted as a probability distribution on the surface of the sphere. This is achieved by doing the replacements $\hat{J}^-\to (N/2)\sin\theta e^{-i\phi}$, $\hat{J}_z\to (N/2)\cos\theta$, leading to
\begin{equation}
    P(\theta,\phi)\propto \left|\frac{1}{\frac{N}{2}\sin\theta e^{i\phi}+\frac{i\Omega_d}{4g^2/\kappa}}\right|^2.
\end{equation}
Expectation values are then obtained by integration with respect to the measure $\sin\theta d\theta d\phi$. The steady state in this regime is  mixed~\cite{Carmichael1980}, with large fluctuations for the spin variables, and therefore it  is  not captured correctly by a mean field treatment. Nevertheless,  it is still captured by a classical probability distribution~\cite{Carmichael1980,Agarwal2024}. Furthermore, the equilibration timescale to this steady state is $(4g^2/\kappa)^{-1}$, which is $N$ times longer than the equilibration timescales deep in the superradiant phase [$\sim (4g^2N/\kappa)^{-1}$], so it is hard to access experimentally. The short time behaviour is instead characterized by persistent oscillations (for all initial conditions) that decay at long times. The mean field equations of motion capture accurately the oscillations but not their decay. 

For Eq.~(\ref{eqn:SuppExperimentMasterEq}) there is no analytical solution to the steady state, so mathematically exact statements are not available. Nevertheless, we expect a similar phenomenology to hold~\cite{Mattes2023}. To begin with, the superradiant steady state from section~\ref{subsec:SuppSuperradiant} no longer exists. The remaining stationary mean field solutions are neither stable nor unstable (i.e., they are centers), so that at short times we can expect persistent oscillations and limit cycles~\cite{Mattes2023}. At longer times the dynamical noise coming from dissipation will induce diffusion between (and within) the oscillatory trajectories, leading to a mixed steady state in the same spirit as Eq.~(\ref{eqn:SuppSolution}).

\section{Numerical simulations in the presence of inhomogeneous broadening}
In this section we provide a mathematical description of the system that includes relevant technical imperfections and derive the full set of mean field equations used to construct the theoretical curves of Figs.~2, 3, 4 and 5 in the main text.

In the presence of inhomogeneous broadening of the atomic transition, the system evolves according to the following master equation
\begin{align}\begin{split}\label{eqn:SI:FullMasterEquation}
    \hspace{-0.09cm}
    \frac{d\hat{\rho}}{dt}&=-i\left[\hat{H}_{\text{tot}},\hat{\rho}\right]+\kappa\,\mathcal{L}_c(\hat{\rho})+\gamma\,\mathcal{L}_{se}(\hat{\rho}).
\end{split}\end{align}
The Hamiltonian is divided into three pieces characterizing the atoms, cavity and atom-light interaction:
\begin{align}\begin{split}\label{eqn:SI:Hamiltonians}
    \hat{H}_{\text{tot}}&=\sum_{k=1}^N\underbrace{(\omega_a+\delta_k)\hat{s}^z_k}_{\text{atoms}}+\underbrace{\omega_c\hat{a}^\dagger\hat{a}-i\frac{\kappa\Omega_d}{4g_{\text{rms}}}(\hat{a}e^{i\omega_d t}-\hat{a}^\dagger e^{-i\omega_d t})}_{\text{cavity}}+\sum_{k=1}^N\underbrace{ g_k(\hat{a}\hat{s}_k^++\hat{a}^\dagger\hat{s}_k^-)}_{\text{interaction}}
\end{split},\end{align}
where $\hat{a}$ ($\hat{a}^\dagger$) are operators for the cavity mode, $\hat{s}_k^{z,\pm}$ are spin $1/2$ operators for atom $k$, $\omega_a$ is the atomic transition frequency, $\omega_c$ is the cavity resonance frequency, $g_k$ is the coupling of atom $k$ to the cavity mode and $g_{\text{rms}}$ is the root-mean-square average of all the $g_k$. A $1$D optical lattice along the cavity axis pins the position of the atoms in the longitudinal direction of the cavity. Confinement along the transverse direction is weaker and relies on the radially varying intensity of the optical lattice laser. As the atoms explore positions away from the peak intensity of the Gaussian beams, they feel different AC Stark shifts to the atomic transition frequency. This is encoded in the inhomogeneous detunings $\delta_k$, which follow a distribution
\begin{equation}\label{eqn:SI:BroadeningDistribution}
    P(\delta)=\left(\frac{U_0}{k_B T\delta_{\text{max}}}\right) \left(\frac{\delta}{\delta_{\text{max}}}\right)^{U_0/(k_B T)-1},\hspace{1cm} 0<\delta<\delta_{\text{max}},
\end{equation}
where $U_0/\hbar= 2\pi\times 1.98$ MHz is the trap depth and $T=15\mu$K is the temperature of the atoms ($U_0=6.34 k_BT$, where $k_B$ is the Boltzmann constant), and the distribution is determined by the combined effect of the Gaussian profile of the laser beam and the thermal atomic distribution in the presence of the confining potential.  The few atoms that are far away from the center of the spot feel almost no intensity and no frequency shift $\delta\approx 0$. However, most atoms are concentrated near the center and feel the maximum possible intensity and suffer a frequency shift $\delta_{\text{max}}\approx 2\pi\times 125(25)$kHz, a number that is inferred from the experimental setup. The physical response of the system is determined mostly by the standard deviation of this distribution, which for $U_0=6.34k_B T$ is about $0.11\delta_{\text{max}}\approx 2\pi\times 14(3)\text{ kHz}$ (roughly the same size as $\gamma$).

The experiment is calibrated in a way that puts the cavity on resonance with the average transition frequency $\omega_c=\omega_a+\bar{\delta}$, where $\bar{\delta}$ is the average of Eq.~(\ref{eqn:SI:BroadeningDistribution}). The laser drive is then put on resonance with the cavity ($\omega_d=\omega_c$) so that in the rotating frame of the drive we have the Hamiltonian
\begin{align}\begin{split}\label{eqn:SI:HamiltoniansRotFram}
    \hat{H}'&=\sum_{k=1}^N(\delta_k-\bar{\delta})\hat{s}^z_k-i\frac{\kappa\Omega_d}{4g_{\text{rms}}}(\hat{a}-\hat{a})+\sum_{k=1}^Ng_k(\hat{a}\hat{s}_k^++\hat{a}^\dagger\hat{s}_k^-).
\end{split}\end{align}
As described in the Methods, in a standing wave cavity the coupling constants have the form $g_k=g_0\cos(\phi_k)$, where $\phi_k=2\pi k \lambda_l/\lambda_c$, $g_0=2\pi\times 10.8$ kHz is the single-photon Rabi frequency at an antinode of the cavity and $\lambda_{l/c}$ are the wavelengths of the lattice/cavity respectively. Since $\lambda_l$ and $\lambda_c$ are incommesurate, we can assume that $\phi_k$ is distributed uniformly in the interval $(0,2\pi]$. In particular, the root-mean-square coupling $g_{\text{rms}}$ is related to $g_0$ by $g_{\text{rms}}=g_0/\sqrt{2}$. We'll express all quantities in terms of $g_{\text{rms}}$ instead of $g_0$. In the main text we defined $g\equiv g_{\text{rms}}$, but here in this Supplementary Material we will keep the rms subscript explicit wherever it appears.

The mean field equations of motion in the rotating frame of the drive are
\begin{align}
    \begin{split}\label{eqn:SI:MeanField}
    \dot{\alpha}&=-\frac{\kappa}{2}\alpha-i\sqrt{N}\left(\frac{1}{N}\sum_k g_k s_k\right)+\frac{\kappa\Omega_d}{4g_{\text{rms}}\sqrt{N}}\\
    \dot{s}_k&=2ig_k\sqrt{N}\alpha z_k-\frac{\gamma}{2}s_k-i(\delta_k-\bar{\delta})s_k\\
    \dot{z}_k&=-ig_k\sqrt{N}(\alpha s_k^*-\alpha^*s_k)-\gamma\left(z_k+\frac{1}{2}\right),
    \end{split}
\end{align}
where $\alpha=\braket{\hat{a}}/\sqrt{N}$, $s_k=\braket{\hat{s}_k^-}$ and $z_k=\braket{\hat{s}_k^z}$. The equation for $\alpha$ involves only a weighted average coherence 
\begin{equation}
    \frac{1}{N}\sum_k g_k s_k\approx\int_0^{2\pi}\left(\sqrt{2}g_{\text{rms}}\cos\phi\right)\frac{d\phi}{2\pi}\int_0^{\delta_{\text{max}}}P(\delta)s(\delta,\phi)\,d\delta,
\end{equation}
where we have traded the index $k$ for the pair of variables $\phi,\delta$ and assumed that the coupling and detuning distributions are uncorrelated because they arise from independent physical effects. We find numerically that a grid of $20$ values of $\phi$ and $40$ values of $\delta$ is sufficient to get convergence in the superradiant phase. To prepare the steady state adiabatically, we use a time dependent drive $\Omega_d(t)$:
\begin{equation}
    \Omega_d(t)=\begin{cases} 
      \Omega_d\left(\frac{t}{T_{\text{ramp}}}\right) & t<T_{\text{ramp}} \\
      \Omega_d &T_{\text{ramp}}<t<T_{\text{hold}}.
   \end{cases}
\end{equation}
The drive strength rises linearly from $0$ to $\Omega_d$ during a time $T_{\text{ramp}}$ and is then held fixed until a time $T_{\text{hold}}$. For the simulations, we use the following parameters
\begin{align}
    \begin{split}
        \kappa&=2\pi\times 153\text{ kHz}\\
        \gamma&=2\pi\times 7.5\text{ kHz}\\
        g_{\text{rms}}&=2\pi\times 7.8\text{ kHz}\\
        \delta_{\text{max}}&=2\pi\times 125\text{ kHz}\\
        N&=10^3-10^4
    \end{split}
\end{align}
The exact atom number $N$ is taken directly from experimental measurements.

Finally, as pointed out in the Methods, the atomic inversion is measured via vaccuum Rabi splitting, which provides a weighted average
\begin{equation}
    \tilde{J}_z=N\frac{\sum_k g_k^2 z_k}{\sum_k g_k^2},
\end{equation}
and the normalization is chosen so that $\tilde{J}_z=-N/2$ when all the atoms are in their ground state. Our numerical simulations account for this weighting directly and the analytical results we will describe in the following sections will also focus on computing $\tilde{J}_z$.
\section{Steady state behaviour}
In this section we consider two broad steady state scenarios: (i) ideal CRF transition, with $\gamma=0$ and $\delta=0$, (ii) first order transition, including nonzero $\gamma$ and $\delta_k$. In a few cases, the mean field equations in Eq.~(\ref{eqn:SI:MeanField}) can be solved analytically.
\subsection{Continuous superradiant transition}
\begin{figure}
    \centering
    \includegraphics[width=0.97\textwidth]{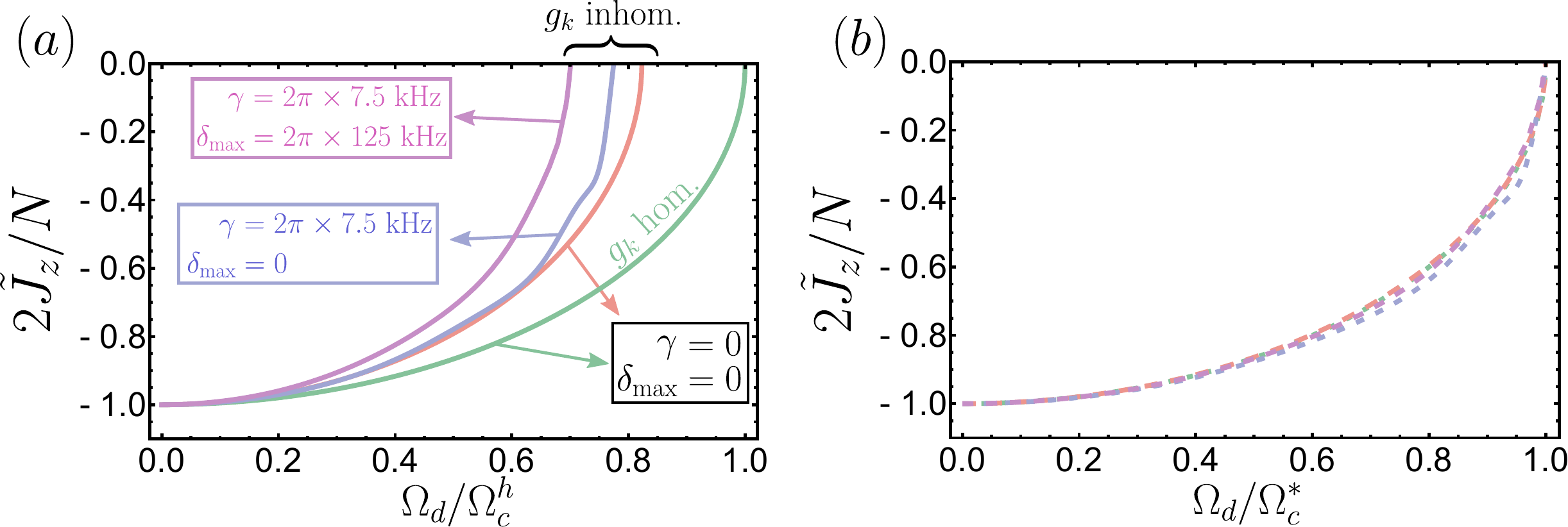}
    \caption{(a) Inversion $\tilde{J}_z$ as a function of $\Omega_d/\Omega_c^h$ for four different cases: homogeneous couplings with $\gamma=0$ and $\delta_k=0$ (green, analytical steady state), inhomogeneous couplings with $\gamma=0$ and $\delta_k=0$ (red, analytical steady state), inhomogenoeus couplings with $\gamma\neq 0$ and $\delta_k=0$ (blue, numerical solution with $T_{\text{hold}}=9.3\mu$s with $T_{\text{ramp}}=5\mu$s), and inhomogeneous couplings with $\gamma\neq 0$ and $\delta_k=0$ (purple, numerical solution with $T_{\text{hold}}=9.3\mu$s with $T_{\text{ramp}}=5\mu$s). (b) Same as panel (a) but after rescaling $\Omega_d$ by each respective critical drive: $\Omega_c^h$ (green), $\Omega_c^{nh}=0.82\Omega_c^h$ (red), $\Omega_c^{nh,*}=0.78\Omega_c^h$ (blue) and $\Omega_c=0.70\Omega_c^h$ (purple). The $\Omega_c^*$ in the horizontal axis label represents each of these different transition frequencies.}
    \label{fig:SI:ShortTimeInversion}
\end{figure}

Here we set $\gamma=0$ and $\delta_k=0$ in Eq.~(\ref{eqn:SI:MeanField}). In this situation there are many possible steady states (at the mean-field level) since the only requirement to get a steady state solution to Eq.~(\ref{eqn:SuppMeanFieldSpin}) with $0$ intracavity field (and hence no atomic dynamics) is that $\sum_k g_k s_k=0$. The relevant configuration for us is the one that we access through the drive, so we need to partly solve the dynamics. If all the atoms start from the ground state (south pole of Bloch sphere), then their individual Bloch vectors will lie on the $yz$ plane, so they can be parameterized as $z_k=-(1/2)\cos\theta_k$ and $s=-(i/2)\sin\theta_k$. This parameterization describes a rotation towards the equator by an angle $\theta_k$ that is measured from $-z$ towards $+y$. Plugging this into Eq.~(\ref{eqn:SI:MeanField}) leads to $\theta_k=g_k \sqrt{N}Q$, where $Q=\int_0^{T} 2\alpha\,dt$ is common to all the spins. The dynamics is thus reduced to two equations
\begin{align}\begin{split}\label{eqn:SI:IdealDynamics}
    \dot{\alpha}&=-\frac{\kappa}{2}\alpha-\sqrt{N}\left[\frac{1}{2N}\sum_k g_k \sin(g_k \sqrt{N}Q)\right]+\frac{\kappa\Omega_d}{4g_{\text{rms}}\sqrt{N}}\\
    \dot{Q}&=2\alpha
\end{split}\end{align}
%cos theta k, sin theta k
There is now a unique steady state solution, characterized by $\alpha_{ss}=0$ and a $Q_{ss}$ that satisfies 
\begin{equation}
    \sqrt{N}\left[\frac{1}{2N}\sum_k g_k \sin(g_k Q_{ss})\right]=\frac{\kappa\Omega_d}{4g_{\text{rms}}\sqrt{N}}
\end{equation}
We distinguish the two-following cases:
\begin{itemize}
    \item \textit{Uniform couplings ($g_k=g_{\text{rms}}$):} The steady state solution is determined by
    \begin{equation}
    \sin(g_{\text{rms}}\sqrt{N}Q_{ss})=\frac{\Omega_d}{2g_{\text{rms}}^2N/\kappa},
\end{equation}
and a solution exists only if $\Omega_d<\Omega_c^h\equiv 2g_{\text{rms}}^2N/\kappa$. In this case the weighted inversion coincides with the true inversion of the system and satisfies
\begin{equation}\label{eqn:SI:tildeSzIdealHomo}
    2\tilde{J}_z/N=-\sqrt{1-\left(\frac{\Omega_d}{\Omega_c^h}\right)^2}
\end{equation}
    \item \textit{Non-uniform couplings ($g_k=\sqrt{2}g_{\text{rms}}\cos\phi_k$):} The steady state is given by
    \begin{equation}
    J_1\left(\sqrt{2}g_{\text{rms}}\sqrt{N} Q_{ss}\right)=\frac{\Omega_d}{\sqrt{2}\Omega_c^h},
\end{equation}
where $J_1$ is a Bessel function and has a solution for $Q_{ss}$ only when $\Omega_d<\Omega_c^{nh}=0.82\Omega_c^h$. The weighted inversion is given by
\begin{equation}\label{eqn:SI:tildeSzIdealInhomo}
    2\tilde{J}_z/N=-\left[J_0\left(\sqrt{2}g_{\text{rms}}\sqrt{N}Q_{ss}\right)-J_2\left(\sqrt{2}g_{\text{rms}}\sqrt{N}Q_{ss}\right)\right]
\end{equation}

Numerical simulations that incorporate the short time effects of spontaneous emission indicate that spontaneous emission further shifts the transition point to $\Omega_c^{nh,*}\equiv0.78\Omega_c^h$ [see Fig.~\ref{fig:SI:ShortTimeInversion}(a)]. Inhomogeneous broadening of the atomic transition further shifts the transition point to $\Omega_c\equiv0.70\Omega_c^h$, which is what is observed experimentally. Furthermore, once we rescale each curve by their respective critical point, we find that they fall on top of each other [Fig.~\ref{fig:SI:ShortTimeInversion}(b)].
\end{itemize}
 %0.777,0.701
\subsection{First order transition}
Here we include the effects of spontaneous emission $\gamma$ and inhomogeneous broadening $\delta_k$. The mean field steady state is then unique and can be obtained explicitly from Eq.~(\ref{eqn:SI:MeanField}):
\begin{align}
\begin{split}
    z_k&=-\frac{1}{2}\left\{1+|\beta|^2\left[\frac{\eta_k^2}{1+\frac{4(\delta_k-\bar{\delta})^2}{\gamma^2}}\right]\right\}^{-1}\\
    s_k&=-i\beta\sqrt{\frac{1}{2}}\left[\frac{\eta_k}{1+\frac{2i(\delta_k-\bar{\delta})}{\gamma}}\right]\left\{1+|\beta|^2\left[\frac{\eta_k^2}{1+\frac{4(\delta_k-\bar{\delta})^2}{\gamma^2}}\right]\right\}^{-1},
\end{split}\end{align}
which are expressed as functions of the normalized intracavity field
\begin{equation}
    \beta=\alpha\left(\sqrt{\frac{8g_{\text{rms}}^2N}{\gamma^2}}\right),
\end{equation}
and of the normalized coupling $\eta_k=g_k/g_{\text{rms}}$. The normalized field $|\beta|$ satisfies the equation
\begin{equation}\label{eqn:SI:IntracavityFieldFull}
    \sqrt{2}\beta\left[\left(\frac{\kappa\gamma}{4g_{\text{rms}}^2 N}\right)+\left(\frac{1}{N}\sum_{k}\left[\frac{\eta_k^2}{1+\frac{2i(\delta_k-\bar{\delta})}{\gamma}}\right]\left\{1+|\beta|^2\left[\frac{\eta_k^2}{1+\frac{4(\delta_k-\bar{\delta})^2}{\gamma^2}}\right]\right\}^{-1}\right)\right]=\frac{\Omega_d}{\Omega_c^h},
\end{equation}
which must be solved for $|\beta|$ as a function of $\Omega_d/\Omega_c^h$. Since $|\beta|\sim 1$ in the superradiant phase and we operate at a large collective cooperativity $NC_{\text{rms}}=4g^2_{\text{rms}}N/(\kappa\gamma)$, we neglect the first term on the left hand side. We then have the following cases (we also include a summary of the results in table~\ref{tab:Supp:CriticalPoint}):
\begin{itemize}
    \item \textit{Uniform couplings ($g_k=g_{\text{rms}}$) and no broadening ($\delta_k-\bar{\delta}=0$):} the equation for the field $|\beta|$ becomes
    \begin{equation}
        \frac{\sqrt{2}\beta}{1+|\beta|^2}=\frac{\Omega_d}{\Omega_c^h},
    \end{equation}
    which has a solution only when $\Omega_d<\Omega_c^h/\sqrt{2}$. In this uniform case, spontaneous emission transforms the second order transition at $\Omega_c^h$ to a first order transition at $\Omega_c^{h,se}\equiv\Omega_c^h/\sqrt{2}$. The weighted inversion can also be calculated, yielding
    \begin{equation}
        2\tilde{J}_z/N=-\frac{1}{2}\left[1+\sqrt{1-2\left(\frac{\Omega_d}{\Omega_c^h}\right)^2}\right].
    \end{equation}
    When $\Omega_d=0$ then $\tilde{J}_z=-N/2$ and when $\Omega_d=\Omega_c^h/\sqrt{2}$ then $\tilde{J}_z=-N/4$. For larger $\Omega_d$, Eq.~(\ref{eqn:SI:IntracavityFieldFull}) indicates that $\tilde{J}_z/N\sim (NC_{\text{rms}})^{-2}\to 0$, so there is a jump from $-N/4$ to $0$ at $\Omega_c^{h,se}=\Omega_c^h/\sqrt{2}$.
    \item \textit{Non-uniform couplings ($g_k=\sqrt{2}g_{\text{rms}}\cos\phi_k$) and no broadening ($\delta_k-\bar{\delta}=0$):} the steady state field in the superradiant phase is given by
    \begin{equation}
        \sqrt{2}\beta\int_0^{2\phi}\frac{d\phi}{2\pi}\left[\frac{2(\cos\phi)^2}{1+2|\beta|^2(\cos\phi)^2}\right]=\frac{\sqrt{2}\beta}{|\beta|^2}\left(1-\frac{1}{\sqrt{1+2|\beta|^2}}\right)=\frac{\Omega_d}{\Omega_c^h},
    \end{equation}
    which has a solution when $\Omega_d<0.6\Omega_c^h$. In this non-uniform coupling case, spontaneous emission transforms the second order transition at $\Omega_c^{nh}=0.82\Omega_c^h$ to a first order transition at $\Omega_c^{nh,se}=0.6\Omega_c^h$.
    \item \textit{Non-uniform couplings ($g_k=\sqrt{2}g_{\text{rms}}\cos\phi_k$) and non-zero broadening ($\delta_k-\bar{\delta}\neq 0$):} This case describes the experimental conditions. The equation for the field $|\beta|$ can only be done analytically to a certain extent
    \begin{align}\begin{split}
        \sqrt{2}\beta\int P(\delta)\,d\delta\int_{0}^{2\pi}\frac{d\phi}{2\pi}\frac{2(\cos\phi)^2}{1+\frac{2i(\delta-\bar{\delta})}{\gamma}}\left\{1+\left[\frac{2|\beta|^2(\cos\phi)^2}{1+\frac{4(\delta_k-\bar{\delta})^2}{\gamma^2}}\right]\right\}^{-1}&=\frac{\Omega_d}{\Omega_c^h}\\[5pt]
        \longrightarrow\frac{\sqrt{2}\beta}{|\beta|^2}\int P(\delta)\,d\delta\left[1-\frac{\sqrt{1+\frac{4(\delta-\bar{\delta})^2}{\gamma^2}}}{\sqrt{1+\frac{4(\delta-\bar{\delta})^2}{\gamma^2}+2|\beta|^2}}\right]\left[1-\frac{2i(\delta-\bar{\delta})}{\gamma}\right]&=\frac{\Omega_d}{\Omega_c^h}
    \end{split}\end{align}
    The numerical solution of the previous equation using Eq.~(\ref{eqn:SI:BroadeningDistribution}) indicates that the first order transition is shifted. When $\delta_{\text{max}}=100$kHz, a solution exists only for $\Omega_d<0.29\Omega_c^h=0.41\Omega_c$. For $\delta_{\text{max}}=125$kHz, a solution exists only for $\Omega_d<0.26\Omega_c^h=0.37\Omega_c$. For $\delta_{\text{max}}=150$kHz, a solution only exists for $\Omega_d<0.24\Omega_c^h=0.33\Omega_c$. We've expressed the previous results also in terms of $\Omega_c=0.71\Omega_c^h$, i.e. the experimentally determined critical point for the collective transition. 
\end{itemize}

\begin{table}[h!]
\setlength{\tabcolsep}{10pt}
\begin{tabular}{lccc}\toprule 
&Uniform $g_k$&Non-uniform $g_k$&Non-uniform $g_k$\\[-1pt]
&$\delta_{\text{max}}=0$&$\delta_{\text{max}}=0$&$\delta_{\text{max}}=2\pi\times 125\text{ kHz}$
\\ \hline
Continuous transition &$\Omega_c^h$&$\Omega_c^{nh}=0.82\Omega_c^h$&$\Omega_c=0.701\Omega_c^h$ (includes $\gamma$)\\
First-order transition &$\Omega_c^{h,se}=0.71\Omega_c^h$&$\Omega_c^{nh,se}=0.6\Omega_c^h$&$0.26\Omega_c^h$\\
\bottomrule
\end{tabular}
\caption{Critical point under different conditions.}\label{tab:Supp:CriticalPoint}
\end{table}

\begin{figure}
    \centering
    \includegraphics[width=0.97\textwidth]{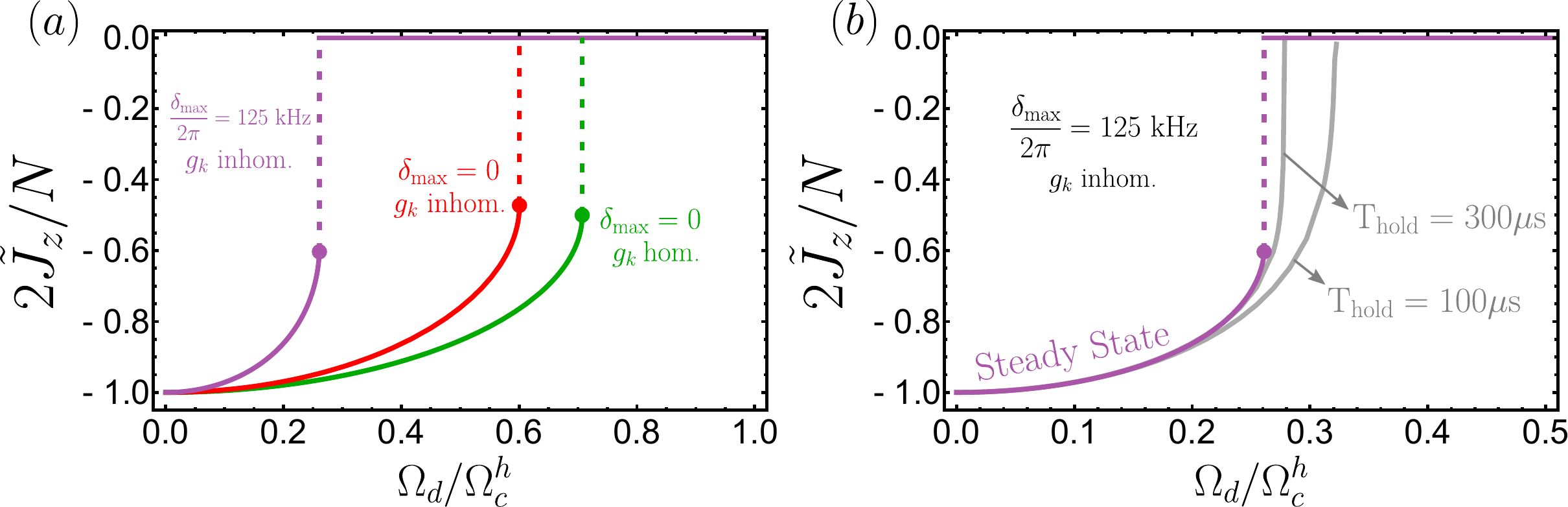}
    \caption{(a) We show the long-time steady state weighted inversion $\tilde{J}_z$ as a function of $\Omega_d/\Omega_c^h$ for three different scenarios: homogeneous couplings and no broadening (green), inhomogeneous couplings and no broadening (red), and inhomogeneous couplings with broadening (purple, $\delta_{\text{max}}=2\pi\times 125$ kHz). All curves incorporate the effects of spontaneous emission. (b) Comparison between the steady state profile for $\delta_{\text{max}}=2\pi\times 125$ kHz and time evolution for $T_{\text{hold}}=100\mu\text{s},\,300\mu\text{s}$.}
    \label{fig:SI:LongTimeInversion}
\end{figure}

In Fig.~\ref{fig:SI:LongTimeInversion}(a) we show $\tilde{J}_z$ as a function of drive strength for the three analyzed cases. Furthermore, in Fig.~\ref{fig:SI:LongTimeInversion}(b) we compare these profiles against $\tilde{J}_z$ obtained via time evolution of Eq.~(\ref{eqn:SI:MeanField}) for long but finite times. The time-dependent calculations do approach the steady state result, but the region around the transition jump takes a very long time to equilibrate.
\subsection{Atom-cavity detuning}
\begin{figure}
    \centering
    \includegraphics[width=0.75\linewidth]{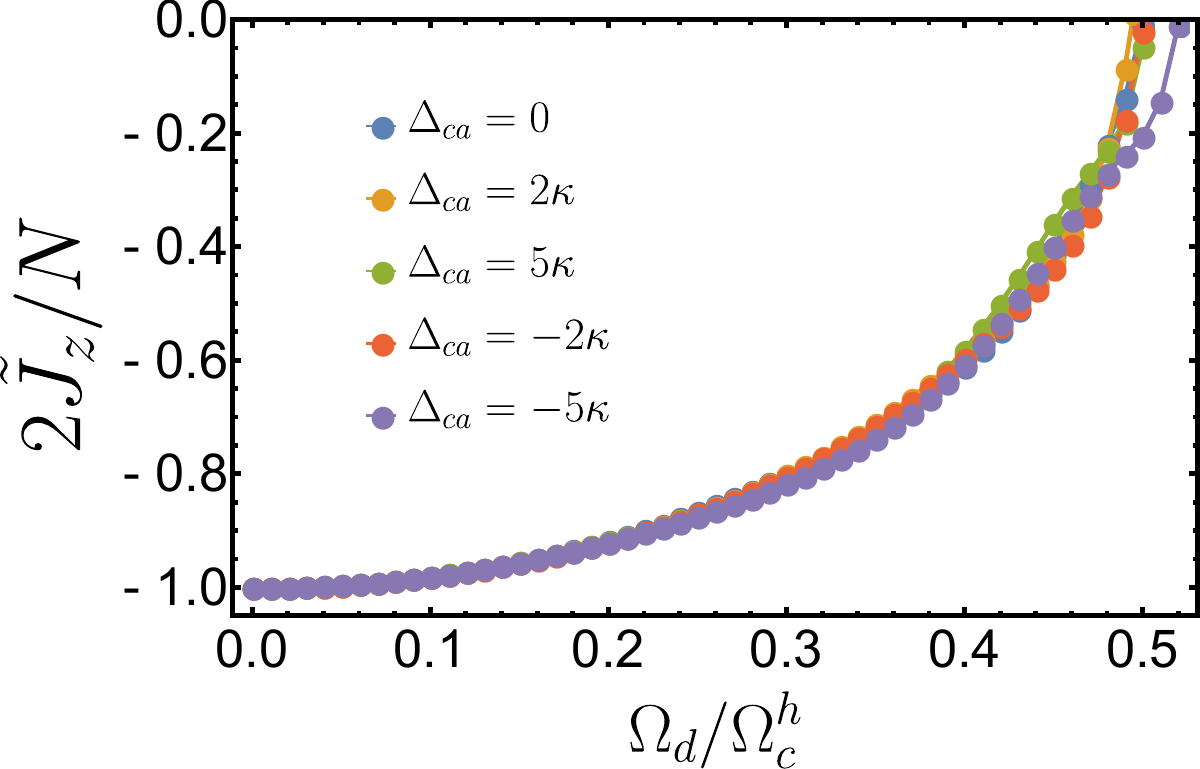}
    \caption{Inversion $\tilde{J}_z$ as a function of normalized Rabi frequency $\Omega_d/\Omega_c^h$ for different values of $\Delta_{ca}$. Simulations were done using a ramp time $T_{\text{ramp}}=5\mu$s and total evolution time $T_{\text{hold}}=9.3\mu$s in the presence of spontaneous emission and inhomogeneous broadening.}
    \label{fig:SI:detplot}
\end{figure}

In this subsection we explore the effects of an atom-cavity detuning on the second-order continuous transition. As described in the main text, we don't expect the profile of $\tilde{J}_z$ as a function of $\Omega_d/\Omega_c^h$ to change substantially. We test this by modifying Eq.~(\ref{eqn:SI:MeanField}) to include an atom-cavity detuning $\Delta_{ca}=\omega_c-\omega_a$:
\begin{equation}
    \dot{\alpha}=-\frac{\kappa}{2}\alpha-i\Delta_{ca}\alpha-i\sqrt{N}\left(\frac{1}{N}\sum_k g_k s_k\right)+\frac{\kappa\Omega_d}{4g_{\text{rms}}\sqrt{N}}
\end{equation}
and simulating the mean field equations of motion for $\Delta_{ca}=0,\pm 2\kappa,\pm 5 \kappa$. We show the results in Fig.~\ref{fig:SI:detplot}, which corroborates our assertion that $\Delta_{ca}$ does not modify the transition point. The highest discrepancy is found at $\Delta_{ca}=-5\kappa$ close to the transition and arises from the finite evolution time. 
\section{Dynamical behaviour}
In the experiment we also examined the dynamical behaviour of the system. For the second order transition, we explored the effect of a quench in the drive $\Omega_d$. For the first order transition, we monitored $\tilde{J}_z$ as it slowly drifted due to spontaneous emission. In this section we provide a simplified theoretical analysis of both effects.
\subsection{Short-time quenches}
\begin{figure}
    \centering
    \includegraphics[width=0.97\textwidth]{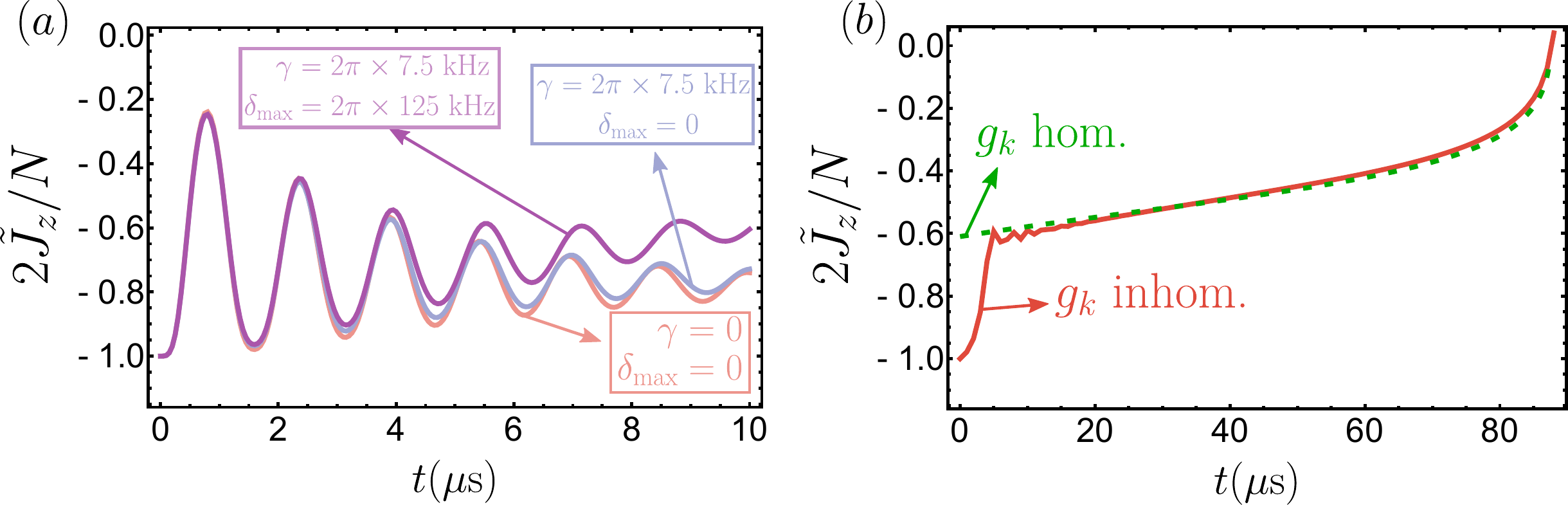}
    \caption{(a) Short time dynamics of $\tilde{J}_z$ after a quench as a function of time for $N=10^4$ and $\Omega=0.52\Omega_c^h$. All curves include inhomogeneous couplings and show the evolution with $\gamma=\delta_{\text{max}}=0$ (red), $\gamma\neq 0$ and $\delta_{\text{max}}=0$ (blue), and $\gamma\neq 0$ and $\delta_{\text{max}}=2\pi\times 125\text{ kHz}$ (purple). (b) Long time dynamics of $\tilde{J}_z$ in the presence of spontaneous emission using Eq.~(\ref{eqn:SI:MeanField}) (red, $\Omega_d=0.64\Omega_c^h$) and Eq.~(\ref{eqn:SI:LongTimeInversionModel}) (dashed green, $\Omega_d=0.75\Omega_c^h$) assuming no broadening ($\delta_{\text{max}}=0$).}
    \label{fig:SI:InversionDynamics}
\end{figure}

If we begin with all $z_k=-1/2$ and $\alpha=0$ and the drive $\Omega_d$ is suddenly turned on, the system has to adjust to the new steady state. This process is described by Eq.~(\ref{eqn:SI:IdealDynamics}) with the initial condition $z_k=-1/2$, $s_k=0$ and $\alpha=0$. Because the equations are nonlinear a general solution is not possible, but we can analyze the behaviour close to the steady state. To do this, we linearize Eq.~(\ref{eqn:SI:IdealDynamics}), leading to
\begin{align}
    \begin{split}
        \delta\dot{\alpha}&=-\frac{\kappa}{2}\delta\alpha-\frac{\delta Q}{2}\left[\sum_k g_k^2\cos(g_k\sqrt{N}Q_{ss})\right]=-\frac{\kappa}{2}\delta\alpha+g_{\text{rms}}^2\tilde{J}_z\delta Q\\
        \delta\dot{Q}&=2\delta\alpha,
    \end{split}
\end{align}
where $\delta Q=Q-Q_{ss}$, $\delta\alpha=\alpha-\alpha_{ss}$ and $\tilde{J}_z$ is defined in Eq.~(\ref{eqn:SI:tildeSzIdealInhomo}). The linearized equation has dynamical eigenvalues
\begin{equation}
    -\frac{\kappa}{4}\pm i\sqrt{2g^2_{\text{rms}}|\tilde{J}_z|-\frac{\kappa^2}{16}}.
\end{equation}
The approach to the steady state is thus characterized by oscillations with frequency $\sim \sqrt{2g^2_{\text{rms}}|\tilde{J}_z|}$ and $1/e$ decay time $4/\kappa\approx 4\mu s$. This is illustrated in Fig.~\ref{fig:SI:InversionDynamics}, which showcases oscillations and their decay. We also include curves with spontanteous emission and $\delta_{\text{max}}=2\pi\times 125\text{ kHz}$ to investigate their effect on the short time dynamics. For the first few oscillations, they don't do much. 
\subsection{Long-time evolution due to spontaneous emission}
In this section, we analyze Eq.~(\ref{eqn:SI:MeanField}) under the assumption that $\gamma,~\sqrt{\text{Variance}(\delta_k)}\ll \kappa, g_{\text{rms}}\sqrt{N}$ and in the regime where the system would equilibrate to the superradiant steady state if $\gamma,\delta_k$ were $0$. As argued in the main text, the physics at intermediate times can be described in terms of the instantaneous cancellation between the drive and the self-radiated field of the atoms. Initially, this cancellation is almost perfect. However, spontaneous emission starts destroying the atomic coherence, so that the field established by the drive is marginally larger than the field generated by the radiating dipole. This field rotates the Bloch vector upward, seeking equilibration, and this process is repeated.

Because of the almost perfect equilibration, both $\sum_k g_k s_k$ (where $\eta_k=g_k/g_{\text{rms}}$) and $\alpha$ should be almost constant. Furthermore, the intracavity field should be almost $0$, but there will be a small correction because the cancellation between drive and self-radiated field is no longer be perfect. This remnant intracavity field can be obtained by enforcing that $\sum_k g_k s_k$ be constant in its equation of motion (enforcing that $\alpha$ be constant leads to the zeroth order result that $\alpha=0$), leading to
\begin{equation}\label{eqn:SIslowIntraField}
    \alpha=\frac{\gamma}{4ig_{\text{rms}}\sqrt{N}\tilde{J}_z}\underbrace{\sum_k\eta_k\left[1+\frac{2i(\delta_k-\bar{\delta})}{\gamma}\right]s_k}_{R},
\end{equation}
where $\eta_k=g_k/g_{\text{rms}}$ and we have defined $R$ as indicated. This remnant field is small because $\gamma/g_{\text{rms}}\sqrt{N}$ is small and so will induce slow dynamics. Replacing this expression for $\alpha$ in the spin equations of motion yields
\begin{align}
    \begin{split}
        \dot{s}_k&\approx \gamma\left[\frac{R}{2\tilde{J}_z}\eta_k z_k-\frac{s_k}{2}\left(1+\frac{2i(\delta_k-\bar{\delta})}{\gamma}\right)\right]\\
        \dot{z}_k&\approx -\gamma\left[\frac{1}{4i\tilde{J}_z}\eta_k(Rs_k^*-R^* s_k)+z_k+\frac{1}{2}\right]
    \end{split}
\end{align}
These equations conserve $\sum_k\eta_kg_k$, but $\tilde{J}_z$ slowly decreases in magnitude, in accord with our expectations. Depending on the value of $\Omega_d$, the system may reach a steady state that sustains a macroscopic dipole moment. Otherwise, $\tilde{J}_z$ will continue drifting towards $0$. Once this happens, the cavity gets populated by photons and the atoms start Rabi flopping. 

To obtain an estimate of the time needed for $\tilde{J}_z$ to reach 0, we work now with the homogeneous system ($g_k=g_{\text{rms}},\,\delta_k=0$). There is now a single equation for $\tilde{J}_z$
\begin{equation}\label{eqn:SI:LongTimeInversionModel}
    \frac{d\tilde{J}_z}{dt}=-\frac{\gamma N}{8\tilde{J}_z/N}\left(\frac{\Omega_d}{\Omega_c^h}\right)^2-\gamma\left(\tilde{J}_z+\frac{N}{2}\right).
\end{equation}
The two terms in the previous equation can be given a direct physical interpretation. The first one is the Rabi rotation caused by the small  remnant intracavity field given by Eq.~(\ref{eqn:SIslowIntraField}) with $\delta_k=0$ and $g_k=g_{\text{rms}}$. This can be seen more clearly by noting that the intracavity field is calculated self-consistently via
\begin{equation}
    \alpha\approx  \frac{\gamma Ns}{4ig_{\text{rms}}\sqrt{N}\tilde{J}_z}
\end{equation}
and hence the Rabi rotation term on the equation for the $z$ component of the spin is given by
\begin{equation}
    ig_{\text{rms}}\sqrt{N}(\bar{\alpha} s-\alpha \bar{s})= -\frac{\gamma}{2 \tilde{J}_z/N}|s|^2=-\frac{\gamma}{8 \tilde{J}_z/N}\left(\frac{\Omega_d}{\Omega_c^h}\right)^2,
\end{equation}
where we are also replacing $s=-i\Omega_d/(2\Omega_c^h)$ since it is constant. This term pushes $\tilde{J}_z$ towards 0. The second term in Eq.~(\ref{eqn:SI:LongTimeInversionModel}) accounts for the repumping of atoms into the ground state caused by spontaneous emission and pushes $\tilde{J}_z$ towards $-1$. When $\Omega_d<\Omega_c^h/\sqrt{2}$ the competition between these two terms leads to equilibration at a finite value of $\tilde{J}_z$. When $\Omega_c^h/\sqrt{2}<\Omega_d<\Omega_c^h$, then $\tilde{J}_z$ drift slowly towards 0. The time is obtained by straightforward integration of Eq.~(\ref{eqn:SI:LongTimeInversionModel})
\begin{align}\begin{split}
    \gamma T&=-\int_{-\frac{2\tilde{J}_z^{ss}}{N}}^0\frac{x\,dx}{\frac{1}{2}\left(\frac{\Omega_d}{\Omega_c^h}\right)^2+x\left(x+1\right)}\\[5pt]
    &=\frac{1}{2}\log\bigg(\frac{2(\Omega_c^h)^2-\Omega_d^2-2\sqrt{(\Omega_c^h)^2-\Omega_d^2}}{\Omega_d^2}\bigg)\\
    &\hspace{0cm}+\frac{\Omega_c^h}{\sqrt{2\Omega_d^2-(\Omega_c^h)^2}}\Bigg[\arctan\bigg(\frac{2\sqrt{(\Omega_c^h)^2-\Omega_d^2}-\Omega_c^h}{\sqrt{2\Omega_d^2-(\Omega_c^h)^2}}\bigg)+\arctan\bigg(\frac{\Omega_c^h}{\sqrt{2\Omega_d^2-(\Omega_c^h)^2}}\bigg)\Bigg],
\end{split}\end{align}
where $\tilde{J}_z^{ss}$ is the steady state value given by Eq.~(\ref{eqn:SI:tildeSzIdealHomo}). When $\Omega_d\to\Omega_c^{h,se}=\Omega_c^h/\sqrt{2}$, this time diverges as $\gamma T\sim \left(\frac{\Omega_d}{\Omega_c^{h,se}}-1\right)^{-1/2}$. For $\Omega_d\approx 1.06\Omega_c^{h,se}$ we have that $T=4.8\gamma^{-1}$, so the macroscopic dipole and the (almost) zero intracavity field are sustained for longer than a typical spontaneous emission $1/e$ time. For $\Omega_d=1.13\Omega_c^{h,se}$, then $T\approx 2\gamma^{-1}$. In Fig.~\ref{fig:SI:InversionDynamics}(b) We compare the effects of the simple model given by Eq.~(\ref{eqn:SI:LongTimeInversionModel}) (with $\Omega_d=0.75\Omega_c^h=1.06\Omega_c^{h,se}$) and the full evolution given by Eq.~(\ref{eqn:SI:MeanField}) (with $\Omega_d=0.637\Omega_c^h=1.06\Omega_c^{nh,se}$). The differing values of $\Omega_d$ were chosen so that the distance from $\Omega_d$ to the respective critical point is the same when measured in units of their respective critical drives.  The curves match up when $\tilde{J}_z(t=\kappa^{-1})$ in Eq.~(\ref{eqn:SI:LongTimeInversionModel}) is chosen to match the value of $\tilde{J}_z$ obtained after equilibration to the superradiant phase at a time $\kappa^{-1}$.
% \bibliography{refs}

% \nocite{*}

% \bibliographystyle{plain} % We choose the "plain" reference style

% \bibliographystyle{biblatex-phys} %
% \bibliography{refs} % Entries are in the refs.bib file

\end{document}